\documentclass[twoside,twocolumn,9pt]{article}
\usepackage{extsizes}
\usepackage[super,sort&compress,comma]{natbib} 
\usepackage[version=3]{mhchem}
\usepackage[left=1.5cm, right=1.5cm, top=1.785cm, bottom=2.0cm]{geometry}
\usepackage{balance}
\usepackage{mathptmx}
\usepackage{sectsty}
\usepackage{graphicx} 
\usepackage{lastpage}
\usepackage{calc}
\usepackage[format=plain,justification=justified,singlelinecheck=false,font={stretch=1.125,small,sf},labelfont=bf,labelsep=space]{caption}
\usepackage{float}
\usepackage{fancyhdr}
\usepackage{fnpos}
\usepackage[english]{babel}
\addto{\captionsenglish}{%
  
}
\usepackage{array}
\usepackage{droidsans}
\usepackage{charter}
\usepackage[T1]{fontenc}
\usepackage[usenames,dvipsnames]{xcolor}
\usepackage{setspace}
\usepackage[compact]{titlesec}
\usepackage{hyperref}

\usepackage{lineno}

\usepackage{epstopdf}%

\definecolor{cream}{RGB}{222,217,201}

\usepackage[utf8]{inputenc}
\usepackage{xcolor}
\usepackage{subcaption}
\usepackage{natbib}
\usepackage{amssymb,amsmath}
\usepackage{xifthen}
\newcommand{\ie}{\textit{i.e.},~}
\newcommand{\eg}{\textit{e.g.},~}
\usepackage{multicol}

\usepackage{appendix}
\usepackage{lmodern}
\usepackage{multirow}
\usepackage{booktabs}
\usepackage{pdfpages}

\newcommand{\bY}{\Delta_c}
\newcommand{\by}{\delta}
\newcommand{\byc}{\by_c}
\newcommand{\bya}{\by_a}
\newcommand{\bysig}{\by_\sigma}
\newcommand{\tby}{\tilde{\by}}
\newcommand{\tbya}{\tby_a}
\newcommand{\tbysig}{\tby_\sigma}

\newcommand{\bx}{\mathbf{x}}

\newcommand{\bxcs}[2]{\mathbf{x}^{(#2)}_{#1}}

\newcommand{\bxsigi}{\bxcs{\sigma}{i}}
\newcommand{\bxc}{\bxcc}
\newcommand{\bxcc}{\bxcs{c}{s}}
\newcommand{\bxar}{\bxcs{a}{s - g}}
\newcommand{\bxcr}{\bxcs{c}{s - g}}
\newcommand{\bxcg}{\bxcs{c}{g}}
\newcommand{\bxm}{\bxmg}

\newcommand{\bxmg}{\bxcs{m}{g}}

\newcommand{\bE}{\mathbf{E}}
\newcommand{\be}{\mathbf{e}}
\newcommand{\bEc}{\bE_c}
\newcommand{\bEm}{\bE_m}
\newcommand{\bec}{\be_c}
\newcommand{\bem}{\be_m}
\newcommand{\bw}{\mathbf{w}}
\newcommand{\bwm}{\bw_m}
\newcommand{\bwc}{\bw_c}
\newcommand{\berror}{\mathbf{\epsilon}}
\newcommand{\berrorm}{\berror_m}
\newcommand{\berrorc}{\berror_c}

\newcommand{\nall}{3'260~}
\newcommand{\ntrain}{2'707~}
\newcommand{\ntrainmol}{3'242~}
\newcommand{\secref}[1]{Sec.~\ref{#1}}
\newcommand{\figref}[1]{Fig.~\ref{#1}}

\renewcommand{\eqref}[1]{Eq.~(\ref{#1})}

\newcommand{\unitz}{kJ/mol}

\def\benzeneN{3'280}
\def\benzeneMinEnergy{-11.27}
\def\benzeneMaxEnergy{18.75}
\def\benzeneMeanEnergy{-2.19}
\def\benzeneStdEnergy{3.0}
\def\COOHN{1'023}
\def\COOHMinEnergy{-26.72}
\def\COOHMaxEnergy{-1.11}
\def\COOHMeanEnergy{-17.17}
\def\COOHStdEnergy{3.82}
\def\waterN{868}
\def\waterMinEnergy{-42.92}
\def\waterMaxEnergy{4.16}
\def\waterMeanEnergy{-24.65}
\def\waterStdEnergy{9.06}
\def\nitroN{2'129}
\def\nitroMinEnergy{-29.9}
\def\nitroMaxEnergy{1.56}
\def\nitroMeanEnergy{-12.76}
\def\nitroStdEnergy{5.66}
\def\allN{46'010}
\def\allMinN{4}
\def\allMaxN{5313}
\def\allMinNMotif{pentazole}
\def\allMaxNMotif{methyl}

\def\ntest{551~}
\def\ntestmol{628~}
\def\ecstd{4.402~}
\def\emstd{4.251~}
\def\ehstd{1.965~}
\def\ecrmse{1.15~}
\def\emrmse{0.727~}
\def\rcmrmse{0.916~}
\def\rhcrmse{0.778~}
\def\rhmrmse{1.101~}
\def\rhrrmse{0.571~}
\def\rhcmrmse{0.671~}

\usepackage{xparse}

\ExplSyntaxOn
\NewDocumentCommand{\DefineDictionary}{mm}
 {
  \arclupus_dict_def:nn { #1 } { #2 }
 }

\seq_new:N \l__arclupus_dict_temp_seq

\cs_new_protected:Nn \arclupus_dict_def:nn
 {
  \prop_gclear_new:c { g_arclupus_#1_dict_prop }
  \clist_map_inline:nn { #2 }
   {
    \__arclupus_dict_add:nn { #1 } { ##1 }
   }
  \cs_new:cpn { #1 } ##1 { \prop_item:cn { g_arclupus_#1_dict_prop } { ##1 } }
 }

\cs_new_protected:Nn \__arclupus_dict_add:nn
 {
  \seq_set_split:Nnn \l__arclupus_dict_temp_seq { / } { #2 }
  \prop_gput:cxx { g_arclupus_#1_dict_prop }
   {
    \seq_item:Nn \l__arclupus_dict_temp_seq { 1 }
   }
   {
    \seq_item:Nn \l__arclupus_dict_temp_seq { 2 }
   }
 }
\cs_generate_variant:Nn \prop_gput:Nnn { cxx }
\ExplSyntaxOff

\DefineDictionary{var}{a/1, b/2, c/3}
\DefineDictionary{citekey}{XAVZUP/XAVZUP, IDUWEJ/IDUWEJ, UVOMIC/UVOMIC, KEDJUB/KEDJUB, VOHBUR/VOHBUR, LINJUN/LINJUN, KEPDIU/KEPDIU, DUPLUV/DUPLUV, UWUFAV/UWUFAV, IJETOG02/IJETOG02, LAYSOV/LAYSOV, HAMFEJ/HAMFEJ, NTSALA/NTSALA, NSBTOA/NSBTOA, BENZAC19/BENZAC19, NOTVAT01/NOTVAT01, AJEREM/AJEREM, MNETAM01/MNETAM01, XOSBIS/XOSBIS, LEBJUX/LEBJUX, PUYVAG/PUYVAG, LACTOS12/LACTOS12, NUZGOG/NUZGOG, AMTETZ/AMTETZ, GIXDIA/GIXDIA, CAHOAM/CAHOAM, KATZEN/KATZEN, CARCAZ/CARCAZ, TIJKEC/TIJKEC, ATZCBX/ATZCBX, NAPDCX/NAPDCX, QNACRD03/QNACRD03, ZAQFAB/ZAQFAB, COFHOW/COFHOW, TATNBZ03/TATNBZ03, DBANQU/DBANQU, ZZZEEU05/ZZZEEU05, JAYDUI06/JAYDUI06, SOWTIH/SOWTIH, VOFVAN23/VOFVAN23, PHENAN14/PHENAN14, EHAJUS/EHAJUS, LAGMUC/LAGMUC, GAFYES/GAFYES, TETYOH/TETYOH, CYGSGS/CYGSGS, HURYUP/HURYUP, QUCMIL/QUCMIL, ACOWEU/ACOWEU, POWBAF/POWBAF, CIWQIK/CIWQIK, KEVXUG/KEVXUG, YUXPEN01/YUXPEN01, XOGSOD/XOGSOD, IPUMAH/IPUMAH, BOFLUF/BOFLUF, GOHHER/GOHHER, XAZNET/XAZNET, JIGCIL/JIGCIL, LAMPUN/LAMPUN, COWXUI/COWXUI, DTUREA10/DTUREA10, FIBHOP/FIBHOP, HILVOO01/HILVOO01, DTURAC01/DTURAC01, DINICA11/DINICA11, VANCIX/VANCIX, AHIHEF/AHIHEF, LEGSIB/LEGSIB, WAMQUW/WAMQUW, CIFSUH/CIFSUH, SEPNUX/SEPNUX, DADYUE/DADYUE, OMOPAH01/OMOPAH01, RIFJOH/RIFJOH, DOSCUK/DOSCUK, TERTUF/TERTUF, YEHPEG/YEHPEG, CEVZIN/CEVZIN, KUMJOR/KUMJOR, WAVDIJ/WAVDIJ, VATSAK03/VATSAK03, UZOJOK/UZOJOK, FUZMUJ/FUZMUJ, MIYPEP/MIYPEP, TARTAC24/TARTAC24, XARYOE/XARYOE, PAVYES/PAVYES, NISMEH/NISMEH, SEHGUI/SEHGUI, RULRUL/RULRUL, CIQXUV/CIQXUV, ACETSC10/ACETSC10, REYGUZ/REYGUZ, XEXYII/XEXYII, FAFDIA/FAFDIA, GOFCUA/GOFCUA, AHPSUL/AHPSUL, DEXBEN02/DEXBEN02, RIXLUH/RIXLUH, QQQFGS10/QQQFGS10, NTRTAC/NTRTAC, XSHCXB10/XSHCXB10, AZENOJ/AZENOJ, XUSGOI/XUSGOI, IPDXTA/IPDXTA, AZAZEI/AZAZEI, QIMFEX/QIMFEX, XEGPAC/XEGPAC, ULANOK/ULANOK, SAWJUW/SAWJUW, BELBUP01/BELBUP01, ILOFOG/ILOFOG, NAMPIB/NAMPIB, BUVYOI/BUVYOI, BEZOLA/BEZOLA, PIMKOL/PIMKOL, AFUHUG/AFUHUG, DAWSEC/DAWSEC, OCAKAG/OCAKAG, SIHHIA/SIHHIA, NUTMOE/NUTMOE, USAHAY/USAHAY, OPERIL/OPERIL, BAMHEC/BAMHEC, ELAVES/ELAVES, GABOBU10/GABOBU10, KOVHIM01/KOVHIM01, OROMIT/OROMIT, NILYAI/NILYAI, HIVWUF/HIVWUF, SIMCOH/SIMCOH, CEDMUV/CEDMUV, EJEVIA/EJEVIA, XUVXIW/XUVXIW, DHNAPH17/DHNAPH17, FUMAAC01/FUMAAC01, YAFFUJ/YAFFUJ, TAFZOQ/TAFZOQ, WUTSEJ/WUTSEJ, MEGDOS/MEGDOS, RATBIZ/RATBIZ, XUDTIZ/XUDTIZ, WACKAM/WACKAM, PUYTOS/PUYTOS, NOPQIT/NOPQIT, YUKXEH/YUKXEH, MOKQIM/MOKQIM, LEVKAY/LEVKAY, EVUMOX/EVUMOX, NABREN/NABREN, XALDUM/XALDUM, LURZUT/LURZUT, CIWYOY/CIWYOY, VIWRAV/VIWRAV, XIPVEY/XIPVEY, LASPRT03/LASPRT03, AZAMIY/AZAMIY, NEHMES01/NEHMES01, FEPYEG/FEPYEG, ARUCEW/ARUCEW, YIWCUC/YIWCUC, WIFKEB/WIFKEB, BUKVAE/BUKVAE, LIVJEG/LIVJEG, HAXWEI/HAXWEI, ABASIG/ABASIG, NOCMEX/NOCMEX, EMOXAF/EMOXAF, GPHXAM/GPHXAM, LUBBUF/LUBBUF, RAGTIF/RAGTIF, TUXFIC/TUXFIC, AQUHAX/AQUHAX, GUHROQ/GUHROQ, DMTTDO/DMTTDO, AZOTEP/AZOTEP, VIYTUU/VIYTUU, WAXNUH/WAXNUH, DUPDAU/DUPDAU, PEMVOS/PEMVOS, XACMOF/XACMOF, TIRJIN/TIRJIN, CAFVUA/CAFVUA, KATREF/KATREF, IFAYEV/IFAYEV, KIVDAV/KIVDAV, DUTTAN10/DUTTAN10, NAVPIL/NAVPIL, ZZZWGK01/ZZZWGK01, JEXLUT/JEXLUT, KOMZIW/KOMZIW, MIDYAA/MIDYAA, FIBYEW/FIBYEW, WATFED/WATFED, GIPVUW03/GIPVUW03, EFAGIB01/EFAGIB01, NODLIC/NODLIC, IQOMIM13/IQOMIM13, ZILKAG01/ZILKAG01, LUNYID/LUNYID, OHOLOO/OHOLOO, PAHZOO/PAHZOO, BAHSUY/BAHSUY, TICMEZ/TICMEZ, SAFGIP/SAFGIP, APANBZ/APANBZ, PUHJAE/PUHJAE, TTDSHD10/TTDSHD10, BEBLEA01/BEBLEA01, VANSUB/VANSUB, GADPEG01/GADPEG01, AXALUH/AXALUH, LIHZUY/LIHZUY, IMAMOX/IMAMOX, GAQWAX/GAQWAX, TANPEE/TANPEE, HUYYAB01/HUYYAB01, TEBHAJ/TEBHAJ, KETVEK03/KETVEK03, QOPKUB/QOPKUB, SECKEQ/SECKEQ, MOJBOD/MOJBOD, YANVUG/YANVUG, SUHDOO/SUHDOO, EKATOB/EKATOB, CEHQEM/CEHQEM, DACXIO/DACXIO, FUTLEN/FUTLEN, TOSVIG/TOSVIG, TUBYIY/TUBYIY, PAMJAO/PAMJAO, VELLEE01/VELLEE01, CPHTRI/CPHTRI, EKUKEC/EKUKEC, SADCUV/SADCUV, VELVUF/VELVUF, DURYAS/DURYAS, UROSIF/UROSIF, ATOMAZ/ATOMAZ, COXXUJ/COXXUJ, LEZRIR/LEZRIR, ZAPDUS/ZAPDUS, RHODIN01/RHODIN01, ZUMKAT/ZUMKAT, GOMXIR/GOMXIR, XANYOB/XANYOB, INAQOD/INAQOD, DAZFLU11/DAZFLU11, QAHVUQ/QAHVUQ, NALCYS11/NALCYS11, ICOPOG/ICOPOG, HODXOP/HODXOP, BEHBOG/BEHBOG, FIDSAN/FIDSAN, CTMTNA06/CTMTNA06, TEGSEE/TEGSEE, JOCLOE/JOCLOE, NUFDOJ/NUFDOJ, GINXIL/GINXIL, IQEYUZ/IQEYUZ, NTRGUA01/NTRGUA01, QOXPEZ/QOXPEZ, VINYLC/VINYLC, TMEASO01/TMEASO01, XESHOT/XESHOT, CIPMAP/CIPMAP, AHEYAO/AHEYAO, LOHQAA/LOHQAA, MCINDE/MCINDE, HEYQUY/HEYQUY, JAYVAG01/JAYVAG01, JINYIP/JINYIP, QAZGIK/QAZGIK, DTBIUR/DTBIUR, FORROU/FORROU, COVJON/COVJON, IZECUL/IZECUL, OXUSUX02/OXUSUX02, YIVKAQ/YIVKAQ, JUKWIV/JUKWIV, DOTTUB/DOTTUB, TAXSOA/TAXSOA, XIHKUU/XIHKUU, SCCHRN03/SCCHRN03, JARRAW/JARRAW, HMALAC01/HMALAC01, THYMMH/THYMMH, ACOTHI/ACOTHI, OCIDUZ/OCIDUZ, PDHTEZ/PDHTEZ, TUDKUZ/TUDKUZ, LOHVEK/LOHVEK, ESEDOW/ESEDOW, RUHCUS/RUHCUS, CECWAK/CECWAK, KUYPAV01/KUYPAV01, BULCUG/BULCUG, VIMRIS01/VIMRIS01, ROMBOL/ROMBOL, UDOZOD/UDOZOD, TZBZTZ/TZBZTZ, FURFON/FURFON, MMCATH01/MMCATH01, AWELOE01/AWELOE01, PHURNC/PHURNC, AMMALA/AMMALA, IHEPIU/IHEPIU, CEYDUF/CEYDUF, MIQKOM01/MIQKOM01, CALVAM/CALVAM, TOLDAM01/TOLDAM01, VENLOP/VENLOP, GEFTIW/GEFTIW, IXEGIB/IXEGIB, MNTDTH/MNTDTH, ROKBEY/ROKBEY, WADKAP/WADKAP, PIFZAG01/PIFZAG01, RIJNON/RIJNON, QEFLAO/QEFLAO, SOQNOC/SOQNOC, BODSAO/BODSAO, REKFET/REKFET, FIXPUX/FIXPUX, DTHPIM/DTHPIM, JOJWOU03/JOJWOU03, VUZHEE/VUZHEE, TEZPDZ10/TEZPDZ10, PUYTEI/PUYTEI, ZERRET01/ZERRET01, GAVWIM/GAVWIM, JOWWIB/JOWWIB, FELRUJ/FELRUJ, BACJEX/BACJEX, XEZBAG/XEZBAG, CYTOSM12/CYTOSM12, YILWEV/YILWEV, HIHZUT/HIHZUT, ODEDOR/ODEDOR, UTUSUZ/UTUSUZ, IKALAJ01/IKALAJ01, AZOBEN01/AZOBEN01, UJIWEQ01/UJIWEQ01, REVJEJ/REVJEJ, TAYWEV/TAYWEV, YAWVAV/YAWVAV, YAMXUF/YAMXUF, JUDZIR/JUDZIR, ZZZRJG01/ZZZRJG01, XICVOV/XICVOV, QIJYEN/QIJYEN, MSORAM10/MSORAM10, HEXTIN/HEXTIN, LITREM/LITREM, KICCOO/KICCOO, JAJROE/JAJROE, HASHAK/HASHAK, JEXZUH/JEXZUH, TNBENZ12/TNBENZ12, CACFIW/CACFIW, JAJGAD/JAJGAD, PUYWEL/PUYWEL, PINXAL/PINXAL, FOCLAM/FOCLAM, ZIPSUO/ZIPSUO, SIFQON/SIFQON, TISGUZ/TISGUZ, RECBEI/RECBEI, LUSNIW/LUSNIW, UTUVAI/UTUVAI, RIQHOO/RIQHOO, XIQCEF/XIQCEF, AMPENA01/AMPENA01, RAHXOP/RAHXOP, KIBFEG/KIBFEG, WADXOQ/WADXOQ, WOVYEL/WOVYEL, LUXPAW/LUXPAW, PEJVOQ/PEJVOQ, FANRAO/FANRAO, XAVMIQ/XAVMIQ, JIPKAV/JIPKAV, LABXES/LABXES, GOHVUU/GOHVUU, MIRKAZ/MIRKAZ, OTAKEB/OTAKEB, HZDCET05/HZDCET05, KEFYAW/KEFYAW, TEPHTH14/TEPHTH14, QOCNAX01/QOCNAX01, YAYBUX/YAYBUX, COFTIA10/COFTIA10, EMOHIY/EMOHIY, EKECAY02/EKECAY02, BESZUU/BESZUU, JIWJEF/JIWJEF, TIHJEA/TIHJEA, DUMQAF/DUMQAF, EVAWUU01/EVAWUU01, PEMBEO/PEMBEO, SADBII/SADBII, XODFON/XODFON, KEKWOM/KEKWOM, QUDNOU/QUDNOU, SOFTEM/SOFTEM, XAFVEH/XAFVEH, MNIMET04/MNIMET04, AYAHUE/AYAHUE, ZIJPIT/ZIJPIT, LUTXOO/LUTXOO, WOBYOD/WOBYOD, THZTAZ/THZTAZ, MTMBCO/MTMBCO, LADLOT/LADLOT, ICPDAZ/ICPDAZ, BUFCAI/BUFCAI, BAZGOY05/BAZGOY05, ULAWAF06/ULAWAF06, AWIWUZ01/AWIWUZ01, RIHJIB/RIHJIB, TEDQAU/TEDQAU, NESTIP/NESTIP, TEKKAV01/TEKKAV01, MUXGIX/MUXGIX, BACXAE10/BACXAE10, BOZGED/BOZGED, CUNTOU/CUNTOU, HIFWOI/HIFWOI, LEFQIY/LEFQIY, YASSIV/YASSIV, MAYXEP/MAYXEP, PYRZIN20/PYRZIN20, REHHER/REHHER, GAKPEN/GAKPEN, GAZGOF/GAZGOF, BIDCEW/BIDCEW, UTAKUX/UTAKUX, GEMMAN/GEMMAN, VEHXAH/VEHXAH, NEXSIU/NEXSIU, SUNFOW/SUNFOW, QETKOP01/QETKOP01, OBUWOA/OBUWOA, DAJXIV/DAJXIV, AMTRAZ/AMTRAZ, DACDEQ10/DACDEQ10, RUKWAV/RUKWAV, ZASWAT/ZASWAT, EPUKOR/EPUKOR, COJMUJ/COJMUJ, CUBTEA/CUBTEA, MIXTCA/MIXTCA, JAKKEL/JAKKEL, YITKAP/YITKAP, VIKKEG/VIKKEG, IMAZOL04/IMAZOL04, WOBTUE/WOBTUE, DUCWEE/DUCWEE, MOQDED/MOQDED, RESRUC/RESRUC, LEPMID/LEPMID, VELWUE/VELWUE, ZIJGOQ/ZIJGOQ, GAPVUO/GAPVUO, SENGAT/SENGAT, TURLAT/TURLAT, KEQPUS/KEQPUS, SIXYAA/SIXYAA, KAJHIM/KAJHIM, VABNAN/VABNAN, POTLEP/POTLEP, UPONEU/UPONEU, IZOWOK/IZOWOK, ROYVUY/ROYVUY, JEMGIS/JEMGIS, YAQLOS/YAQLOS, XUBLEN/XUBLEN, RIWSAT/RIWSAT, SUSDAL/SUSDAL, VOMYED/VOMYED, TELWOW/TELWOW, PENCEN/PENCEN, KECXIZ/KECXIZ, YIKQOY/YIKQOY, YOBBEW/YOBBEW, IHANUB/IHANUB, WICTUX01/WICTUX01, REKBUE01/REKBUE01, WAKNUS/WAKNUS, FUTMUE/FUTMUE, HYNPHD/HYNPHD, PORXEA/PORXEA, UFUXOI/UFUXOI, LCYSTN24/LCYSTN24, HAGLUZ/HAGLUZ, HERHUI/HERHUI, GACLEE/GACLEE, QAGRIC/QAGRIC, EXUQAP/EXUQAP, TMTHOY/TMTHOY, DIOXAZ10/DIOXAZ10, WEZCAG/WEZCAG, DIVRIJ/DIVRIJ, BIXNEB10/BIXNEB10, MTKETZ/MTKETZ, JAZFOF01/JAZFOF01, PAMPAX/PAMPAX, DMEGLY03/DMEGLY03, NALCYS26/NALCYS26, GUMMUW04/GUMMUW04, DUFBOV11/DUFBOV11, TEDROL/TEDROL, BIVTUV01/BIVTUV01, AMTETA/AMTETA, NIKVUZ/NIKVUZ, BAWVUQ01/BAWVUQ01, TRIRED/TRIRED, TRITAN10/TRITAN10, LUHSIR/LUHSIR, KIKYIM/KIKYIM, FORMAC01/FORMAC01, AJIGAB/AJIGAB, DUROQN10/DUROQN10, FORAMO01/FORAMO01, DMOCDO10/DMOCDO10, TAPZAN/TAPZAN, DIHIXL10/DIHIXL10, REGTAA/REGTAA, FEPGEM01/FEPGEM01, PIWDIJ/PIWDIJ, LEFLUE/LEFLUE, ARCLAM02/ARCLAM02, VICREE/VICREE, DIKETE11/DIKETE11, MEUREA/MEUREA, EMIMUI/EMIMUI, CYURAC13/CYURAC13, ACETAC09/ACETAC09, PRUVAC/PRUVAC, CIZSEL/CIZSEL, NIPYAZ/NIPYAZ, SEYJAH/SEYJAH, TAQNEH/TAQNEH, IMTAZO01/IMTAZO01, DUNVEN/DUNVEN, UHUMEP01/UHUMEP01, MUYHIY/MUYHIY, HUVGIQ/HUVGIQ, GLYCIN35/GLYCIN35, MAZGAX/MAZGAX, NOACTD/NOACTD, MAYDIB/MAYDIB, FUNSOW/FUNSOW, GUZWII/GUZWII, QAGDEK/QAGDEK, CYHEXO/CYHEXO, ADARAZ/ADARAZ, HOQHAW/HOQHAW, BUSFUQ/BUSFUQ, URACIL/URACIL, NENZOX/NENZOX, HABTEJ/HABTEJ, GEJVOH/GEJVOH, HIZTEQ/HIZTEQ, AKOVOL01/AKOVOL01, RUXMED/RUXMED, FOHPOI/FOHPOI, ZEQVIC/ZEQVIC, YEJSUD/YEJSUD, WOLNOC/WOLNOC, GLYCIN81/GLYCIN81, QOXVEE/QOXVEE, LIVSEQ/LIVSEQ, PAMPAU/PAMPAU, XOSJUK/XOSJUK, TADMES/TADMES, XEXXAZ/XEXXAZ, GEFBAV/GEFBAV, BDTOLE13/BDTOLE13, RAZYID/RAZYID, MELAMI01/MELAMI01, EZOPEP/EZOPEP, YAPMIM/YAPMIM, IMAZOL33/IMAZOL33, BOQQUT08/BOQQUT08, FIFHUX01/FIFHUX01, DUFMIA/DUFMIA, FACYUE/FACYUE, WEVVOJ/WEVVOJ, SOYJAR/SOYJAR, SALOXM06/SALOXM06, GEKXOJ/GEKXOJ, ZIKMOV01/ZIKMOV01, HIXHIF02/HIXHIF02, PHGLOL01/PHGLOL01, AXOSOW03/AXOSOW03, THIEPO10/THIEPO10, BAPCUS/BAPCUS, TRMTRZ/TRMTRZ, DIFJAF/DIFJAF, SEDTUQ09/SEDTUQ09, NAKWUR/NAKWUR, MEPCHX/MEPCHX, OGEGIR/OGEGIR, DOCNIS/DOCNIS, ZEYBAG/ZEYBAG, RIZZEH/RIZZEH, TOAZCD/TOAZCD, GIPPAW/GIPPAW, MUHFOK/MUHFOK, IPELOF/IPELOF, MAMGEN/MAMGEN, AXOSEN/AXOSEN, FADYIV01/FADYIV01, POVLAP/POVLAP, ANIMUH/ANIMUH, VIPKIO03/VIPKIO03, YEJQOT/YEJQOT, KAVCAL/KAVCAL, DZTNON02/DZTNON02, ZOGPAJ/ZOGPAJ, HATYEG02/HATYEG02, WIPVEW/WIPVEW, DMDFHZ/DMDFHZ, HAZQUV/HAZQUV, BARBAC01/BARBAC01, VEDTAZ/VEDTAZ, DZCDON/DZCDON, JOHBUD01/JOHBUD01, YARZUN03/YARZUN03, BAKQUA/BAKQUA, CUTTOB/CUTTOB, GOYLOV/GOYLOV, VASDUO/VASDUO, GAJTEQ/GAJTEQ, OMAXOQ/OMAXOQ, LAKDAD/LAKDAD, QEJPUQ/QEJPUQ, UCONIJ/UCONIJ, YIHHON05/YIHHON05, CBUDCX/CBUDCX, BASBON/BASBON, CINMER04/CINMER04, XAGWIM02/XAGWIM02, QUFPAK/QUFPAK, FUDYOS01/FUDYOS01, BEMYEY/BEMYEY, RUWRAC/RUWRAC, LEGZII/LEGZII, EVOQEL/EVOQEL, HYXBUR10/HYXBUR10, EFUMAU02/EFUMAU02, UHIVEM/UHIVEM, XODLEJ/XODLEJ, DALGON03/DALGON03, XEMMUZ01/XEMMUZ01, LIXGOO/LIXGOO, JUXHUF/JUXHUF, AMTZTZ/AMTZTZ, XOQHER/XOQHER, XOJXOL06/XOJXOL06, TEKQAB01/TEKQAB01, EKAVIX/EKAVIX, FOXWAR/FOXWAR, XEZFIT/XEZFIT, CAGJEC/CAGJEC, VAYCAB/VAYCAB, MEJXUV/MEJXUV, CSURCD10/CSURCD10, PIMELA07/PIMELA07, YAMFOI/YAMFOI, HMHOCN04/HMHOCN04, DZHPDO/DZHPDO, AMXPYZ/AMXPYZ, DOKGUG01/DOKGUG01, TUPWEI/TUPWEI, FUZYUU/FUZYUU, BISFUG/BISFUG, AHXGLP/AHXGLP, BIRGEO/BIRGEO, CREATH07/CREATH07, TADNOC/TADNOC, BELASS/BELASS, VIOLME05/VIOLME05, FORRUB/FORRUB, QIHHIY/QIHHIY, BABBUB/BABBUB, KAGNUB/KAGNUB, PIXRET/PIXRET, PEBWOI/PEBWOI, VENREN/VENREN, QEGSOL/QEGSOL, ROHBUL/ROHBUL, PEVKUX/PEVKUX, TCIMDT20/TCIMDT20, BAZYAC01/BAZYAC01, RUNJIV/RUNJIV, DIMEDO03/DIMEDO03, REDUCA/REDUCA, NANQUO02/NANQUO02, FOBSEU/FOBSEU, AHOMAM/AHOMAM, WECCAK/WECCAK, DLMAND01/DLMAND01, SECDAI/SECDAI, RATDEV/RATDEV, IMIWOQ/IMIWOQ, YODHIJ01/YODHIJ01, BEXGAM/BEXGAM, UTETIX/UTETIX, KUDZAL/KUDZAL, TOVTED/TOVTED, ZARJEJ/ZARJEJ, EADPEE/EADPEE, NIJHUJ/NIJHUJ, ACXTPY/ACXTPY, WOSWAC/WOSWAC, YITBOS/YITBOS, ZZZGEO03/ZZZGEO03, DODNOZ/DODNOZ, ZEFXIR01/ZEFXIR01, AZALIX/AZALIX, UGEBOZ/UGEBOZ, XOZBUK/XOZBUK, SIVNAM/SIVNAM, UFIYUD/UFIYUD, DAFLUO/DAFLUO, DLSERN26/DLSERN26, IRETEE/IRETEE, NUVXIM02/NUVXIM02, CICROV/CICROV, MOAZCD/MOAZCD, QUSMEW/QUSMEW, COXYAR01/COXYAR01, KOVHAE01/KOVHAE01, JEMFUD/JEMFUD, PYRCOX01/PYRCOX01, YELZOF/YELZOF, VIYMOH/VIYMOH, HXNIPA10/HXNIPA10, ISAJUI/ISAJUI, LULRUG/LULRUG, MOVLUG/MOVLUG, QIKDAP/QIKDAP, DELMOX/DELMOX, HAMFEH/HAMFEH, TICHUI04/TICHUI04, AXUDED01/AXUDED01, ZUTKEG/ZUTKEG, LAQDUE/LAQDUE, BIBREL/BIBREL, MAPCIQ/MAPCIQ, JECYUM/JECYUM, JACPOS01/JACPOS01, FEMCEF/FEMCEF, MENOGU/MENOGU, GUKVIT/GUKVIT, ZUHQOI/ZUHQOI, UDAYUT01/UDAYUT01, RUMNAQ/RUMNAQ, DOTNON/DOTNON, HMPYTH/HMPYTH, AGESEK/AGESEK, WUHYIH/WUHYIH, HAJMOU/HAJMOU, QAZFOP/QAZFOP, NIPQUA/NIPQUA, SURHIW/SURHIW, LATZAH/LATZAH, PYMCOX/PYMCOX, KEDXEZ/KEDXEZ, WAKFEU/WAKFEU, SALOXM/SALOXM, LAGRAM/LAGRAM, DETSBR01/DETSBR01, JIXGUS/JIXGUS, HABFUL/HABFUL, YUHTIE/YUHTIE, HYQUIN/HYQUIN, XEDDAM/XEDDAM, RUWKEZ/RUWKEZ, JICLAI/JICLAI, IVAGUH/IVAGUH, ZUGHEO/ZUGHEO, SUCCIN/SUCCIN, BIFYIY/BIFYIY, EJITAU/EJITAU, LIBVIC/LIBVIC, MEHTAZ/MEHTAZ, TUQDAM/TUQDAM, LEFYUS/LEFYUS, ADOJEK/ADOJEK, TAURIN09/TAURIN09, PYRGAL03/PYRGAL03, NEMWUX01/NEMWUX01, GAQSEY/GAQSEY, GLURAC07/GLURAC07, FATQOJ/FATQOJ, DUJHEV/DUJHEV, TIBLUM/TIBLUM, NPHDZB/NPHDZB, VEDLOG/VEDLOG, YORLAT/YORLAT, SAXHEH/SAXHEH, HACMUT/HACMUT, JALZOL/JALZOL, LALNIN28/LALNIN28, CUGMUO/CUGMUO, FEGHAA11/FEGHAA11, JIPFAQ/JIPFAQ, PUFFOL01/PUFFOL01, TITLUE/TITLUE, QEBCUY/QEBCUY, MAMPOL02/MAMPOL02, ZZZKPE06/ZZZKPE06, FOCZUT/FOCZUT, MMXPDZ10/MMXPDZ10, ZENPAJ/ZENPAJ, KUTCOS/KUTCOS, EQICEN/EQICEN, VEKGOH/VEKGOH, NEDNOZ01/NEDNOZ01, LOZPAR/LOZPAR, AYUZEZ/AYUZEZ, BEPNIT01/BEPNIT01, LUFQOS/LUFQOS, TAHFUD/TAHFUD, ERIVAC/ERIVAC, TROPOL10/TROPOL10, SENMAB/SENMAB, FIGMAJ/FIGMAJ, DOBYOJ/DOBYOJ, XAGDOY/XAGDOY, TBFRAC01/TBFRAC01, AYIFET/AYIFET, VEXREV/VEXREV, VAVVET/VAVVET, DUJQOP/DUJQOP, SOXBMB/SOXBMB, DODGEI/DODGEI, ACRLAC/ACRLAC, FIJKEP/FIJKEP, DUCYOQ/DUCYOQ, JEMJUG/JEMJUG, OVEQAJ/OVEQAJ, FUYREW/FUYREW, UYIVAB/UYIVAB, GOTSOX/GOTSOX, EACLEZ/EACLEZ, REWBOL/REWBOL, NOJVOX/NOJVOX, SURKOF/SURKOF, UKIWAN/UKIWAN, BIRTUR/BIRTUR, LIWQAJ01/LIWQAJ01, CETQOI/CETQOI, SEGDUF/SEGDUF, OMOKUW/OMOKUW, NALXOP/NALXOP, XAQNUZ/XAQNUZ, SALYOT/SALYOT, MIYFEF/MIYFEF, COYMUY/COYMUY, HAVKOE/HAVKOE, VUWSUB/VUWSUB, EQUKAE/EQUKAE, DUMHIC/DUMHIC, ECPDTC/ECPDTC, ICRFRA10/ICRFRA10, DAXPEZ01/DAXPEZ01, NIMGAR/NIMGAR, DIPICA10/DIPICA10, DUWRIW/DUWRIW, DEVRUR/DEVRUR, UMUZOR/UMUZOR, MSULIN10/MSULIN10, YEKSOX/YEKSOX, EKAQOY/EKAQOY, SUQDEN/SUQDEN, AMMPRA01/AMMPRA01, CACGUI/CACGUI, KADBAU/KADBAU, SOJCUP/SOJCUP, HMUSCM/HMUSCM, MEHFIP/MEHFIP, NINZUF/NINZUF, ATDZDX/ATDZDX, PELKEY/PELKEY, VUYVES/VUYVES, BUVCIF/BUVCIF, FEPNAP/FEPNAP, WAXPIW/WAXPIW, BOGRUL/BOGRUL, QIWJEM/QIWJEM, NOCREC/NOCREC, UJUKOZ/UJUKOZ, MFCBXA/MFCBXA, WONFUB/WONFUB, YAHQAB/YAHQAB, UGECOA01/UGECOA01, HALMAJ/HALMAJ, MFZCHZ/MFZCHZ, MDTBVL/MDTBVL, FADTIQ/FADTIQ, NIRBAS/NIRBAS, ZATDUW/ZATDUW, WOLGOV/WOLGOV, KABLOQ/KABLOQ, GALDEC02/GALDEC02, ULAVEK/ULAVEK, DAVRIE/DAVRIE, BERMYA/BERMYA, TPCYPR10/TPCYPR10, DOKYAD/DOKYAD, ADIPAM11/ADIPAM11, PAPVAD/PAPVAD, ZZZDRQ01/ZZZDRQ01, SURDOX/SURDOX, OHILIC/OHILIC, UDIXIP/UDIXIP, MOZSEZ/MOZSEZ, FEMMOA/FEMMOA, UCOTEM/UCOTEM, ISOHON/ISOHON, PEPHUN09/PEPHUN09, NALYOQ/NALYOQ, GOKNOK/GOKNOK, WUFTUO/WUFTUO, GUDTAC/GUDTAC, JOYHOU/JOYHOU, SALCAN05/SALCAN05, ADPTHZ/ADPTHZ, FEXKOI/FEXKOI, TABBEE/TABBEE, HEYJOK01/HEYJOK01, COQWIO/COQWIO, ROGTOY/ROGTOY, CIGBAV/CIGBAV, JAHPOX/JAHPOX, HOZPET01/HOZPET01, FIHNIV/FIHNIV, BECJEY/BECJEY, FUSVAQ01/FUSVAQ01, ABAQIG/ABAQIG, JEVPIJ/JEVPIJ, XOSCAK/XOSCAK, MACTUH/MACTUH, SIGCAM/SIGCAM, QUSLAR/QUSLAR, GUFJAS/GUFJAS, NAVZAO/NAVZAO, VIRGOS01/VIRGOS01, WEHYAK/WEHYAK, WIBCAL/WIBCAL, MEUREA01/MEUREA01, WAGMOH01/WAGMOH01, SUZPEI/SUZPEI, LSERMH16/LSERMH16, THXMAM02/THXMAM02, DPTHDO/DPTHDO, HTIEPO/HTIEPO, WURMOM01/WURMOM01, WOCQEK/WOCQEK, XUMXEI/XUMXEI, HERMEX/HERMEX, JOKVOW/JOKVOW, THIPAC01/THIPAC01, NACXEU01/NACXEU01, VAZCEG/VAZCEG, REBCUX/REBCUX, GLULAD/GLULAD, TARKAZ/TARKAZ, LEGLEQ/LEGLEQ, APAJIL/APAJIL, UCOJUS/UCOJUS, WOWFOD/WOWFOD, ZABKIZ/ZABKIZ, LEWDUN/LEWDUN, PICANO04/PICANO04, TAZZEA/TAZZEA, YOTSUV/YOTSUV, QEFBUY01/QEFBUY01, EQATUN/EQATUN, IFOFUE/IFOFUE, KAMLIT/KAMLIT, WINZUO/WINZUO, NATTAD/NATTAD, LAGTOD/LAGTOD, OKUXUP/OKUXUP, GADZAN/GADZAN, LEGHAH/LEGHAH, FOACAM/FOACAM, TXBNON/TXBNON, FIMWOO/FIMWOO, DZOCTO/DZOCTO, HOQCOF/HOQCOF, DOVCAT/DOVCAT, ANNULE01/ANNULE01, ISUSIY/ISUSIY, RABVIC/RABVIC, GONQIL/GONQIL, FOLXIN/FOLXIN, ABULIT03/ABULIT03, JOSXUK/JOSXUK, APEGOS/APEGOS, BIBVAK/BIBVAK, ADMANN20/ADMANN20, LINMOK/LINMOK, VABJOX10/VABJOX10, KUKTEQ/KUKTEQ, SUJQIZ/SUJQIZ, YIJVES02/YIJVES02, PUQJOC/PUQJOC, TABRET/TABRET, XARHII/XARHII, ZOSPOM/ZOSPOM, MOPPUC01/MOPPUC01, XURCIW/XURCIW, DAFJOK/DAFJOK, YODRUF/YODRUF, PUXDAN/PUXDAN, HAKNAJ/HAKNAJ, FUJTUZ/FUJTUZ, LISTAJ/LISTAJ, YOVJUP01/YOVJUP01, WUZROZ01/WUZROZ01, SEGCAI01/SEGCAI01, EVARID/EVARID, WANVEP/WANVEP, EMITEZ/EMITEZ, LIKLAU/LIKLAU, ANQTDZ/ANQTDZ, FULPIM01/FULPIM01, LACFON01/LACFON01, PELJOG/PELJOG, COZYOG/COZYOG, CINSOH/CINSOH, AVULIM/AVULIM, BAHMAB/BAHMAB, SUFWIB/SUFWIB, GAFMAB/GAFMAB, NOCPAW/NOCPAW, IJARES/IJARES, FORTAH/FORTAH, QOYJOD06/QOYJOD06, CDECOL11/CDECOL11, NEKZEI/NEKZEI, ELIMUJ/ELIMUJ, QECYOO/QECYOO, KONDEW/KONDEW, RESKIK/RESKIK, PAXMOS/PAXMOS, KEDWUL/KEDWUL, SAFKAL/SAFKAL, IZANOM01/IZANOM01, DAHHUS/DAHHUS, DAYDEP/DAYDEP, LIYSAN/LIYSAN, QAYJAE/QAYJAE, ZIYGUL/ZIYGUL, EPSULF/EPSULF, IGEDON/IGEDON, KULMUA/KULMUA, DZBASK/DZBASK, CASHOT02/CASHOT02, HANJOX/HANJOX, SETHEE/SETHEE, TAJWAC/TAJWAC, ATTDAZ01/ATTDAZ01, NIPGAW/NIPGAW, YEPXAS/YEPXAS, NESZAN/NESZAN, XANHYD/XANHYD, RHXFUR/RHXFUR, WOLVUP/WOLVUP, JAXYEN/JAXYEN, PAZRUF/PAZRUF, GAKQEP/GAKQEP, FEPWAY/FEPWAY, MUTVEE/MUTVEE, BAZVEE/BAZVEE, JORBUN/JORBUN, EMUFAU/EMUFAU, FEJYIC/FEJYIC, TEKVIO/TEKVIO, ITIFAT/ITIFAT, WANSAG02/WANSAG02, CHONDR/CHONDR, GISDIX/GISDIX, ACIDOH/ACIDOH, LIJXOR/LIJXOR, HUXJOZ/HUXJOZ, UXILUK/UXILUK, NEHQAU/NEHQAU, NAXDOF/NAXDOF, DOMDUE/DOMDUE, QEYFUW/QEYFUW, AYAMAP/AYAMAP, BEMFUU/BEMFUU, CEJWOD/CEJWOD, TOLUEN10/TOLUEN10, FORMAM03/FORMAM03, LESFAS/LESFAS, XAZTID/XAZTID, XESTUM/XESTUM, AZASER11/AZASER11, IWOMAJ/IWOMAJ, QORPOD/QORPOD, HAMHUZ/HAMHUZ, TAZOLD02/TAZOLD02, FEZHAV/FEZHAV, FINTAZ02/FINTAZ02, JUVKAM/JUVKAM, AYOZAR01/AYOZAR01, XURYOY/XURYOY, DOHGOW/DOHGOW, KUVKER/KUVKER, PUVVUX/PUVVUX, REZMOY/REZMOY, MOQCAX/MOQCAX, KEPMAT/KEPMAT, AMHPYR/AMHPYR, CEGKII/CEGKII, FAYPAX/FAYPAX, GEZQOR/GEZQOR, ACICAR/ACICAR, ERERAU/ERERAU, RUVXUB/RUVXUB, JEZZUJ/JEZZUJ, YOGCAZ/YOGCAZ, GIXLOQ/GIXLOQ, BAWHOW/BAWHOW, WUBJAG/WUBJAG, NIXPUF/NIXPUF, MALNAC02/MALNAC02, IZAWUB/IZAWUB, WOGFAZ/WOGFAZ, ZIVMIA/ZIVMIA, TSCRBZ20/TSCRBZ20, VOPTEZ/VOPTEZ, XAZSID/XAZSID, HOHLAR/HOHLAR, CECPIK/CECPIK, GATYEG/GATYEG, ADPRLA/ADPRLA, SOBMAY01/SOBMAY01, FEDKIJ/FEDKIJ, POVKEQ/POVKEQ, WAZDEG/WAZDEG, PEQFOG/PEQFOG, DEMRAO/DEMRAO, KEMDOW/KEMDOW, FEZLAX/FEZLAX, EMASIV/EMASIV, TOSQUN/TOSQUN, EJEHEG/EJEHEG, TIHJOK/TIHJOK, TOBRUX01/TOBRUX01, HIYNUY/HIYNUY, KUQTEX/KUQTEX, BAGQIJ01/BAGQIJ01, NBONAN10/NBONAN10, GACGEY/GACGEY, VUHFIP/VUHFIP, NTPYRO03/NTPYRO03, SICZUZ/SICZUZ, CPCOHA/CPCOHA, GOFPUN02/GOFPUN02, SAJMOH01/SAJMOH01, PHGLYA/PHGLYA, EFAMII/EFAMII, UQALUU/UQALUU, WIPKUB01/WIPKUB01, KEPKEV/KEPKEV, HYTBPH/HYTBPH, DMSCPY/DMSCPY, EVIQAB/EVIQAB, DOXHPX/DOXHPX, OVOSID/OVOSID, PODFIZ/PODFIZ, TEDQUO/TEDQUO, MYTOLD01/MYTOLD01, SOKGAA/SOKGAA, RICXAC/RICXAC, ZOZYUI/ZOZYUI, MOSYUO/MOSYUO, USUMOL/USUMOL, FEFDOM/FEFDOM, DETYLE/DETYLE, MELWUW/MELWUW, WEDSOO/WEDSOO, DUYRIY/DUYRIY, VAZMEP/VAZMEP, DEMQIW01/DEMQIW01, IXANOK/IXANOK, HIRXAI/HIRXAI, MALEHY01/MALEHY01, ESILID02/ESILID02, RIGLAU/RIGLAU, NUTYOQ/NUTYOQ, THACEM02/THACEM02, COXZEU/COXZEU, HIWJIG01/HIWJIG01, BARBAC03/BARBAC03, ULEZES/ULEZES, MOLJED/MOLJED, DTSUCC/DTSUCC, MUKLIO/MUKLIO, PIHVIL/PIHVIL, XANAZH01/XANAZH01, FOGZEG/FOGZEG, DPHSOX02/DPHSOX02, LIJFUG/LIJFUG, TETZOL02/TETZOL02, RIKWAJ/RIKWAJ, AYAYOP/AYAYOP, FERSUR/FERSUR, LAGHOQ/LAGHOQ, KOPYIX/KOPYIX, VEFMEZ/VEFMEZ, PABNAJ/PABNAJ, FEDNUY/FEDNUY, VEBTIG/VEBTIG, WOKPAO/WOKPAO, METHOL04/METHOL04, ZDGLPN/ZDGLPN, IDOQUO/IDOQUO, COVSEM/COVSEM, FIGYID01/FIGYID01, YUYCOK/YUYCOK, METACM02/METACM02, KUBCOA/KUBCOA, MHXBEN/MHXBEN, DAQNIT/DAQNIT, GAWNAW01/GAWNAW01, XIXQOL/XIXQOL, XALTUZ/XALTUZ, BAWHUD/BAWHUD, DOXDAN/DOXDAN, JISVEM01/JISVEM01, RIWJEM/RIWJEM, SEQCOG/SEQCOG, YUCNER/YUCNER, NAYNUX/NAYNUX, QADRET/QADRET, COXZAS04/COXZAS04, HIVGAW/HIVGAW, XULHIV/XULHIV, JEGYAW/JEGYAW, YIGTOY/YIGTOY, COVVAK/COVVAK, FUNGUS/FUNGUS, RIWQET01/RIWQET01, ODEKIR/ODEKIR, WIKBAT/WIKBAT, GOGXUV/GOGXUV, HXOXAM10/HXOXAM10, JAZQIN/JAZQIN, ADOFEF/ADOFEF, METTOP10/METTOP10, QIHPUS/QIHPUS, TIJLAZ/TIJLAZ, BEYZIO01/BEYZIO01, GESNUN/GESNUN, DAWCIQ/DAWCIQ, ARAGUV/ARAGUV, NEBPUF/NEBPUF, UVALIN01/UVALIN01, JAYDUJ/JAYDUJ, REBZAZ/REBZAZ, ZAKMIH02/ZAKMIH02, MASKUP/MASKUP, QQQFBP01/QQQFBP01, UZEWIH/UZEWIH, VEBQIF/VEBQIF, DIALAC02/DIALAC02, BUKTAC/BUKTAC, DIHPEQ/DIHPEQ, PAJYOR/PAJYOR, HXANTH10/HXANTH10, AFIHON/AFIHON, PEYZUQ/PEYZUQ, DTHPDT/DTHPDT, DUZZAZ/DUZZAZ, ZENMUA/ZENMUA, QUIDON/QUIDON, CIGYIA/CIGYIA, QOZFIV/QOZFIV, RIVKAJ/RIVKAJ, THIOUR08/THIOUR08, NSCYHX/NSCYHX, NUMKOW/NUMKOW, XEYMIX/XEYMIX, IFAPIP/IFAPIP, LANXAB/LANXAB, KEMCAH/KEMCAH, KUHMEF/KUHMEF, YANREM01/YANREM01, BECVEN/BECVEN, GIRZIR/GIRZIR, MTHPHE01/MTHPHE01, DOMQUS/DOMQUS, ILOLOL/ILOLOL, BIJSAO/BIJSAO, QUGGAA/QUGGAA, FONMOL/FONMOL, XAYMEP/XAYMEP, FITZAJ/FITZAJ, YOKROF/YOKROF, GIXFID/GIXFID, DEDWER/DEDWER, BOVNEF/BOVNEF, ZUHQUQ/ZUHQUQ, MOXCIL/MOXCIL, MOPPAJ/MOPPAJ, FELCIK/FELCIK, MOAOSP10/MOAOSP10, NUXGOC/NUXGOC, LIWCUP/LIWCUP, ACACOX/ACACOX, KIGYOP/KIGYOP, OHIKOH/OHIKOH, CAZDOX/CAZDOX, TICYOT/TICYOT, OPUSIC/OPUSIC, LIGWIH/LIGWIH, TAYHUX/TAYHUX, HOQZEU/HOQZEU, PYMDON/PYMDON, AMOXAC/AMOXAC, ALXANM10/ALXANM10, CUBNOC/CUBNOC, EFONAP/EFONAP, SEGCOW/SEGCOW, NADNIP/NADNIP, CAKNOU/CAKNOU, SOJNAG02/SOJNAG02, FABPIH10/FABPIH10, HEVLUP/HEVLUP, FEFXUM/FEFXUM, RAYLAF/RAYLAF, EXIHAV/EXIHAV, OXAZIL10/OXAZIL10, FUGJUM01/FUGJUM01, HACGAT/HACGAT, YESNIT/YESNIT, SEBJOY/SEBJOY, KAKHEL/KAKHEL, HEHQOC/HEHQOC, OGIWAD/OGIWAD, XUHREX/XUHREX, ETYTAC/ETYTAC, LAMHUD/LAMHUD, LUGPAF/LUGPAF, ROJQEO/ROJQEO, AFUTDZ10/AFUTDZ10, NUPBIL/NUPBIL, QOBHAS/QOBHAS, METMSX/METMSX, FASGUB/FASGUB, XOLQUL/XOLQUL, RECZII/RECZII, WIDJOI/WIDJOI, QAHCUA/QAHCUA, IPMUDS/IPMUDS, PYRIDO04/PYRIDO04, XIZRII/XIZRII, QEDWUR/QEDWUR, AFESUA/AFESUA, QOHCIA/QOHCIA, WOQPEY/WOQPEY, SAPDAQ/SAPDAQ, OLOBIA/OLOBIA, KUPHIM01/KUPHIM01, OREHOK/OREHOK, SAQZOA/SAQZOA, ATUWAN/ATUWAN, MEPYRZ/MEPYRZ, GANHUY/GANHUY, WADSOL/WADSOL, CYAMPD01/CYAMPD01, EDAWIP/EDAWIP, TCTDTO20/TCTDTO20, KICPOB/KICPOB, RUVSEG/RUVSEG, ILOCUH/ILOCUH, ADALAU/ADALAU, ZAJJEZ/ZAJJEZ, NOTPET/NOTPET, DILNUH/DILNUH, ZOSRUU/ZOSRUU, UVIZEG/UVIZEG, ALOVUS/ALOVUS, MOLQAH/MOLQAH, OPISIQ/OPISIQ, RASHOL/RASHOL, TIYLEU/TIYLEU, UPAQIM/UPAQIM, GAMNEO/GAMNEO, KAPCUZ/KAPCUZ, ZENQDX/ZENQDX, MENNAV/MENNAV, XYLTOL03/XYLTOL03, BOVBAP/BOVBAP, DIVKOK/DIVKOK, SANQII/SANQII, URMALN/URMALN, UCULIN/UCULIN, LEBDUR/LEBDUR, KOXRIY/KOXRIY, SOCGUM/SOCGUM, HODLES/HODLES, TAZCAZ/TAZCAZ, KUNXEW/KUNXEW, NITPOV/NITPOV, XDHURC/XDHURC, ETANOL/ETANOL, ITOYIA/ITOYIA, DIZVOZ/DIZVOZ, HMDNOB/HMDNOB, OGUREN/OGUREN, HOMGOG/HOMGOG, DOXXAP/DOXXAP, BECYEP/BECYEP, ISEDOB/ISEDOB, FURDCB01/FURDCB01, WOGHAD/WOGHAD, CEGXES/CEGXES, CEKVIX/CEKVIX, ZIJMEM/ZIJMEM, JURQAO/JURQAO, CASJIS/CASJIS, MEWSUE/MEWSUE, EXUNEQ/EXUNEQ, LOLJAX/LOLJAX, KOMKEE/KOMKEE, PYZTAZ/PYZTAZ, HIJFIP/HIJFIP, ETAJOY/ETAJOY, PUSYIM/PUSYIM, SEFZIM/SEFZIM, QOWFUF/QOWFUF, DAHFIC/DAHFIC, YALQEK/YALQEK, QOTWUR/QOTWUR, VALIDL03/VALIDL03, RAPDOD/RAPDOD, POVMES/POVMES, KERWIO/KERWIO, IMIPID/IMIPID, KUVVOO/KUVVOO, AJAPIL04/AJAPIL04, PTZCNB/PTZCNB, QEDROG/QEDROG, RESHAY/RESHAY, COXYUJ/COXYUJ, THCYTO10/THCYTO10, AMIHAH/AMIHAH, JAXSOR/JAXSOR, VOPTOL/VOPTOL, DHBFZO/DHBFZO, FOWVIX/FOWVIX, WAXCUV/WAXCUV, GUGJOH/GUGJOH, WABFAG/WABFAG, YAXJIS/YAXJIS, LIPWEM01/LIPWEM01, CAKWUH/CAKWUH, DNBENZ11/DNBENZ11, AFEMIH/AFEMIH, QARQUY/QARQUY, YEGXOX10/YEGXOX10, HONSEI/HONSEI, NUVYIM/NUVYIM, CAJWUG01/CAJWUG01, JULZAR01/JULZAR01, NRURAM11/NRURAM11, HOKSOR/HOKSOR, RHODHY/RHODHY, BOPXOT/BOPXOT, XIMQUG/XIMQUG, DUROQN11/DUROQN11, POHZUJ/POHZUJ, AMSMPI/AMSMPI, WAHBOX/WAHBOX, ZZZEYQ01/ZZZEYQ01, CANXIA/CANXIA, SETVES10/SETVES10, RALDEQ/RALDEQ, PRONAC02/PRONAC02, YAFMOI01/YAFMOI01, JEVRAE/JEVRAE, SIKJIF/SIKJIF, JABMAD/JABMAD, RIZFAI01/RIZFAI01, PAGPOF/PAGPOF, AFIFUR/AFIFUR, CIJMEO/CIJMEO, MOCTUU/MOCTUU, KUVBOT/KUVBOT, VEFMOI/VEFMOI, BCBDNT/BCBDNT, RAMNID/RAMNID, DAFWOZ/DAFWOZ, PANVUV/PANVUV, YEBHAQ/YEBHAQ, NAGHOT03/NAGHOT03, HORSEO/HORSEO, HICCIF/HICCIF, OJOLAC/OJOLAC, COHWIF02/COHWIF02, ROFYER/ROFYER, YUZTET/YUZTET, XOZJAZ/XOZJAZ, KAVYIP/KAVYIP, CEJJIM/CEJJIM, HAWSOO/HAWSOO, AVIMEY/AVIMEY, CAWRAU/CAWRAU, MAZADN/MAZADN, MUDZER/MUDZER, SOWJET/SOWJET, VALSEG02/VALSEG02, QULWOL/QULWOL, MUKJIL/MUKJIL, QILZIU/QILZIU, GUMCOI/GUMCOI, BUGKOE/BUGKOE, AMITDZ/AMITDZ, NISGEB/NISGEB, ZEVLAN/ZEVLAN, GLYCIN18/GLYCIN18, MIOZPO01/MIOZPO01, RAVNUX/RAVNUX, XIBTUY/XIBTUY, QOZFOB/QOZFOB, FEBKAZ/FEBKAZ, COWSUC/COWSUC, NEVYUI/NEVYUI, KANSIC/KANSIC, LICROE/LICROE, METAMI/METAMI, WEKXAL/WEKXAL, FUGYAI/FUGYAI, HEZHOL/HEZHOL, YIRYEE01/YIRYEE01, FOHMUK/FOHMUK, IYAQOP01/IYAQOP01, FEWBOA/FEWBOA, ZOXHID/ZOXHID, HAMGOT/HAMGOT, PUSZOU/PUSZOU, SIXYEE/SIXYEE, NMHXNT/NMHXNT, OZALUY/OZALUY, NUJHIK/NUJHIK, YUQDEU/YUQDEU, VEZCAE/VEZCAE, DEBSOS/DEBSOS, QATTIO/QATTIO, CERMOB/CERMOB, AMTHTZ/AMTHTZ, NAXDUL01/NAXDUL01, GAVLUK/GAVLUK, GAQZOP/GAQZOP, MESCZB/MESCZB, RUNGAJ/RUNGAJ, IXAQAB/IXAQAB, KODXEI/KODXEI, MAKDIL/MAKDIL, FIGNEO/FIGNEO, QUKSOF/QUKSOF, HTHSOC/HTHSOC, HEKWAV01/HEKWAV01, WUDJOU/WUDJOU, MENPDL/MENPDL, ETCYPY01/ETCYPY01, SEGBUB/SEGBUB, DAZTUW/DAZTUW, SABMAL/SABMAL, DSCARA01/DSCARA01, ACEOXM01/ACEOXM01, GIPVOQ/GIPVOQ, BALNIN04/BALNIN04, QICGAM/QICGAM, DOKQIE/DOKQIE, JOPBAT/JOPBAT, DAYHIV/DAYHIV, VELNOR/VELNOR, DIGDIG01/DIGDIG01, ASIYOR/ASIYOR, VOZSEI/VOZSEI, XOVFEV/XOVFEV, GECLUX/GECLUX, YINDUU/YINDUU, ASAPUH/ASAPUH, THHYDT/THHYDT, MUVPUO/MUVPUO, GUMMUW/GUMMUW, BIXJAU/BIXJAU, WACJIT/WACJIT, MULPEO/MULPEO, IPIXUB10/IPIXUB10, XOCJEE/XOCJEE, MEGLOZ/MEGLOZ, TMENBZ02/TMENBZ02, VAMRAE/VAMRAE, DHXANT11/DHXANT11, PEDERL/PEDERL, EVOPOV/EVOPOV, WIFPAC01/WIFPAC01, WAQNUY/WAQNUY, VOFDEZ/VOFDEZ, ITIPEG/ITIPEG, ERABIJ/ERABIJ, COPGUK/COPGUK, DILCUY/DILCUY, LIMFUI/LIMFUI, BUDPIB/BUDPIB, VOHJEH/VOHJEH, FUFBAJ/FUFBAJ, XAYYUS/XAYYUS, REHYIM/REHYIM, YEFXAJ/YEFXAJ, PANGOB/PANGOB, ATZTHD10/ATZTHD10, FEMFUY01/FEMFUY01, SICTOO/SICTOO, CUZWEA/CUZWEA, AKEZIB05/AKEZIB05, XEFSIK/XEFSIK, MPYDCX/MPYDCX, PSULHZ11/PSULHZ11, JIHVEB/JIHVEB, XEDTOP/XEDTOP, CRPACX10/CRPACX10, ELATUG01/ELATUG01, UDEGUH/UDEGUH, KEPKIZ/KEPKIZ, MUCHUP/MUCHUP, CUGBIP/CUGBIP, FACVAH/FACVAH, IZILIO/IZILIO, IYAWAG/IYAWAG, FOWTER/FOWTER, YUHNOG/YUHNOG, TBDOTH/TBDOTH, FITBAM/FITBAM, VABFEL/VABFEL, HICHUY/HICHUY, TSCARB01/TSCARB01, ANISIC04/ANISIC04, QIPVEQ/QIPVEQ, NILCER/NILCER, SARCEU/SARCEU, DMTHTN/DMTHTN, QODQIK/QODQIK, WAKXAI/WAKXAI, QIQCUQ/QIQCUQ, GIPKUL/GIPKUL, VEQCIE/VEQCIE, ZZZUEU04/ZZZUEU04, UXUKUV/UXUKUV, GEJCUU/GEJCUU, MEZDUS/MEZDUS, QAHBAF/QAHBAF, SITQET/SITQET, IJUMEG03/IJUMEG03, WAHYIP/WAHYIP, DMTRZT/DMTRZT, REJTUW01/REJTUW01, ACETAC01/ACETAC01, SAZCEC/SAZCEC, ODEDAC/ODEDAC, CEJSEP/CEJSEP, PEWMIN/PEWMIN, CULPOO01/CULPOO01, DHMTTX10/DHMTTX10, TUPVAB01/TUPVAB01, NURWOM03/NURWOM03, LAPVIK/LAPVIK, YOKCIK01/YOKCIK01, QIVCUU/QIVCUU, AMPYRE01/AMPYRE01, VUCBAW/VUCBAW, BIJJIP/BIJJIP, MALIAC12/MALIAC12, VOPJEP/VOPJEP, TRAZOL04/TRAZOL04, DAYQEZ/DAYQEZ, KIJPOI/KIJPOI, BOHLIT01/BOHLIT01, QAFFIM01/QAFFIM01, QACPOA/QACPOA, DOGPAQ/DOGPAQ, OWONUL/OWONUL, COROSE10/COROSE10, ZUHKIW/ZUHKIW, HINDEN/HINDEN, DIPMUK/DIPMUK, OQIXET/OQIXET, XOCQIQ/XOCQIQ, NOGUNA01/NOGUNA01, IRILEB/IRILEB, DNNEPH10/DNNEPH10, CUJVOS10/CUJVOS10, QEPLON/QEPLON, ODIKAO/ODIKAO, XZBTZO/XZBTZO, CIDROW/CIDROW, GUKCUM/GUKCUM, EGAKII/EGAKII, FEFKUY/FEFKUY, PUBLUT/PUBLUT, FUSJOT/FUSJOT, KABJAA/KABJAA, DAGRAF/DAGRAF, PILQUX/PILQUX, HXQUIO03/HXQUIO03, TRPHAM/TRPHAM, TEGRUV/TEGRUV, HTDZDX10/HTDZDX10, ZOYCOH/ZOYCOH, HOSHEC/HOSHEC, QELLOI/QELLOI, DIWCIV/DIWCIV, MTDZDT/MTDZDT, FUBLOE/FUBLOE, DAPSUO16/DAPSUO16, OWEKEH/OWEKEH, BOXKEE/BOXKEE, RUPBAG/RUPBAG, NITPOL09/NITPOL09, BZCBUO/BZCBUO, GIMLUL/GIMLUL, PUVSOO/PUVSOO, VEQVAP/VEQVAP, VECXIL01/VECXIL01, OTANAC/OTANAC, LEJPAS/LEJPAS, POKNUY/POKNUY, JUBBUD/JUBBUD, FADBOB/FADBOB, EYIFEZ/EYIFEZ, KOFKAR/KOFKAR, COWNEJ/COWNEJ, GAZHAT/GAZHAT, EVIXIQ/EVIXIQ, CADSUV/CADSUV, PEXZOI01/PEXZOI01, FIQCUD/FIQCUD, VIXRID/VIXRID, HIQVAE/HIQVAE, MIZSAR01/MIZSAR01, CIHWUL/CIHWUL, YIGJOP/YIGJOP, NAVZIT/NAVZIT, ZILQOA01/ZILQOA01, CEXQIG/CEXQIG, QUVFIX/QUVFIX, VANXIS01/VANXIS01, EWIXIT02/EWIXIT02, OGAHEJ/OGAHEJ, LEQVUY02/LEQVUY02, QERTUD/QERTUD, NIVJIM/NIVJIM, UDECAK/UDECAK, ADEJIC/ADEJIC, GEBVIU/GEBVIU, HOJLIC/HOJLIC, FEYDUJ/FEYDUJ, PARDAO/PARDAO, OMIWIQ/OMIWIQ, EFOQOG/EFOQOG, LIKYOV/LIKYOV, COKBEJ/COKBEJ, EPAHEJ/EPAHEJ, DMTCUN10/DMTCUN10, DALGON/DALGON, MUKGAA/MUKGAA, IQIXIP/IQIXIP, FALHEI/FALHEI, MAMHEO/MAMHEO, UVOKAT/UVOKAT, PYTCBM/PYTCBM, QOXMOG/QOXMOG, GLUTAM09/GLUTAM09, GOMMUQ/GOMMUQ, BETYIK/BETYIK, NAZTEN/NAZTEN, KOYLIT/KOYLIT, BOXGEA/BOXGEA, PYRTHA10/PYRTHA10, EHUJIB/EHUJIB, URICAC/URICAC, ZAJGUM/ZAJGUM, WATHUU/WATHUU, NUZMEA/NUZMEA, ADEZUF/ADEZUF, NETROT/NETROT, AHGLPY13/AHGLPY13, FOHXEF/FOHXEF, GLICAC01/GLICAC01, TAUCYM01/TAUCYM01, DAYQOK/DAYQOK, THCHYD/THCHYD, FUPGII/FUPGII, WOSGIU/WOSGIU, NTRACD/NTRACD, UBAYON/UBAYON, HMITZC/HMITZC, QEJKOF/QEJKOF, GLYGLY18/GLYGLY18, PYRAZI06/PYRAZI06, SUMKIU/SUMKIU, XOGBEB/XOGBEB, ZAVTEV/ZAVTEV, QIBGUD/QIBGUD, LAJZAZ/LAJZAZ, NERJOK/NERJOK, SEQBUN/SEQBUN, OSIXEV/OSIXEV, HAXPAX/HAXPAX, KEQJEV/KEQJEV, PAWCOI/PAWCOI, JOJBIT/JOJBIT, VECSAX/VECSAX, THGUAN10/THGUAN10, ZOJSEY/ZOJSEY, BZAMID06/BZAMID06, BUWCAX/BUWCAX, EHAXOB/EHAXOB, KADPOU02/KADPOU02, VAXNEP/VAXNEP, BYZMTS10/BYZMTS10, LINCIU/LINCIU, APAKUZ/APAKUZ, FOVVAN/FOVVAN, HARGOY/HARGOY, XEGPII/XEGPII, NUMHIN/NUMHIN, VIZCEO/VIZCEO, NOFBIT/NOFBIT, LIWPOY/LIWPOY, OXAZDO/OXAZDO, EFURIH04/EFURIH04, CEYGIW/CEYGIW, FAMTZT10/FAMTZT10, SAHDIQ/SAHDIQ, GIHJEO/GIHJEO, CIXWIP/CIXWIP, WOPJUI/WOPJUI, SOGUAN20/SOGUAN20, WUKJUH/WUKJUH, PARBAC04/PARBAC04, NIMFOE02/NIMFOE02, COUMAR11/COUMAR11, HIFBOP/HIFBOP, DHURAC10/DHURAC10, ROGSOW/ROGSOW, SEZJEN/SEZJEN, GUDVUY/GUDVUY, JAJWUM/JAJWUM, VUWDUO/VUWDUO, PADQAN/PADQAN, NIPPEJ/NIPPEJ, UQABUL/UQABUL, NURQEW/NURQEW, NUGSUE/NUGSUE, SOSFIQ/SOSFIQ, BUDPOG/BUDPOG, LUYKOF01/LUYKOF01, THBZPS10/THBZPS10, WERTIW/WERTIW, DOCMEO/DOCMEO, SINMEI/SINMEI, LANXON/LANXON, EBUSIG/EBUSIG, CPENAD/CPENAD, GIPLOG/GIPLOG, TAJNIB/TAJNIB, TRIZIN06/TRIZIN06, CIWVAH01/CIWVAH01, YOWMAZ/YOWMAZ, NEXJUW/NEXJUW, WOKGIM/WOKGIM, WAPVOY/WAPVOY, CUVWEV/CUVWEV, BAKFUO/BAKFUO, QURYUZ01/QURYUZ01, HEXFUL/HEXFUL, TUDQIS/TUDQIS, AMPTOP10/AMPTOP10, REQQUA/REQQUA, PAVCUM/PAVCUM, FEGFUV/FEGFUV, DTHDOM/DTHDOM, CURJUV/CURJUV, EWUKIQ01/EWUKIQ01, YUFPOE01/YUFPOE01, HEFTOE/HEFTOE, KEHYAZ/KEHYAZ, UQIXUO01/UQIXUO01, BUXWIA/BUXWIA, USIWEZ/USIWEZ, OKESAY/OKESAY, RAHNOG/RAHNOG, LARZEM/LARZEM, KAWQEF/KAWQEF, COVXAN01/COVXAN01, XAYQOD/XAYQOD, HOWBAW/HOWBAW, SOQGAI/SOQGAI, TMETEY/TMETEY, MENFXN/MENFXN, EZUYIJ/EZUYIJ, IGISIA01/IGISIA01, EJIQEU/EJIQEU, VEXNIV/VEXNIV, TOMBAA/TOMBAA, EDASEH/EDASEH, AJUXEI/AJUXEI, ITUBEF/ITUBEF, CAKREM/CAKREM, HEMLAM/HEMLAM, VIGPEG/VIGPEG, JAKPER/JAKPER, IPUVIA/IPUVIA, FUTFIJ/FUTFIJ, AJUGOC/AJUGOC, CXMCYS/CXMCYS, BZDMAZ07/BZDMAZ07, JATFUF02/JATFUF02, MEMJEU/MEMJEU, SAVRIT/SAVRIT, FOJRAX/FOJRAX, HEWQOQ/HEWQOQ, PILQUW/PILQUW, BAGMEC/BAGMEC, IHUFAS/IHUFAS, TOQMUH01/TOQMUH01, KIBGEH/KIBGEH, PATXAJ/PATXAJ, NIGNOH/NIGNOH, EBIROX/EBIROX, NIFFEN/NIFFEN, VEBPEX/VEBPEX, DESJOA/DESJOA, UZAWOJ/UZAWOJ, JIZFAZ/JIZFAZ, BASVOG/BASVOG, IJULII/IJULII, REGLUL/REGLUL, KOKVUD/KOKVUD, YAHMEB/YAHMEB, TRZTHO/TRZTHO, TCUNDO/TCUNDO, LIBWUQ/LIBWUQ, KEKHEO01/KEKHEO01, CAGNUW/CAGNUW, ULEWIR/ULEWIR, WUVBAS/WUVBAS, PAPKOG/PAPKOG, REWHOR/REWHOR, ACRLAC04/ACRLAC04, IHOZUA/IHOZUA, CAFMIF/CAFMIF, ESIWUZ/ESIWUZ, TACMSO/TACMSO, MPIMSA/MPIMSA, WUMDIS/WUMDIS, LSERIN21/LSERIN21, PUYROQ/PUYROQ, BONGES/BONGES, IYIMUY/IYIMUY, FAGQOV/FAGQOV, FIFSIY/FIFSIY, COSLOL/COSLOL, XUPFUJ/XUPFUJ, MTHPYO/MTHPYO, PUBMUU23/PUBMUU23, RUNSIE/RUNSIE, PIRFIF/PIRFIF, HEQZEK/HEQZEK, NUYJEW02/NUYJEW02, SETFUS/SETFUS, NUJVOE/NUJVOE, SETRIS/SETRIS, DOSBAO/DOSBAO, GISSAD/GISSAD, KAGBID/KAGBID, MNHCHA/MNHCHA, YUGVUT/YUGVUT, AHENAC/AHENAC, NUZHUL/NUZHUL, TURACD/TURACD, XILMUA/XILMUA, IWUGIQ01/IWUGIQ01, XAYVIE/XAYVIE, REFTAX/REFTAX, TOCYHX/TOCYHX, ILIVAC/ILIVAC, VALDIV/VALDIV, ECIHAZ/ECIHAZ, AFEXUF/AFEXUF, LOBHIV/LOBHIV, TANTAD/TANTAD, DIBGEA01/DIBGEA01, QACXUP/QACXUP, EPEDEK/EPEDEK, OXTPTZ/OXTPTZ, BABXUA/BABXUA, TECNUL/TECNUL, TADMUJ/TADMUJ, DICHEC/DICHEC, IXUVAY/IXUVAY, OQOZOK/OQOZOK, FODLER/FODLER, XAZKIU/XAZKIU, HOXKUA/HOXKUA, GUVBAC/GUVBAC, BZDMAZ29/BZDMAZ29, GEBMUV/GEBMUV, KIKVAC01/KIKVAC01, IMICOW/IMICOW, MACNHS/MACNHS, BOSPUU/BOSPUU, VIGTUA/VIGTUA, FITXIP/FITXIP, FORIZT10/FORIZT10, TEHMUO/TEHMUO, ZEXKEU/ZEXKEU, LIJQAW/LIJQAW, DATYII/DATYII, LIJJIY/LIJJIY, TIBGUG/TIBGUG, KUQCAB/KUQCAB, MEDBUS/MEDBUS, PHOXTZ/PHOXTZ, YUGFAJ/YUGFAJ, VEDVUY/VEDVUY, HEGSUK/HEGSUK, CUYXAV/CUYXAV, BEBWEK/BEBWEK, QULBIJ/QULBIJ, YUWMUY/YUWMUY, MNPYDO25/MNPYDO25, PHTHAO01/PHTHAO01, DACLOL/DACLOL, REHVUV/REHVUV, OMEQUS/OMEQUS, DEDCUN01/DEDCUN01, MAJTIB/MAJTIB, NAXTIQ/NAXTIQ, MUGNUX/MUGNUX, XOPGEQ/XOPGEQ, GOVKEH/GOVKEH, BACYEJ/BACYEJ, NUNPOB/NUNPOB, WOCVUF/WOCVUF, OSOKUD/OSOKUD, GOCHAI/GOCHAI, SEYDIJ01/SEYDIJ01, ZORHUJ/ZORHUJ, WEMTIS/WEMTIS, IQUJEK/IQUJEK, MUDDEU/MUDDEU, FENWIG/FENWIG, OZAHEE/OZAHEE, TECPOG/TECPOG, GAMGIL/GAMGIL, GUKVOX/GUKVOX, FOQKIF/FOQKIF, FACREG/FACREG, TZANDT/TZANDT, RAJDEM/RAJDEM, GAVRAX/GAVRAX, NILPIJ/NILPIJ, JOPDEX01/JOPDEX01, FUDYUZ/FUDYUZ, DIFDUS/DIFDUS, VIDVEL/VIDVEL, NABKIN/NABKIN, AFIXUK/AFIXUK, ETIMQO01/ETIMQO01, CASYIG/CASYIG, JARMIB/JARMIB, SITRIY/SITRIY, KINFES/KINFES, RIBVUU/RIBVUU, PIHCAK/PIHCAK, COSNII/COSNII, RIYZEF/RIYZEF, AKEQEN/AKEQEN, TRZPUR/TRZPUR, RAMVUW/RAMVUW, SUNDEK/SUNDEK, TEMDOG/TEMDOG, HOFGUF/HOFGUF, IFIJOW/IFIJOW, SSOXAM02/SSOXAM02, AVIBEN/AVIBEN, REGYEH/REGYEH, ZIYNUS/ZIYNUS, CACPUS/CACPUS, RELLAV/RELLAV, EKEKOU/EKEKOU, WOWFUL/WOWFUL, DUTYUN/DUTYUN, BAQSUK/BAQSUK, CALRAI/CALRAI, LORMOV/LORMOV, LDOPAS03/LDOPAS03, REVYAT/REVYAT, DUPNOT/DUPNOT, LACJEF/LACJEF, ACOWOE/ACOWOE, SIKKOO/SIKKOO, OYURUX/OYURUX, VABLUF/VABLUF, HICJIO01/HICJIO01, ZIVRON/ZIVRON, YOJDOS/YOJDOS, JOFHOB/JOFHOB, NUXTAC01/NUXTAC01, HIRGUL/HIRGUL, LIKJUL/LIKJUL, EFOZAB03/EFOZAB03, TOTGIU/TOTGIU, DOXSEQ/DOXSEQ, DIVSOQ01/DIVSOQ01, IBPRAC15/IBPRAC15, XOQBEK/XOQBEK, AYUBUS/AYUBUS, YIHTUG/YIHTUG, POWBAE/POWBAE, TASVAL/TASVAL, BTPETS/BTPETS, PYZPYT01/PYZPYT01, TEJROP/TEJROP, GISKEA/GISKEA, LILLEY/LILLEY, PAGYUS/PAGYUS, MEXCUP/MEXCUP, DIYJAW/DIYJAW, FUXWIF/FUXWIF, SUVWAJ/SUVWAJ, FIYNIM/FIYNIM, QEGBUZ/QEGBUZ, DUFBOV10/DUFBOV10, KIZDAA/KIZDAA, JEVNEE/JEVNEE, PITZAT/PITZAT, HXBNZM/HXBNZM, LECFUV/LECFUV, CABCUD25/CABCUD25, SINZIY/SINZIY, BATWOK/BATWOK, JUPPAL/JUPPAL, TAYMOY/TAYMOY, KODTIG01/KODTIG01, HETAUR10/HETAUR10, DAQFEG01/DAQFEG01, YANSIR/YANSIR, IDOREY/IDOREY, FAJGUS/FAJGUS, RAMGUH/RAMGUH, HTMXTC10/HTMXTC10, DAGTUA/DAGTUA, DEHNOW/DEHNOW, DAMHEE/DAMHEE, PIKYEN/PIKYEN, YIDBAO/YIDBAO, FEZRAD/FEZRAD, POBQIG/POBQIG, NAHMUE/NAHMUE, ABINOS01/ABINOS01, KAMFAH/KAMFAH, FOGVIG01/FOGVIG01, GAJCOM/GAJCOM, NTROMA01/NTROMA01, WUVXUG/WUVXUG, RUWFIY/RUWFIY, QOTWAX/QOTWAX, COKJER/COKJER, TILJOP/TILJOP, QUJTOE/QUJTOE, IJUCIB04/IJUCIB04, CEZTAD/CEZTAD, BAZKEV01/BAZKEV01, AKIHIN/AKIHIN, PEJBAJ/PEJBAJ, LERGOG/LERGOG, FADSAF/FADSAF, WITROI/WITROI, FABDOE/FABDOE, GIJCAD/GIJCAD, DOXYOE/DOXYOE, ESAXED/ESAXED, FOJMOG01/FOJMOG01, ZIPRUN/ZIPRUN, OXHEPA03/OXHEPA03, GUHZEQ/GUHZEQ, MILYIP/MILYIP, SALWAD/SALWAD, NUFCOH/NUFCOH, FENCIK/FENCIK, ATDZSA03/ATDZSA03, TAFLET/TAFLET, BIPPEV/BIPPEV, COSXIT/COSXIT, ABIRUZ/ABIRUZ, LERWOU/LERWOU, COGHUC/COGHUC, IXEBAQ/IXEBAQ, BUKNEA/BUKNEA, DOGQUL/DOGQUL, YOFROB/YOFROB, INICAC03/INICAC03, SAZGOP/SAZGOP, VALWOW/VALWOW, ZACSII/ZACSII, YESGAF/YESGAF, TALQAB/TALQAB, MOKPUX/MOKPUX, KEQSOQ/KEQSOQ, KISFEY/KISFEY, PACCEB/PACCEB, ALYTUR/ALYTUR, VOMMOZ/VOMMOZ, LAHMOX/LAHMOX, GEHPUE/GEHPUE, IFIKAK/IFIKAK, RESVAN/RESVAN, WOGMIQ/WOGMIQ, EXAROK/EXAROK, IYUCOU/IYUCOU, FRANAC02/FRANAC02, KALXOK/KALXOK, FOWFII/FOWFII, CEYLUN/CEYLUN, MEDVID/MEDVID, BAHPAC/BAHPAC, KOXBOO/KOXBOO, BUYZAW/BUYZAW, KOYMES/KOYMES, VADJAM/VADJAM, NEDQES/NEDQES, XAPQUD/XAPQUD, CAFVOW/CAFVOW, OXTTCD/OXTTCD, OYOJES/OYOJES, GUTCIH/GUTCIH, SUWBUH/SUWBUH, QEYCOM/QEYCOM, UPAWEP/UPAWEP, LONJAA/LONJAA, UFOFAW/UFOFAW, OGEQIB/OGEQIB, TAZPYR/TAZPYR, AMAPAH/AMAPAH, TIBMOG/TIBMOG, ROKPAK/ROKPAK, MEFNIW/MEFNIW, FIBXUL/FIBXUL, NOBLOG/NOBLOG, KABWUF/KABWUF, FORTCH/FORTCH, FIZJON01/FIZJON01, DMMSSI10/DMMSSI10, DNCOOC/DNCOOC, BOWGAV/BOWGAV, BUFMOG/BUFMOG, NOZKES/NOZKES, BUFGAK/BUFGAK, IVEZEO/IVEZEO, JORLOS01/JORLOS01, WUJQEY/WUJQEY, HIWHAY/HIWHAY, FIZDIB/FIZDIB, CBOHAZ05/CBOHAZ05, LEVNEF/LEVNEF, OBEZEC/OBEZEC, HXIBAC/HXIBAC, PUVVUZ/PUVVUZ, LECZOL/LECZOL, VEZHOY/VEZHOY, ASALUD/ASALUD, CIGMUB/CIGMUB, ETTHUR04/ETTHUR04, SOVZAE/SOVZAE, FUSFIJ/FUSFIJ, MLEICA/MLEICA, HARNEV/HARNEV, VALFEW/VALFEW, KATVEJ/KATVEJ, CILWUP11/CILWUP11, GIKTUP/GIKTUP, DMOXAM/DMOXAM, ETINOK/ETINOK, MOKJEC/MOKJEC, GIRKOI/GIRKOI, IHURUA/IHURUA, TMFZNO/TMFZNO, FUMRAM01/FUMRAM01, ILIQEZ/ILIQEZ, CEDSAH/CEDSAH, DELZOK/DELZOK, EHIRET/EHIRET, HEXDAQ/HEXDAQ, PAVBEU01/PAVBEU01, CEZSEF/CEZSEF, FACQOP/FACQOP, ABEWAG/ABEWAG, HILHIT/HILHIT, PUHHOQ/PUHHOQ, LEYXIW/LEYXIW, DISBUE/DISBUE, OWOJOA/OWOJOA, DADCUJ/DADCUJ, ODNPDS11/ODNPDS11, YANFAX/YANFAX, HEQZAG/HEQZAG, IYINAF01/IYINAF01, BIGUAN01/BIGUAN01, UXUYUJ/UXUYUJ, NUPVUR/NUPVUR, WIDBAO/WIDBAO, JEXRUZ/JEXRUZ, ZAYGEO/ZAYGEO, QEMJOH/QEMJOH, JOHJIB03/JOHJIB03, GARZUX/GARZUX, CEDGAV/CEDGAV, SETBAV01/SETBAV01, AHEKUT/AHEKUT, ELUKIH/ELUKIH, MPYAZO10/MPYAZO10, TUMZAE/TUMZAE, XOGROA/XOGROA, GALCAX01/GALCAX01, EROMUU/EROMUU, HAMTEX/HAMTEX, WOJPUI/WOJPUI, LEPWEI/LEPWEI, LCYSTI11/LCYSTI11, PONVUK/PONVUK, RAFSID/RAFSID, JANGIP/JANGIP, GIVXUE/GIVXUE, MEOXAL/MEOXAL, DIYLII/DIYLII, GELWOL/GELWOL, UMIQEO/UMIQEO, ZAQWUL/ZAQWUL, DUSBUQ/DUSBUQ, WINMOV/WINMOV, ZOBXOD/ZOBXOD, YUYWEU/YUYWEU, XANYOC/XANYOC, ROHDEX/ROHDEX, DOZKOS/DOZKOS, PIDHIU/PIDHIU, EMUFAV/EMUFAV, JAPFAI/JAPFAI, VUKFOW/VUKFOW, KERYIP/KERYIP, VINXEX/VINXEX, KOMHAV01/KOMHAV01, QORROF/QORROF, SULAMD07/SULAMD07, MUGFUR/MUGFUR, MUVZUZ/MUVZUZ, GEKTOF/GEKTOF, RIHVIP/RIHVIP, XAVGUZ/XAVGUZ, TARCOE/TARCOE, ALOPUR/ALOPUR, WAFPOK/WAFPOK, WAQHED/WAQHED, SAGLAQ/SAGLAQ, DAPJUA/DAPJUA, FIFBAY/FIFBAY, LAKVOI/LAKVOI, PROLON/PROLON, KABLUX/KABLUX, AVEBUA/AVEBUA, YOYSUC/YOYSUC, NAWSUB/NAWSUB, MXPACX/MXPACX, UKIKEF/UKIKEF, WERVEU/WERVEU, MOQLUA/MOQLUA, VAYJIO/VAYJIO, IKEGAH/IKEGAH, COSKUR/COSKUR, LESPUW/LESPUW, YIQZUU/YIQZUU, DOVDIC/DOVDIC, GOHYUX/GOHYUX, QQQGEM01/QQQGEM01, CEPNAN/CEPNAN, FILHAJ/FILHAJ, FUWMIT/FUWMIT, EHIXOI/EHIXOI, BZTROP11/BZTROP11, KIPPAA/KIPPAA, ATZCXA/ATZCXA, GEMZAZ/GEMZAZ, QOZFER/QOZFER, TIMCHX/TIMCHX, CARYAV/CARYAV, FECVEP/FECVEP, SUPFOY/SUPFOY, YIHLEI/YIHLEI, EJEQUF/EJEQUF, SEDHUG/SEDHUG, JAXVEK/JAXVEK, FOYSUJ/FOYSUJ, QUNHIQ/QUNHIQ, WACYEF/WACYEF, YULZUA/YULZUA, KEKQEY/KEKQEY, OFUKOR/OFUKOR, DALQOW/DALQOW, IZALUQ/IZALUQ, KOWTEV/KOWTEV, KANVAX/KANVAX, LIHJAP/LIHJAP, VEVLAL/VEVLAL, XIPQOC/XIPQOC, HOYMAJ/HOYMAJ, PUMQIX/PUMQIX, SEYDEF/SEYDEF, XULDUD01/XULDUD01, NEKREA/NEKREA, JOTSIW/JOTSIW, SOLLOV02/SOLLOV02, DIFPEO/DIFPEO, GOGYUX/GOGYUX, QOYNUN/QOYNUN, HAFLIJ/HAFLIJ, YILPAM/YILPAM, KEXNIK/KEXNIK, OLORUC/OLORUC, KOBFUD/KOBFUD, ONAVEF/ONAVEF, ZZZNWQ01/ZZZNWQ01, KUZROO/KUZROO, MAZPAL/MAZPAL, MUQCAC/MUQCAC, VOBDOF/VOBDOF, GETKOF10/GETKOF10, FEXGUL/FEXGUL, RONVEV/RONVEV, BZTZAD01/BZTZAD01, KETZIS/KETZIS, NUQPOF/NUQPOF, OXTHDC01/OXTHDC01, LONMAD/LONMAD, IJEPOE/IJEPOE, ANONIN11/ANONIN11, GADPAE/GADPAE, BOXBEV/BOXBEV, AHIMOU/AHIMOU, JIWZAQ/JIWZAQ, LAWNIJ/LAWNIJ, VISYEB/VISYEB, HOMCOD/HOMCOD, VAJHIY/VAJHIY, BUXCEC/BUXCEC, ULUTIE/ULUTIE, CTMTNA04/CTMTNA04, UREAOH01/UREAOH01, HOWYEZ/HOWYEZ, RUVZEN/RUVZEN, CIKXOJ/CIKXOJ, JEDSUG/JEDSUG, ZBCNON01/ZBCNON01, SUCANH16/SUCANH16, CEVGIT/CEVGIT, GAKTAN/GAKTAN, QOBFUI/QOBFUI, KEHWUR/KEHWUR, CAZGIU/CAZGIU, VUPTAC01/VUPTAC01, APITEA/APITEA, GEJSET/GEJSET, TEJPEE/TEJPEE, PHTHAN/PHTHAN, XAHLOJ/XAHLOJ, ATOJAW/ATOJAW, YILMIR/YILMIR, KOWYEA11/KOWYEA11, BUWJAE/BUWJAE, KIWVIW01/KIWVIW01, MEADEN04/MEADEN04, EBELUV/EBELUV, AMFURZ/AMFURZ, HETPAL/HETPAL, RASVAI/RASVAI, TAJHUH/TAJHUH, PEVSIS/PEVSIS, RINXOD/RINXOD, VIKWAN/VIKWAN, GASCOS02/GASCOS02, VETKEL/VETKEL, DIKCOP01/DIKCOP01, RAHDIN/RAHDIN, TZBFZO/TZBFZO, IMUWIX/IMUWIX, CALFOM/CALFOM, OKINAY/OKINAY, BUSLIL/BUSLIL, TUPRBN01/TUPRBN01, ROFFEY/ROFFEY, QAJQAV01/QAJQAV01, GESNIB/GESNIB, OMIWEM/OMIWEM, RIGYEM/RIGYEM, ZUHQAW/ZUHQAW, HOMZEQ/HOMZEQ, ETDIAM17/ETDIAM17, SAQZAK/SAQZAK, NAPMEW/NAPMEW, FEKVOG/FEKVOG, TAHHOC/TAHHOC, NPCBCP/NPCBCP, FUVDOP/FUVDOP, DAWZOS/DAWZOS, BIRFEN01/BIRFEN01, QEKQOM/QEKQOM, BONJIY/BONJIY, ZIVTOP/ZIVTOP, AMCYTS/AMCYTS, HEPHUF/HEPHUF, ZIKWEV/ZIKWEV, GOHWAB/GOHWAB, XIXTUU/XIXTUU, COTTAI/COTTAI, EFOTEY/EFOTEY, ZOTDAP/ZOTDAP, GODPUK/GODPUK, YOTYOW/YOTYOW, JAYDUI/JAYDUI, TAHMUN/TAHMUN, HIQMAV/HIQMAV, JOPNAD/JOPNAD, YICMIG/YICMIG, PEZDUU/PEZDUU, ISONAH/ISONAH, JAGMEM/JAGMEM, IPRPOL03/IPRPOL03, ZIKQOZ/ZIKQOZ, ACTCBZ/ACTCBZ, YAXWAX01/YAXWAX01, FUSDII/FUSDII, LIFNOE/LIFNOE-BISMEV10, BISMEV10/LIFNOE-BISMEV10, CUPYUJ/CUPYUJ-CUPYOD, CUPYOD/CUPYUJ-CUPYOD, DOJKUJ/DOJKUJ-DOJNAS, DOJNAS/DOJKUJ-DOJNAS, FIGYUR/FIGYUR-FIGYOL, FIGYOL/FIGYUR-FIGYOL, UCOKUV/UCOKUV-UCIYEN, UCIYEN/UCOKUV-UCIYEN, METAMI04/METAMI04-NAWYES-CEKGUU02, NAWYES/METAMI04-NAWYES-CEKGUU02, CEKGUU02/METAMI04-NAWYES-CEKGUU02, DAXSON/DAXSON-DAXSAZ, DAXSAZ/DAXSON-DAXSAZ, CINJAM/CINJAM-ZIWXEK, ZIWXEK/CINJAM-ZIWXEK, ESOZAP/ESOZAP-ESOYUI, ESOYUI/ESOZAP-ESOYUI, WEWPUJ/WEWPUJ-WEWRAR, WEWRAR/WEWPUJ-WEWRAR, ACEMID02/ACEMID02-FORMAM01, FORMAM01/ACEMID02-FORMAM01, PUTWOS/PUTWOS-PUTXAF, PUTXAF/PUTWOS-PUTXAF, OFIYUZ/OFIYUZ-WIFKOL01-OFIZEK, WIFKOL01/OFIYUZ-WIFKOL01-OFIZEK, OFIZEK/OFIYUZ-WIFKOL01-OFIZEK, OPIROV/OPIROV-OPISAI, OPISAI/OPIROV-OPISAI, YUWFUT02/YUWFUT02-XEXVEB01, XEXVEB01/YUWFUT02-XEXVEB01, ODEZEE/ODEZEE-ODEYAZ-ODEXOM, ODEYAZ/ODEZEE-ODEYAZ-ODEXOM, ODEXOM/ODEZEE-ODEYAZ-ODEXOM, BCBANN06/BCBANN06-BCBANN02, BCBANN02/BCBANN06-BCBANN02, VEYXUT/VEYXUT-VEYYII, VEYYII/VEYXUT-VEYYII, ZEFPUX/ZEFPUX-ZEFPAD, ZEFPAD/ZEFPUX-ZEFPAD, UREAXX28/UREAXX28-UREAXX24, UREAXX24/UREAXX28-UREAXX24, AJAQAE/AJAQAE-AJAROT-AJAPOR01-AJAREJ, AJAROT/AJAQAE-AJAROT-AJAPOR01-AJAREJ, AJAPOR01/AJAQAE-AJAROT-AJAPOR01-AJAREJ, AJAREJ/AJAQAE-AJAROT-AJAPOR01-AJAREJ, UFORUE01/UFORUE01-WIKCIE, WIKCIE/UFORUE01-WIKCIE, AHUFOZ/AHUFOZ-AHUFUF, AHUFUF/AHUFOZ-AHUFUF, QARVOV/QARVOV-QARWAI, QARWAI/QARVOV-QARWAI, KAGMIO/KAGMIO-KAGMUA, KAGMUA/KAGMIO-KAGMUA, COWJAB/COWJAB-COWHUT, COWHUT/COWJAB-COWHUT, MUMZEA/MUMZEA-MUMYUP, MUMYUP/MUMZEA-MUMYUP, KABJON/KABJON-KABJED, KABJED/KABJON-KABJED, BOZKUW/BOZKUW-BOZKOQ, BOZKOQ/BOZKUW-BOZKOQ, IFEVAQ/IFEVAQ-IFETUI, IFETUI/IFEVAQ-IFETUI, CUZJEO/CUZJEO-NOLDOJ01, NOLDOJ01/CUZJEO-NOLDOJ01, SIYBEJ/SIYBEJ-SIYBUZ, SIYBUZ/SIYBEJ-SIYBUZ, FEFBIE/FEFBIE-FEFBOK, FEFBOK/FEFBIE-FEFBOK, EDEFUQ/EDEFUQ-BAWKEP11, BAWKEP11/EDEFUQ-BAWKEP11, JUVMUI/JUVMUI-JUVMOC, JUVMOC/JUVMUI-JUVMOC, HORZAQ/HORZAQ-HORZEU-HORZIY, HORZEU/HORZAQ-HORZEU-HORZIY, HORZIY/HORZAQ-HORZEU-HORZIY, NABMUY/NABMUY-NABMIM, NABMIM/NABMUY-NABMIM, CUKCAM22/CUKCAM22-GAUTAM15, GAUTAM15/CUKCAM22-GAUTAM15, ZZZKAY02/ZZZKAY02-ZZZKAY05, ZZZKAY05/ZZZKAY02-ZZZKAY05, QEKYIO/QEKYIO-QEKXUZ, QEKXUZ/QEKYIO-QEKXUZ, ROGWOB/ROGWOB-ROGVIU, ROGVIU/ROGWOB-ROGVIU, APENUG/APENUG-APENIU, APENIU/APENUG-APENIU, JEFPEQ/JEFPEQ-JEFPIU, JEFPIU/JEFPEQ-JEFPIU, KIXCUP01/KIXCUP01-MUTPOI, MUTPOI/KIXCUP01-MUTPOI, JARDEM/JARDEM-JARDUC, JARDUC/JARDEM-JARDUC, LUVPEY/LUVPEY-LUVNAS, LUVNAS/LUVPEY-LUVNAS, CIZGEX/CIZGEX-CIZJIE, CIZJIE/CIZGEX-CIZJIE, FOGJAM/FOGJAM-FOGHOY-FOGHUE, FOGHOY/FOGJAM-FOGHOY-FOGHUE, FOGHUE/FOGJAM-FOGHOY-FOGHUE, LIQVUC/LIQVUC-LIQWIR-LIQWEN, LIQWIR/LIQVUC-LIQWIR-LIQWEN, LIQWEN/LIQVUC-LIQWIR-LIQWEN, JUXJET/JUXJET-JUXLEV, JUXLEV/JUXJET-JUXLEV, WETFUX/WETFUX-WETGAE, WETGAE/WETFUX-WETGAE, MTCZME/MTCZME-DTCZME, DTCZME/MTCZME-DTCZME, KEMZIL01/KEMZIL01-AXOSIQ, AXOSIQ/KEMZIL01-AXOSIQ, GLYCIN56/GLYCIN56-GLYCIN62, GLYCIN62/GLYCIN56-GLYCIN62, IFUWIR/IFUWIR-IFUVOW, IFUVOW/IFUWIR-IFUVOW, VOBJEB01/VOBJEB01-PYRDNA06, PYRDNA06/VOBJEB01-PYRDNA06, NADQOZ/NADQOZ-NADQUF, NADQUF/NADQOZ-NADQUF, JEMSUP/JEMSUP-JEMTAW, JEMTAW/JEMSUP-JEMTAW, PIWXEY06/PIWXEY06-PIWXEY02, PIWXEY02/PIWXEY06-PIWXEY02, RATMEF01/RATMEF01-TEDNIB, TEDNIB/RATMEF01-TEDNIB, DMETSO04/DMETSO04-DMETSO06, DMETSO06/DMETSO04-DMETSO06, MOGYIR/MOGYIR-MOGYOX, MOGYOX/MOGYIR-MOGYOX, HOPLAA/HOPLAA-HOPKUT, HOPKUT/HOPLAA-HOPKUT, KAQPUN/KAQPUN-KAQPOH, KAQPOH/KAQPUN-KAQPOH, KICCOO12/KICCOO12-CYSTAC05, CYSTAC05/KICCOO12-CYSTAC05, DAHDAR/DAHDAR-DAHDIZ, DAHDIZ/DAHDAR-DAHDIZ, NOLCIA/NOLCIA-NOLCUM, NOLCUM/NOLCIA-NOLCUM, NOYWED/NOYWED-NOYWAZ, NOYWAZ/NOYWED-NOYWAZ, PYRZOL06/PYRZOL06-PYRZOL32, PYRZOL32/PYRZOL06-PYRZOL32, TIPVIZ/TIPVIZ-TIPTUJ, TIPTUJ/TIPVIZ-TIPTUJ, NAHNIT/NAHNIT-BICVIS01, BICVIS01/NAHNIT-BICVIS01, NMALAM01/NMALAM01-MAJNES, MAJNES/NMALAM01-MAJNES, HIJCAE/HIJCAE-HIJCEI, HIJCEI/HIJCAE-HIJCEI, WILBEY/WILBEY-WIKZUL, WIKZUL/WILBEY-WIKZUL, PUPBAD/PUPBAD-PUPBAD02, PUPBAD02/PUPBAD-PUPBAD02, ASOPOP/ASOPOP-ASOPEF, ASOPEF/ASOPOP-ASOPEF, LONDUN/LONDUN-LONDOH, LONDOH/LONDUN-LONDOH, PEXMEK/PEXMEK-PEXQAK, PEXQAK/PEXMEK-PEXQAK, UJIBIZ/UJIBIZ-UJICIA-UJIBAR, UJICIA/UJIBIZ-UJICIA-UJIBAR, UJIBAR/UJIBIZ-UJICIA-UJIBAR, SAWMIO/SAWMIO-SAWNAH, SAWNAH/SAWMIO-SAWNAH, QONRAO/QONRAO-QONRUI, QONRUI/QONRAO-QONRUI, KOVGOR/KOVGOR-KOVGIL, KOVGIL/KOVGOR-KOVGIL, HUWSOI02/HUWSOI02-HUWSOI04, HUWSOI04/HUWSOI02-HUWSOI04, TAFKAM/TAFKAM-TAFKUG-TAFKIU, TAFKUG/TAFKAM-TAFKUG-TAFKIU, TAFKIU/TAFKAM-TAFKUG-TAFKIU, XOCBAT/XOCBAT-XOBZUK, XOBZUK/XOCBAT-XOBZUK, JUMXIA/JUMXIA-JUMXEW-JUMWUL, JUMXEW/JUMXIA-JUMXEW-JUMWUL, JUMWUL/JUMXIA-JUMXEW-JUMWUL, FUHZOY/FUHZOY-FUDWUX, FUDWUX/FUHZOY-FUDWUX, UQAFEZ/UQAFEZ-UQAFAV, UQAFAV/UQAFEZ-UQAFAV, CIKRET/CIKRET-BITTED10, BITTED10/CIKRET-BITTED10, TUBCIC/TUBCIC-TUBCOI, TUBCOI/TUBCIC-TUBCOI, DAJWIV/DAJWIV-DAJVUG01, DAJVUG01/DAJWIV-DAJVUG01, NEZNAH/NEZNAH-NEZNEL-NEZMUA, NEZNEL/NEZNAH-NEZNEL-NEZMUA, NEZMUA/NEZNAH-NEZNEL-NEZMUA, KABFID/KABFID-KABDOH, KABDOH/KABFID-KABDOH, MUGZUJ/MUGZUJ-MUHBOG, MUHBOG/MUGZUJ-MUHBOG, UMERAF/UMERAF-UMEREJ, UMEREJ/UMERAF-UMEREJ, OTOGEL/OTOGEL-BULVAL04, BULVAL04/OTOGEL-BULVAL04, QQQCIG26/QQQCIG26-QQQCIG30, QQQCIG30/QQQCIG26-QQQCIG30, GEDXUJ/GEDXUJ-GEDYAQ-GEDXOD, GEDYAQ/GEDXUJ-GEDYAQ-GEDXOD, GEDXOD/GEDXUJ-GEDYAQ-GEDXOD, NICGAH/NICGAH-NICGEL, NICGEL/NICGAH-NICGEL, NEFJUG/NEFJUG-NEFJEQ, NEFJEQ/NEFJUG-NEFJEQ, ZUNTAD/ZUNTAD-ZUNSUW, ZUNSUW/ZUNTAD-ZUNSUW, REVMIP/REVMIP-RATMIJ-RATJOM, RATMIJ/REVMIP-RATMIJ-RATJOM, RATJOM/REVMIP-RATMIJ-RATJOM, BIDJON/BIDJON-BIDJIH, BIDJIH/BIDJON-BIDJIH, HINNID/HINNID-HINNEZ, HINNEZ/HINNID-HINNEZ, ROKTAO/ROKTAO-ROKSUH, ROKSUH/ROKTAO-ROKSUH, CADBER/CADBER-CADBOB, CADBOB/CADBER-CADBOB, DIRMOG/DIRMOG-DIRMIA, DIRMIA/DIRMOG-DIRMIA, DOZXEW/DOZXEW-DOZXIA, DOZXIA/DOZXEW-DOZXIA, XODBAU/XODBAU-XOCZUL, XOCZUL/XODBAU-XOCZUL, YECVUX/YECVUX-YECWAE, YECWAE/YECVUX-YECWAE, NOPKEI/NOPKEI-WATKEH01, WATKEH01/NOPKEI-WATKEH01, YAFPAZ/YAFPAZ-YAFDER, YAFDER/YAFPAZ-YAFDER, SIPGIH/SIPGIH-SIPGON, SIPGON/SIPGIH-SIPGON, YOKNET/YOKNET-YOKJOZ, YOKJOZ/YOKNET-YOKJOZ, TUPBUD/TUPBUD-TUPBIR, TUPBIR/TUPBUD-TUPBIR, ATDZPY/ATDZPY-AFZPYM, AFZPYM/ATDZPY-AFZPYM, MIHJIW/MIHJIW-MIHKET, MIHKET/MIHJIW-MIHKET, UHAMOF/UHAMOF-KANQAR01, KANQAR01/UHAMOF-KANQAR01, NPSELD10/NPSELD10-NPSELC10, NPSELC10/NPSELD10-NPSELC10, DASCAC/DASCAC-DATMAN, DATMAN/DASCAC-DATMAN, HUKROU/HUKROU-HUKRUA, HUKRUA/HUKROU-HUKRUA, SUFJAG/SUFJAG-SUFHUY, SUFHUY/SUFJAG-SUFHUY, FEHVIY/FEHVIY-FEHVUK, FEHVUK/FEHVIY-FEHVUK, FIPZIN01/FIPZIN01-NUYQEE, NUYQEE/FIPZIN01-NUYQEE, VUBVAR/VUBVAR-VUBQAM01, VUBQAM01/VUBVAR-VUBQAM01, IQIZEO/IQIZEO, VEQMUA/VEQMUA, RICTIG/RICTIG, OHEWOP/OHEWOP, HAMTIZ/HAMTIZ, XEZFUF/XEZFUF, BAPPUF/BAPPUF, GADSIO/GADSIO, MENDAL01/MENDAL01, SUHFEH/SUHFEH, PEXPEN/PEXPEN, EMIPUM/EMIPUM, KOFKAR/KOFKAR, KAHJEK/KAHJEK, IVABEO/IVABEO, PILFIB/PILFIB, AHOWOL/AHOWOL, HISNII/HISNII, ZEMHOO/ZEMHOO, IDURIJ/IDURIJ, OTAKAW/OTAKAW, WIZWOS/WIZWOS, MAQWIM23/MAQWIM23, XAZYIG/XAZYIG, EVILEB/EVILEB, GASXON/GASXON, SAWVET/SAWVET, GAWFEQ/GAWFEQ, PYAZAC/PYAZAC, ODEJAJ01/ODEJAJ01, IQULUC/IQULUC, COSPEG/COSPEG, ROGRIQ/ROGRIQ, DUZLUF/DUZLUF, LIXQEO/LIXQEO, BEHWER/BEHWER, NORFUW/NORFUW, XINLAJ/XINLAJ, IVEZAK/IVEZAK, MOBNUM/MOBNUM, AZIDES/AZIDES, UBUXOG/UBUXOG, CINCHO10/CINCHO10, KUTKAL/KUTKAL, MEJQEY/MEJQEY, ESESEA/ESESEA, FACZIU/FACZIU, AROKUN/AROKUN, VAFPAV/VAFPAV, KOTMUB/KOTMUB, MUBBAN/MUBBAN, GERPOJ01/GERPOJ01, XASHUW/XASHUW, AYUNEO/AYUNEO, ROHJED/ROHJED, GAQJUF/GAQJUF, WAGTII/WAGTII, PELXAG10/PELXAG10, AQEYAW/AQEYAW, YIDTIQ/YIDTIQ, TAVJAD/TAVJAD, WAWQUH/WAWQUH, JEBQOW/JEBQOW, CBMZPN21/CBMZPN21, GUTZOM/GUTZOM, PUNFAH/PUNFAH, BOLGOZ/BOLGOZ, ZIGBAS/ZIGBAS, CXMTUN/CXMTUN, ZOYMOP07/ZOYMOP07, RECYIH/RECYIH, QOMVUK/QOMVUK, PIWBUS/PIWBUS, ELAWIX/ELAWIX, YIXPUR/YIXPUR, ZOFNUD/ZOFNUD, VEZCUY/VEZCUY, IYASUW/IYASUW, UTICIK/UTICIK, SATPEI02/SATPEI02, JOQTUE/JOQTUE, FAHLAB/FAHLAB, KETYUF/KETYUF, HODLOC/HODLOC, RIQSER/RIQSER, RUKTAU/RUKTAU, CUMZAM/CUMZAM, AKUBIT/AKUBIT, NAPTPR/NAPTPR, EDAXOW/EDAXOW, QEXKUA/QEXKUA, UQAMIK/UQAMIK, SIMYOE/SIMYOE, KEZNUZ/KEZNUZ, PANLEZ10/PANLEZ10, TOPRIB/TOPRIB, AXAWIG/AXAWIG, QUBQOT/QUBQOT, SATPUZ/SATPUZ, VOCHUR/VOCHUR, IPINIE/IPINIE, SAJCAJ/SAJCAJ, BEGDIB01/BEGDIB01, YAWWOK/YAWWOK, LEVSIO/LEVSIO, HIWYIV/HIWYIV, DOMNEY/DOMNEY, BERSOG/BERSOG, BIKNUE/BIKNUE, DAJZEU/DAJZEU, RUVPIJ/RUVPIJ, EBAXOW/EBAXOW, PUPGUD/PUPGUD, GADVAJ/GADVAJ, OCATOC/OCATOC, WOKJOV/WOKJOV, PIJREF/PIJREF, EXEWEJ/EXEWEJ, XUJKUK/XUJKUK, GOVQOX/GOVQOX, VUHZEE/VUHZEE, EVIHUM02/EVIHUM02, COBHUW/COBHUW, ENIMET/ENIMET, UWOCAM/UWOCAM, AMUQOQ/AMUQOQ, PEPLAX/PEPLAX, YODPAJ/YODPAJ, NECFIM/NECFIM, YUXCUP/YUXCUP, EMODUG/EMODUG, AFIKAC/AFIKAC, XUHZOR/XUHZOR, SULROG/SULROG, IQUBZA/IQUBZA, WECZEJ/WECZEJ, SEZXIG/SEZXIG, GIXKOP/GIXKOP, SOMNIT/SOMNIT, NIQTAJ/NIQTAJ, QUFCEZ/QUFCEZ, XIMGAB/XIMGAB, EZISUC/EZISUC, ODOROO/ODOROO, DAFTAF/DAFTAF, UJUKIT/UJUKIT, VILPUB/VILPUB, PEWNIQ/PEWNIQ, HOMZUG/HOMZUG, XULNOI/XULNOI, SUWKEC/SUWKEC, OMABEK/OMABEK, YOXRIO/YOXRIO, PETRAH/PETRAH, BAJCIY03/BAJCIY03, VASLOR/VASLOR, EXEYUD/EXEYUD, GUCJUK/GUCJUK, POLJEF/POLJEF, VOGDIE/VOGDIE, EBOVEX/EBOVEX, SAZFOO/SAZFOO, ITOFEE/ITOFEE, YOCWUK/YOCWUK, NUKSAO/NUKSAO, IMEQUO/IMEQUO, HUVWOL/HUVWOL, SUKNIW02/SUKNIW02, WOBRIP/WOBRIP, EKAHOP/EKAHOP, UQOLIW/UQOLIW, EDIZUM/EDIZUM, DOLBIR10/DOLBIR10, MOTNUF/MOTNUF, SAFQAU/SAFQAU, ACRDIN07/ACRDIN07, VINZUP/VINZUP, REJVAE/REJVAE, ALOSEZ/ALOSEZ, MEHNAP/MEHNAP, GUJGEX/GUJGEX, UMUKUJ/UMUKUJ, FESNEW05/FESNEW05, OFEVOL/OFEVOL, PMPZOL/PMPZOL, HAXREE/HAXREE, OJICUF/OJICUF, RACGEK/RACGEK, BUFNEV01/BUFNEV01, ZOGTIY/ZOGTIY, QUFJUY/QUFJUY, ANOSAY/ANOSAY, DIZWEQ/DIZWEQ, HUDHEU/HUDHEU, QAKDAJ/QAKDAJ, BAWRAT/BAWRAT, ITINEG/ITINEG, EABZBU/EABZBU, SIGSAD/SIGSAD, FOSLEG/FOSLEG, UBUVIY/UBUVIY, CORTPY/CORTPY, AMEXOH/AMEXOH, VIDMAX02/VIDMAX02, APUPIK/APUPIK, SENKUR/SENKUR, PEFGIS/PEFGIS, YOWYOY02/YOWYOY02, QUYZIV/QUYZIV, DENXUP02/DENXUP02, WEGPEF/WEGPEF, XOHMAI/XOHMAI, IROZIY/IROZIY, NOPHKN/NOPHKN, WEVVEZ/WEVVEZ, ZULCEQ/ZULCEQ, EVIQEF/EVIQEF, AHATEK/AHATEK, BIXQEF/BIXQEF, WOYTAH/WOYTAH, KAKHEL/KAKHEL, POQSEU/POQSEU, AXADAF/AXADAF, QQQAMS02/QQQAMS02, YERTIZ01/YERTIZ01, HIMSUS/HIMSUS, JEKPIZ/JEKPIZ, ARONOM/ARONOM, RAKTOO/RAKTOO, FIHLEO/FIHLEO, HUVZUV/HUVZUV, MEJDOU/MEJDOU, CUTCUQ/CUTCUQ, EHAHAY/EHAHAY, RELJAU/RELJAU, KOJTOT/KOJTOT, AFIQUC/AFIQUC, MALSOH/MALSOH, GUFYOX/GUFYOX, WEPBAV/WEPBAV, YEGGIA/YEGGIA, EXUVUP/EXUVUP, XOFFEF/XOFFEF, QUGWUK/QUGWUK, NURZOP/NURZOP, QECNAP/QECNAP, VORMEV/VORMEV, RAVFOK/RAVFOK, ROJHOP/ROJHOP, QUWFIZ/QUWFIZ, XIZVAD/XIZVAD, THYDIN05/THYDIN05, RIFZAI/RIFZAI, AZOBEN12/AZOBEN12, DOVWAM/DOVWAM, SIHCES/SIHCES, NEQPEG/NEQPEG, UNURIF/UNURIF, DOHPEV/DOHPEV, YUNYUC/YUNYUC, RAZVUJ/RAZVUJ, MOSLAI/MOSLAI, MISDAT/MISDAT, XINHIL/XINHIL, LUQDOS/LUQDOS, COLBAG/COLBAG, HECNOS/HECNOS, HIGCIK01/HIGCIK01, ZZZFFY01/ZZZFFY01, HABNED/HABNED, IZAKOK/IZAKOK, ZICKIF/ZICKIF, ROJXOD/ROJXOD, WAFBIQ/WAFBIQ, ORADUH/ORADUH, ZIWMOJ/ZIWMOJ, OCAWOF/OCAWOF, PHBZAC01/PHBZAC01, QIQCOI/QIQCOI, PUPBAD01/PUPBAD01, EWOBIB/EWOBIB, UBASAT/UBASAT, LILDEP/LILDEP, SIHZAM/SIHZAM, PUMQEV/PUMQEV, LADNEL/LADNEL, UCANIV/UCANIV, FEPTID/FEPTID, XAVQOB/XAVQOB, LIZHEJ/LIZHEJ, IDUJEW/IDUJEW, EFIBAX/EFIBAX, ZZZBPY10/ZZZBPY10, LAFHEH/LAFHEH, ISIJIE/ISIJIE, DASNIV/DASNIV, YIPPOC/YIPPOC, FAJDEC/FAJDEC, DUTKOU/DUTKOU, ONBZAM/ONBZAM, ALEXEW/ALEXEW, BOMSIH/BOMSIH, ZAYPOE/ZAYPOE, VOXNOL/VOXNOL, VIDDAO/VIDDAO, MELAMI05/MELAMI05, BASNOZ/BASNOZ, HOMKIF/HOMKIF, EKAWAQ/EKAWAQ, FADHOJ/FADHOJ, KEMFIS/KEMFIS, AXOSOW03/AXOSOW03, LAVSIL/LAVSIL, RIZBAF/RIZBAF, UNAMOL/UNAMOL, APODUG/APODUG, MESQOR/MESQOR, KUZQIG/KUZQIG, ZOXYOA/ZOXYOA, NASZAJ/NASZAJ, NEZFON/NEZFON, NUPPOD/NUPPOD, ECODUV/ECODUV, HUYYOP/HUYYOP, TIWZUV/TIWZUV, FUPWES/FUPWES, HONKEC/HONKEC, DOTFOI/DOTFOI, SEGROL/SEGROL, GIZFEB/GIZFEB, HUVPAR/HUVPAR, DORKOK/DORKOK, PEDHAJ/PEDHAJ, XUVBAT/XUVBAT, NBZOAC11/NBZOAC11, JULGOO/JULGOO, WUCVIB/WUCVIB, NCUBEB10/NCUBEB10, LUDZIT/LUDZIT, UKUROJ/UKUROJ, HUMNEK/HUMNEK, SAVREN/SAVREN, AQAGII/AQAGII, YODXAS/YODXAS, VUGWAX/VUGWAX, WIQZOL/WIQZOL, WIHBEW/WIHBEW, AJIXUM/AJIXUM, RIHFIY/RIHFIY, HESTOO/HESTOO, POHCAS/POHCAS, JIPCUG10/JIPCUG10, JESHIZ/JESHIZ, BEDLEB01/BEDLEB01, SOGCUN/SOGCUN, VETJIO/VETJIO, RULDAF/RULDAF, BZAMID08/BZAMID08, XOWJUP/XOWJUP, EKOGAO/EKOGAO, XICCAO/XICCAO, FAHXUH/FAHXUH, OGIMIC/OGIMIC, EMISUQ/EMISUQ, XACTEC/XACTEC, FEMXOK/FEMXOK, XAQTUF/XAQTUF, FELDOR/FELDOR, EVINII/EVINII, ZIKQIT/ZIKQIT, YOFTOE/YOFTOE, GIZRUE/GIZRUE, SORFIQ/SORFIQ, KIVZIZ/KIVZIZ, DIWWEN/DIWWEN, COWPUZ/COWPUZ, HUDYUA/HUDYUA, OCIPAR/OCIPAR, MEHLER/MEHLER, CIKSAQ/CIKSAQ, DITZOX/DITZOX, ECACER/ECACER, YAZDEI/YAZDEI, KUJZIY/KUJZIY, LUXSAY/LUXSAY, BUMNOM/BUMNOM, TEMKAZ/TEMKAZ, ZIYYUD/ZIYYUD, DXCYTD/DXCYTD, COCYAW/COCYAW, RULHOX/RULHOX, ITIREI/ITIREI, OXUJUN/OXUJUN, MATQOO/MATQOO, MUTWON/MUTWON, HODKEQ/HODKEQ, MAHPUJ/MAHPUJ, KUZJIA/KUZJIA, SEBVAW/SEBVAW, ARUZUK/ARUZUK, POKKAD10/POKKAD10, BZTROP11/BZTROP11, XIYTIJ/XIYTIJ, HIZHOP/HIZHOP, INAVIC/INAVIC, GIDHUW/GIDHUW, WIFQEI/WIFQEI, EYASAZ/EYASAZ, MBPHOL02/MBPHOL02, ASPARM10/ASPARM10, ANAHII/ANAHII, OMSTER01/OMSTER01, BAQNEM/BAQNEM, BUGQUQ/BUGQUQ, ZEBXOV/ZEBXOV, TOPXUT/TOPXUT, MEYBIB/MEYBIB, UWEZED/UWEZED, ZEMNAG/ZEMNAG, NEVDOH/NEVDOH, FEZLUT/FEZLUT, NUZPUT01/NUZPUT01, VANFEV/VANFEV, QIYLAM/QIYLAM, KUYWEH/KUYWEH, BUZJIR/BUZJIR, PIHBOZ/PIHBOZ, EMEFOT/EMEFOT, XANJUS/XANJUS, BAYPAT/BAYPAT, DILDUZ/DILDUZ, TAJSOM/TAJSOM, PACWAU/PACWAU, YIMPOB/YIMPOB, KOGWUZ/KOGWUZ, UNUVEF/UNUVEF, IJEZUS/IJEZUS, VONNOB/VONNOB, MUJGEE/MUJGEE, COYBOJ/COYBOJ, DILKIT/DILKIT, REKMEZ/REKMEZ, QEPRIO/QEPRIO, WULTUT/WULTUT, ZATDOP/ZATDOP, SEYCUU/SEYCUU, MAJJUD/MAJJUD, ONOTIW/ONOTIW, XOTFAN/XOTFAN, NANJIW/NANJIW, FALHIJ/FALHIJ, PUWNIG/PUWNIG, VEMBAS/VEMBAS, IBEHII/IBEHII, RUCNOU/RUCNOU, SOYMEZ/SOYMEZ, VUDDUV/VUDDUV, SEFNOG/SEFNOG, OXAROV/OXAROV, LOPLUZ/LOPLUZ, AHOXOL/AHOXOL, RIMHEC/RIMHEC, EPHEDR01/EPHEDR01, AWAVEZ/AWAVEZ, NAJLUF/NAJLUF, ZOSVEI/ZOSVEI, KURZOO/KURZOO, DAGMEF/DAGMEF, EZUMUJ/EZUMUJ, NILSAE01/NILSAE01, AVULUY/AVULUY, HOTNUA/HOTNUA, AKOFUD/AKOFUD, UCAYUS/UCAYUS, LAWCIW/LAWCIW, VEKHAX/VEKHAX, CUSTAN/CUSTAN, KELNIA06/KELNIA06, RAMJIB/RAMJIB, ADITAI/ADITAI, SEYMEO/SEYMEO, VUJHOX/VUJHOX, ZOYPIM/ZOYPIM, LANQIC/LANQIC, GUFVIN/GUFVIN, TIPLEK/TIPLEK, MAXVUF/MAXVUF, YUPXUD/YUPXUD, KUSVOK/KUSVOK, DUGGET/DUGGET, NAPKAR/NAPKAR, GAQHIR01/GAQHIR01, OFAQIV/OFAQIV, TNONDX/TNONDX, WIWZAF/WIWZAF, ZAPJUX/ZAPJUX, CEPSIZ/CEPSIZ, COMLOH/COMLOH, OJUYAV/OJUYAV, PUGNUB/PUGNUB, ZZZPUS09/ZZZPUS09, AJOGUC/AJOGUC, WIJDIE/WIJDIE, CIDYIZ/CIDYIZ, LAWTAH/LAWTAH, PARJUQ/PARJUQ, ZOHNER/ZOHNER, VAHKIZ/VAHKIZ, LODWAE/LODWAE, GIKQIA/GIKQIA, KISXOB/KISXOB, LAQPOK/LAQPOK, TEYDUW/TEYDUW, HIRWIP/HIRWIP, METZHY/METZHY, AKEVOC01/AKEVOC01, YIHTUG/YIHTUG, JEBLEK/JEBLEK, UNOPAQ/UNOPAQ, HULQIO/HULQIO, NINREI/NINREI, MADTET/MADTET, VAKJAV/VAKJAV, LABVIU/LABVIU, RICXEG/RICXEG, ABEQII/ABEQII, MPAZTP/MPAZTP, MACMOT/MACMOT, LEGREW/LEGREW, QOKVUI/QOKVUI, PAHDIM/PAHDIM, GOLXAH/GOLXAH, MIPLUT/MIPLUT, VEBPUQ/VEBPUQ, SUXYOZ/SUXYOZ, FEPJER/FEPJER, PDTHON/PDTHON, HAGXET/HAGXET, JARDUC/JARDUC, GAVPAX/GAVPAX, ALAQEK/ALAQEK, CACYIQ/CACYIQ, SIDBUD/SIDBUD, HIQPEC/HIQPEC, MOVBAA/MOVBAA, TIHJOK/TIHJOK, YOFTIY/YOFTIY, QIFDIT01/QIFDIT01, PIKGEV01/PIKGEV01, TIYXII/TIYXII, ISULUF/ISULUF, SODGID/SODGID, NAJBAB/NAJBAB, FOGVIG04/FOGVIG04, VINNEM/VINNEM, IZALUQ/IZALUQ, PHENAN08/PHENAN08, TICKAT/TICKAT-TICLIC, TICLIC/TICKAT-TICLIC, JIFHAK/JIFHAK, DUKXAJ/DUKXAJ, FENQEX/FENQEX-ODIDEN01, ODIDEN01/FENQEX-ODIDEN01, ORIKIL/ORIKIL-ORIKOR-ORIKUX-ORILAE-ORILEI, ORIKOR/ORIKIL-ORIKOR-ORIKUX-ORILAE-ORILEI, ORIKUX/ORIKIL-ORIKOR-ORIKUX-ORILAE-ORILEI, ORILAE/ORIKIL-ORIKOR-ORIKUX-ORILAE-ORILEI, ORILEI/ORIKIL-ORIKOR-ORIKUX-ORILAE-ORILEI, QULLUF/QULLUF-QULLUF01-QULLUF02, QULLUF01/QULLUF-QULLUF01-QULLUF02, QULLUF02/QULLUF-QULLUF01-QULLUF02, REHSAA/REHSAA-REHSII-REHSUU-REHTAB, REHSII/REHSAA-REHSII-REHSUU-REHTAB, REHSUU/REHSAA-REHSII-REHSUU-REHTAB, REHTAB/REHSAA-REHSII-REHSUU-REHTAB, TIWPIB/TIWPIB-TIWPOH, TIWPOH/TIWPIB-TIWPOH, VAKTOS/VAKTOS-VAKTOS01, VAKTOS01/VAKTOS-VAKTOS01, VUHFIO/VUHFIO-VUHFIO01, VUHFIO01/VUHFIO-VUHFIO01, WUZHOP/WUZHOP-WUZHOP01-WUZJEH-WUZJIL-WUZJOR-WUZJUX-WUZKAE-WUZKEI, WUZHOP01/WUZHOP-WUZHOP01-WUZJEH-WUZJIL-WUZJOR-WUZJUX-WUZKAE-WUZKEI, WUZJEH/WUZHOP-WUZHOP01-WUZJEH-WUZJIL-WUZJOR-WUZJUX-WUZKAE-WUZKEI, WUZJIL/WUZHOP-WUZHOP01-WUZJEH-WUZJIL-WUZJOR-WUZJUX-WUZKAE-WUZKEI, WUZJOR/WUZHOP-WUZHOP01-WUZJEH-WUZJIL-WUZJOR-WUZJUX-WUZKAE-WUZKEI, WUZJUX/WUZHOP-WUZHOP01-WUZJEH-WUZJIL-WUZJOR-WUZJUX-WUZKAE-WUZKEI, WUZKAE/WUZHOP-WUZHOP01-WUZJEH-WUZJIL-WUZJOR-WUZJUX-WUZKAE-WUZKEI, WUZKEI/WUZHOP-WUZHOP01-WUZJEH-WUZJIL-WUZJOR-WUZJUX-WUZKAE-WUZKEI, ODICUC/ODICUC-ODIDEN, ODIDEN/ODICUC-ODIDEN}
\newcommand{\CCDC}[1]{CSD Ref.~#1\cite{\citekey{#1}}}
\newcommand{\ccdc}[1]{#1\cite{\citekey{#1}}}

\begin{document}
\pagestyle{fancy}
\thispagestyle{plain}
\fancypagestyle{plain}{
\renewcommand{\headrulewidth}{0pt}
}
\makeFNbottom
\makeatletter
\renewcommand\LARGE{\@setfontsize\LARGE{15pt}{17}}
\renewcommand\Large{\@setfontsize\Large{12pt}{14}}
\renewcommand\large{\@setfontsize\large{10pt}{12}}
\renewcommand\footnotesize{\@setfontsize\footnotesize{7pt}{10}}
\makeatother

\renewcommand{\thefootnote}{\fnsymbol{footnote}}
\renewcommand\footnoterule{\vspace*{1pt}%
\color{cream}\hrule width 3.5in height 0.4pt \color{black}\vspace*{5pt}} 
\setcounter{secnumdepth}{5}

\makeatletter 
\renewcommand\@biblabel[1]{#1}            
\renewcommand\@makefntext[1]%
{\noindent\makebox[0pt][r]{\@thefnmark\,}#1}
\makeatother 
\renewcommand{\figurename}{\small{Fig.}~}
\sectionfont{\sffamily\Large}
\subsectionfont{\normalsize}
\subsubsectionfont{\bf}
\setstretch{1.125} %
\setlength{\skip\footins}{0.8cm}
\setlength{\footnotesep}{0.25cm}
\setlength{\jot}{10pt}
\titlespacing*{\section}{0pt}{4pt}{4pt}
\titlespacing*{\subsection}{0pt}{15pt}{1pt}
\fancyfoot{}
\fancyfoot[LO,RE]{\vspace{-7.1pt}\includegraphics[height=9pt]{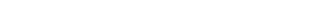}}
\fancyfoot[CO]{\vspace{-7.1pt}\hspace{13.2cm}\includegraphics{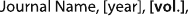}}
\fancyfoot[CE]{\vspace{-7.2pt}\hspace{-14.2cm}\includegraphics{head_foot/RF}}
\fancyfoot[RO]{\footnotesize{\sffamily{1--\pageref{LastPage} ~\textbar  \hspace{2pt}\thepage}}}
\fancyfoot[LE]{\footnotesize{\sffamily{\thepage~\textbar\hspace{3.45cm} 1--\pageref{LastPage}}}}
\fancyhead{}
\renewcommand{\headrulewidth}{0pt} 
\renewcommand{\footrulewidth}{0pt}
\setlength{\arrayrulewidth}{1pt}
\setlength{\columnsep}{6.5mm}
\setlength\bibsep{1pt}
\makeatletter 
\newlength{\figrulesep} 
\setlength{\figrulesep}{0.5\textfloatsep} 

\newcommand{\topfigrule}{\vspace*{-1pt}%
\noindent{\color{cream}\rule[-\figrulesep]{\columnwidth}{1.5pt}} }

\newcommand{\botfigrule}{\vspace*{-2pt}%
\noindent{\color{cream}\rule[\figrulesep]{\columnwidth}{1.5pt}} }

\newcommand{\dblfigrule}{\vspace*{-1pt}%
\noindent{\color{cream}\rule[-\figrulesep]{\textwidth}{1.5pt}} }

\makeatother
\twocolumn[
  \begin{@twocolumnfalse}
{\includegraphics[height=30pt]{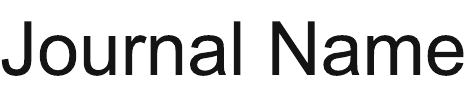}\hfill\raisebox{0pt}[0pt][0pt]{\includegraphics[height=55pt]{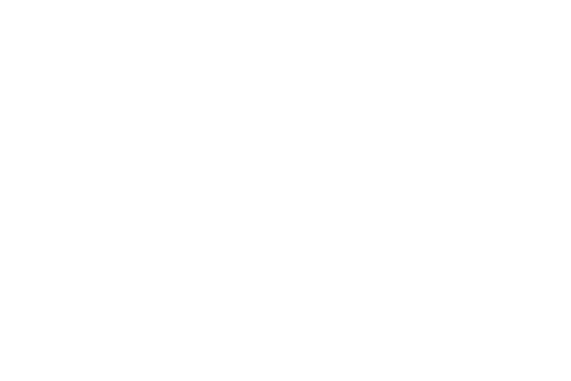}}\\[1ex]
\includegraphics[width=18.5cm]{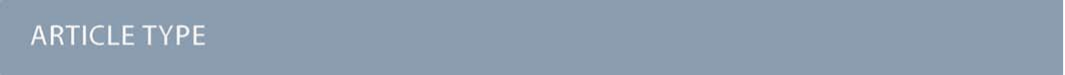}}\par
\vspace{1em}
\sffamily
\begin{tabular}{m{4.5cm} p{13.5cm} }

\includegraphics{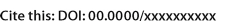} & \noindent\LARGE{\textbf{A data-driven interpretation of the stability of organic molecular crystals}}\\
\vspace{0.3cm} & \vspace{0.3cm} \\ %

 & \noindent\large{Rose K. Cersonsky,$^{\ast}$\textit{$^{a}$} Maria Pakhnova,\textit{$^{a}$} Edgar A. Engel,\textit{$^{b}$} and Michele Ceriotti\textit{$^{a}$}} \\
\vspace{-0.25cm}
\includegraphics{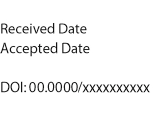} & \noindent\normalsize{Due to the subtle balance of intermolecular interactions that govern structure-property relations, predicting the stability of crystal structures formed from molecular building blocks is a highly non-trivial scientific problem.
A particularly active and fruitful approach involves classifying the different combinations of interacting chemical moieties, as understanding the relative energetics of different interactions enables the design of molecular crystals and fine-tuning their stabilities. While this is usually performed based on the empirical observation of the most commonly encountered motifs in known crystal structures, we propose to apply a combination of supervised and unsupervised machine-learning techniques to automate the construction of an extensive library of molecular building blocks.
We introduce a structural descriptor tailored to the prediction of the binding (lattice) energy and apply it to a curated dataset of organic crystals and exploit its atom-centered nature to obtain a data-driven assessment of the contribution of different chemical groups to the lattice energy of the crystal.
We then interpret this library using a low-dimensional representation of the structure-energy landscape and discuss selected examples of the insights into crystal engineering that can be extracted from this analysis, providing a complete database to guide the design of molecular materials.} \\

\end{tabular}
\end{@twocolumnfalse} \vspace{0.6cm}
]
\renewcommand*\rmdefault{bch}\normalfont\upshape
\rmfamily
\section*{}
\vspace{-1cm}

\footnotetext{\textit{$^{a}$~Laboratory of Computational Science and Modeling (COSMO), École Polytechnique Fédérale de Lausanne, Lausanne, Switzerland}}
\footnotetext{\textit{$^{b}$~TCM Group, Trinity College, Cambridge University, Cambridge, UK}}
\footnotetext{\textit{$^{*}$~Present address: Rose.Cersonsky@wisc.edu}}

\section{Introduction}
Understanding molecular crystallization is critical to many fields of chemical sciences -- from anticipating pharmaceutical stability and solubility \cite{yousef_pharmaceutical_2019,dudek_along_2022,iuzzolino_use_2017,datta_crystal_2004, beyer_prediction_2001} to preventing\cite{azrain_analysis_2018} or fostering\cite{mei_aggregation-induced_2015} aggregation in organic electronics to understanding complex formation in biological macromolecules\cite{KRISSINEL2007774,WOS:000310196200046}.

Yet, molecular crystallization is a complex process that involves multiple cooperative and competing forces.
Initial nucleation is typically motivated by strong interactions between functional groups\cite{ganguly_long-range_2010, davey_concerning_2006}. 
The structural patterns associated with these guiding interactions (deemed ``supramolecular synthons'') and their hierarchies are often the focus of experimental and computational studies in crystal structure prediction\cite{dey_crystal_2005, sarma_supramolecular_2002}. 
Nevertheless, once molecules have moved within closer range, many factors, including weaker interactions, the expulsion of solvent molecules, and geometric packing, will then determine the short- and potentially long-range order, leading to many potentially-stable polymorphs for a given stoichiometry.
In the past decades, there has been a growing push to develop a ``holistic'' view of molecular crystallization\cite{desiraju_crystal_2007,corpinot_practical_2019}, not only taking into account the nearest-neighbor contacts but also the interplay of these interactions with other components of the molecular assembly. 

Thus, molecular crystallization has emerged as a hotbed for computational inquiry. While it is simpler to rationalize single-site interactions, the interplay of many competing interactions necessitates diverse, high-throughput studies\cite{desiraju_crystal_2007}.
This focus has led to considerable theoretical and software developments for qualitative and quantitative analyses, including those tailored to crystal structure prediction (CSP)\cite{bruno_new_2002, bruno_retrieval_2004, neumann2008major,doi:10.1126/sciadv.aau3338} and the representation of electrostatic surfaces and molecular geometry\cite{spackman_hirshfeld_2009, spackman_fingerprinting_2002}.
Even more recently, machine learning has been used to understand the individual configurational and energy landscapes of molecules\cite{egorova_multifidelity_2020, musil_machine_2018,wengert_hybrid_2022,wengert_data-efficient_2021, kapil_complete_2022,seko_representation_2017,bereau_transferable_2015}; however, such techniques have yet to be applied in the general, holistic vein required to extract the qualitative insights that can be used to support crystal design efforts.

To study molecular crystallization in this broad lens, we have curated a dataset of roughly \nall C+H+N+O+S-containing molecular crystals from those reported in \citet{cordova_shiftml2_2022}. In \citet{cordova_shiftml2_2022}, these crystals were initially selected by querying the Cambridge Structural Database (CSD) to identify a diverse set of synthesizable molecular assemblies, including those originally experimentally stabilized at extreme conditions. The experimental properties of the full dataset are summarized in Appendix A3.

\indent The stability of molecular crystals is traditionally studied through the binding (lattice) energy, which is computationally determined by computing the ground-state energies for both the crystal and its molecular components in the dilute gas limit, here computed using DFT-PBE-D2 calculations of each crystal and its relaxed molecular components at ambient conditions.

\begin{figure*}
    \centering
    \includegraphics[width=0.8\linewidth]{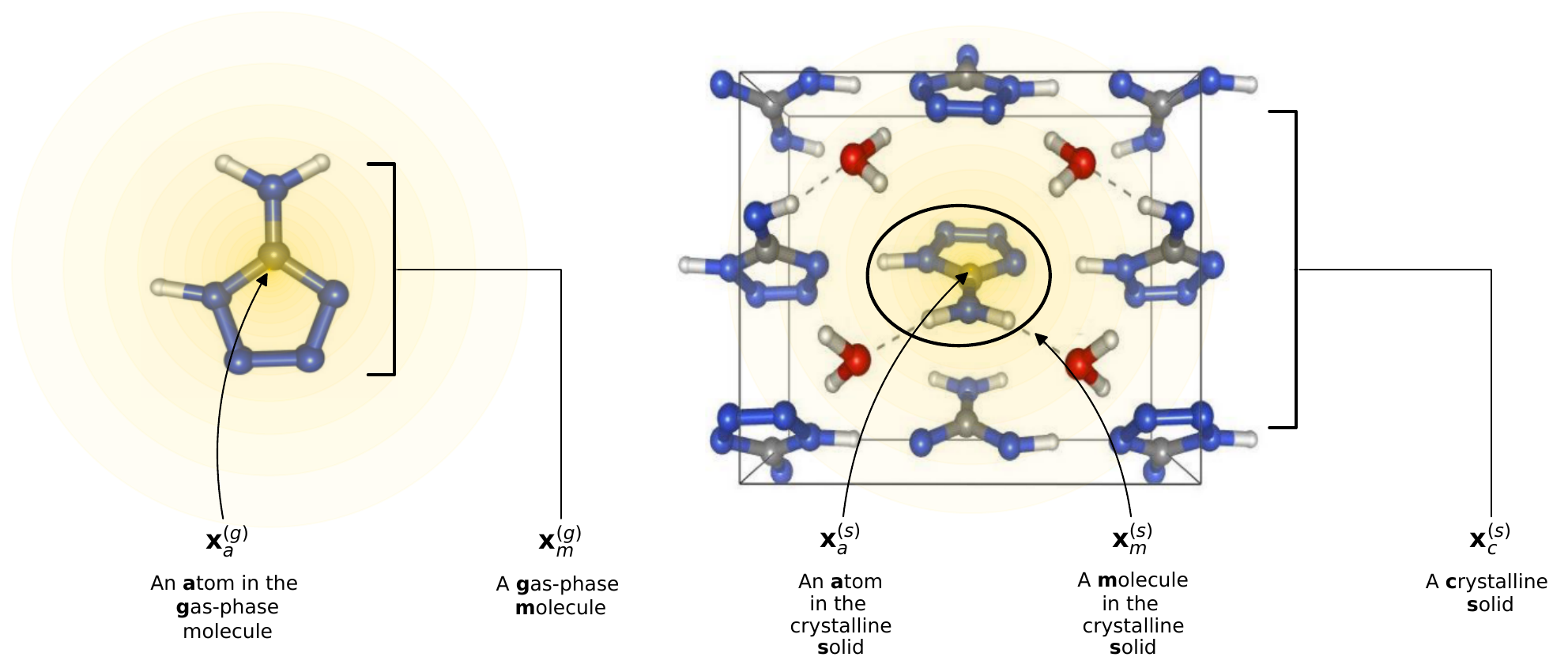}
    \caption{\textbf{Visualization of Descriptor Notation,} as described in \secref{sec:desc},  visualized for 5-Aminotetrazole Monohydrate (\CCDC{AMTETZ}). Each descriptor contains the information of an atom and its neighborhood (shown in yellow shading).}
    \label{fig:desc}
\end{figure*}

From here, we build an atom-centered regression model for this lattice energy, demonstrating the improvements in accuracy and reduction in model complexity from using a physics-informed approach. This atom-centered approach, wherein we represent each molecular crystal using the average ML descriptor for each of its atomic constituents, facilitates estimating the contribution of each atom, or combination of atoms, to the binding energy.
Then, employing a combination of supervised and unsupervised machine learning models, we can determine and interpret each molecular moiety's intermolecular interactions. Using these approaches, we show how physically-motivated machine learning models can not only ``rediscover'' the known maxims of crystal engineering, but provide insight and guidance for crystal design. We have made our datasets, and analyses openly-available through the Materials Cloud\cite{talirz_materials_2020}%
, with interactive components aimed to guide future molecular design and narrower or targeted studies.

\section{Notation}
In this study, we employ atom-centered descriptors\cite{willatt_atom_2019} to identify the contributions of specific collections of atoms to the binding of a crystal.
Given the many atomic and energetic entities (atoms, molecules, crystals, total energy versus lattice energy), we rely on many numerical representations and equations; hence we start by establishing a consistent notation we will use throughout the text.

\subsection{Descriptors}
\label{sec:desc}
To reflect the physics of atomic interactions, we use symmetry-adapted descriptors to encode/describe the geometric arrangement of atoms in their atomistic configurations, specifically the 3-body SOAP descriptors outlined in Appendix B1. 
Each of these input descriptors is written as $\bxsigi$, where the subscript $\sigma$ signifies the collection of atoms being described, including the entire crystal ($c$), a molecule ($m$), or an atom ($a$). $n_\sigma$ is the number of atoms in the given collection. Thus it follows that $n_c$ is the number of atoms in a given crystal, and $n_m$ is the number of atoms in a given molecule. Because we discuss analogous atoms or molecules in both the solid and gas phases, we use the superscript $(i)$ to denote the phase (crystalline solid $(s)$ or dilute gas $(g)$). 

The descriptor for a given collection should be assumed as the average of the descriptors for the constituent atoms:
\begin{equation}
\bxsigi = \frac{1}{n_\sigma}\sum_{a \in \sigma} \bxcs{a}{i}.
\label{eq:xsig}
\end{equation}
For example, the descriptor for the atoms in a molecule in a dilute gas is $\bxcs{m}{g} = \frac{1}{n_m}\sum_{a \in m} \bxcs{a}{g}$. If we were to look at the same molecule in the crystalline solid we would get $\bxcs{m}{s} = \frac{1}{n_m}\sum_{a \in m} \bxcs{a}{s}$. A schematic of these concepts is shown in \figref{fig:desc}, using the co-crystal 5-Aminotetrazole Monohydrate (\CCDC{AMTETZ}) as an example.

\subsection{Energies and Regressions} 

We use $\bE_\sigma$ to denote the total energy of a collection of atoms. In this study, the total energies of the crystals $\bEc$ are taken from those reported in \citet{cordova_shiftml2_2022} and the total energies of the molecules $\bEm$ are determined by DFT-PBE-D2 calculations, as described in Appendix A2. We use $\be_\sigma \equiv \bE_\sigma/n_\sigma$ to indicate the per-atom energy. Note that we express all energies in \unitz{}, where $\be_\sigma$ is to be interpreted as having the units of \unitz{}e of atoms.
Constructing a linear regression amounts to the ansatz
\begin{equation}
    {\bf e}_\sigma = {\bf x}_\sigma {\bf w}_\sigma + \epsilon_\sigma 
    \label{eq:genreg}
\end{equation}
where ${\bf w}_\sigma$ is the regression weights and $\epsilon_\sigma$ the residual errors. The \emph{lattice} energy (also referred to as the \emph{binding} or \emph{cohesive} energy in literature) of a molecular crystal is given by $\bY$, where 
\begin{equation}
    \bY = \bEc - \sum_{m\in c} \bEm.
    \label{eq:bY}
\end{equation}
With the average lattice energy per atom given by
\begin{equation}
    \byc \equiv \bY / n_c = \bec - \sum_{m \in c} \frac{n_m}{n_c} \bem
    \label{eq:by}
\end{equation}
Later, we will use our regression model to determine the atomic contributions to the lattice energy, which we will denote $\bya$, where $\byc = \frac{1}{n_c}\sum_{a \in c} \bya$. We will also consider the contributions for different collection of atoms, and will denote the average lattice energy contribution as $\bysig = \frac{1}{n_\sigma}\sum_{a \in \sigma} \bya$. When we regularize these contributions using a Gaussian filter (discussed in \secref{sec:filter} and Appendix B2), we will use a tilde to give $\tby$.

\section{Results and Discussion}

In the following, we consider crystals and gas-phase molecules, both of which have been geometry-optimized by minimizing their configurational energies with respect to the atomic positions, as described in Appendix A. Unless stated otherwise, we use as our featurization the 3-body SOAP vectors (as described in Appendix B1) and build a regularized ridge regression models using \texttt{scikit-learn}\cite{pedregosa_scikit-learn_2011}. 
All models were trained on the same training set of \ntrain crystals (or the corresponding \ntrainmol molecules). We report errors on a mutually-exclusive set of \ntest crystals (or the corresponding \ntestmol molecules). When interpreting the results, it is important to consider that the test set has been selected at random, and is therefore representative of the makeup of the CSD, while the training structures were selected with Farthest Point Sampling\cite{imbalzano2018, cersonsky2021} to maximize the diversity, and therefore contain a large fraction of unstable, ``extreme'' cases.

\subsection{Building a Model for the Lattice Energy}

\begin{table}
\centering
\begin{tabular}{|l l c c|}
\hline
 \bf{Regression Equation} & \bf{Eq.} & \bf{RMSE} & \bf{MAE} \\
\hline
\hline
 $\bec = \bxc\bwc$ & (\ref{eq:genreg}) & 1.15 & 0.863\\
\hline
\hline
 $\bem = \bxm\bwm$ & (\ref{eq:genreg}) & 0.727 & 0.563\\
\hline
\hline
 $\!\begin{aligned}[t]
\byc &= \bxc\bwc \\
&- \sum_{m\in c} \frac{n_m}{n_c}\left( \bxm\bwm\right)
\end{aligned}$ & (\ref{eq:bysum}) & 0.916 & 0.652\\
\hline
 $\byc = \bxc\bw$ &  & 0.778 & 0.552\\
\hline
 $\byc = \bxcg\bw$ &  & 1.101 & 0.723\\
\hline
 $\byc = \bxcr\bw$ & (\ref{eq:bxcrw}) & 0.571 & 0.404\\
\hline
 $\byc = \{\bxc, \bxcg\}\bw$ &  & 0.671 & 0.461\\
\hline
\end{tabular}
\caption{\textbf{Results of Linear Regression Exercises. }In each linear regression, an independent, 5-fold cross-validated model was build on \ntrain crystals (or the \ntrainmol coinciding molecules). Here we report the errors (in \unitz{}) on a separate set of \ntest crystals (or the coinciding \ntestmol molecules). In each regression equation $\bw$ is unique to that regression.}
\label{tab:regr}
\end{table}

One can estimate the atomic contributions to a target property (and thereby assess the contributions of specific molecular motifs) by building a robust machine learning model on an atom-centered descriptor\cite{helfrecht_new_2019}.
Suppose we have a descriptor 
\begin{equation}
\bx_\sigma = \frac{1}{n_\sigma}\sum_{a \in \sigma} \bx_a
\end{equation}
and train a regression model on some target $\mathbf{y}$ such that $\mathbf{y} = \bx_\sigma \bw + \berror$, where $\bw$ is the regression weight and $\berror$ is the residual error from the regression. We can then estimate the approximate contribution of each atom by computing $\mathbf{y}_a = \bx_a \bw$.

\paragraph*{Combining models of $\bec$ and $\bem$} Given ~\eqref{eq:by}, it is possible to build a model for the lattice energy from two separate models for crystal and molecular energy, replacing each energy $\be$ with its approximation via linear regression (\eqref{eq:genreg})
\begin{equation}
    \byc = \bxc\bwc + \berrorc - \sum_{m\in c} \frac{n_m}{n_c}\left( \bxm\bwm + \berrorm\right).
    \label{eq:bysum}
\end{equation}

\noindent \eqref{eq:bysum} may then be rewritten as:
\begin{align}
\begin{split}
    \byc &= \bxc \bwc -\bxcg \bwm + \berror
\end{split}
\label{eq:yw}
\end{align}
where we have defined $\berror \equiv \berrorc - \sum_{m\in c}\frac{n_m}{n_c}\berrorm$. In this scheme, the regression of the lattice energy is implicitly limited by the errors of the independent regressions; therefore, if we obtained a good fit for $\bec$ and $\bem$, this should be a fairly robust way to predict the lattice energy. 

When we predict the crystal and molecular atomic energies $\bec$ and $\bem$, we obtain RMSEs of \ecrmse \unitz{} and \emrmse \unitz{}, respectively, which are acceptably small compared to the intrinsic variance of the baselined\footnote{To improve the regressions of crystal and molecular energies, we subtract a \textit{baseline} determined by linear regression of the atomic composition on the total energies.} target energies of the test set, which have standard deviations of \ecstd \unitz{} and \emstd \unitz{}, respectively. However, the intrinsic variance of the lattice energies is smaller (\ehstd \unitz{}); therefore, the resulting RMSE of \rcmrmse \unitz{} from \eqref{eq:bysum} is very unsatisfactory and suggests that the errors in the independent regressions generally overlap with the lattice energy contributions.

\paragraph*{Building a model directly on $\byc$} With the reduced variance of the target (lattice energy), it thus makes sense to construct the regression model directly on our target. Building a regression on the gas-phase descriptors $\bxcg$, while conceptually nonsensical (the descriptors of the molecules contain no information on the intermolecular interactions), yields an RMSE of \rhmrmse \unitz{}. 
Regressing on the solid-phase descriptors $\bxcc$ improves the regression substantially, achieving an RMSE of \rhcrmse \unitz{}. 

Yet, conceptually, neither of these two representations ($\bxcc$ and $\bxcg$) contain the full set of relevant information -- the molecular descriptor $\bxcg$ is missing information on intermolecular interactions, and the crystal descriptor $\bxcc$ is unaware of the conformational changes that the molecules undergo upon crystallization. The necessity of this missing information is confirmed when we regress on concatenated descriptors $\lbrace\bxcc, \bxcg\rbrace$ and our RMSE drops to \rhcmrmse \unitz{}.

\begin{figure*}
    \centering
    \includegraphics[width=0.95\linewidth]{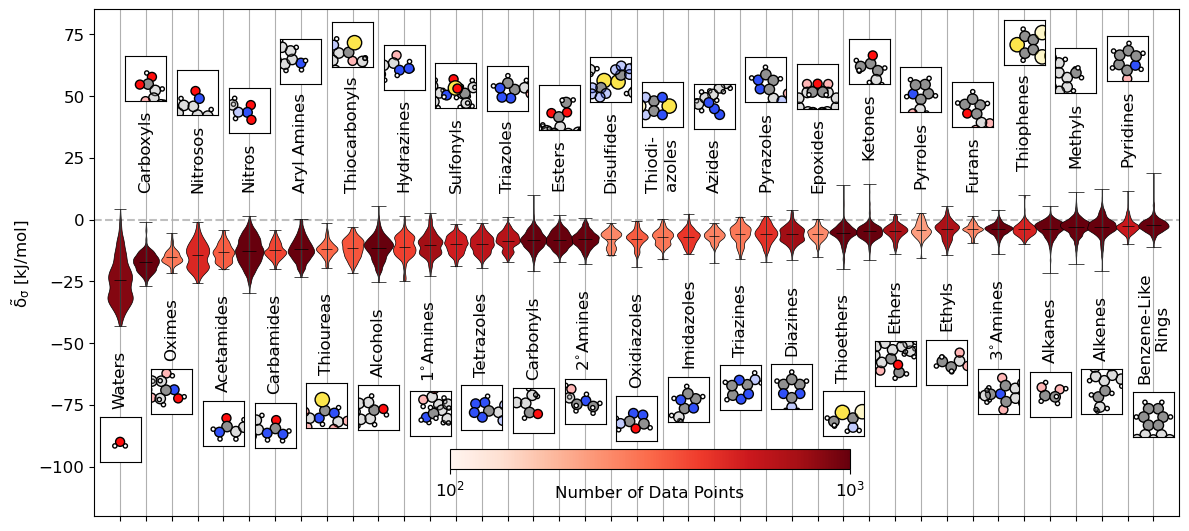}
    \caption{\textbf{Distribution of Energetic Contributions for Different Functional Groups.} For each functional group, we have taken the averaged remnant descriptor $\bx_\sigma^{(s-g)}$ and computed the estimated contribution to the binding energy $\tbysig$ using the regressions detailed in ~\eqref{eq:bya} and the filtering procedure in  ~\eqref{eq:tbya}}. We have arranged the functional groups in order of average contribution, with a representative example is shown above or below the violin plot with the functional group highlighted. We have limited this figure to those functional groups with more than 200 instances in the dataset (see Figure S4 for all groups). The lines on each plot denote each group's extreme and mean contributions. The plots are colored by the number of examples within the dataset, ranging from \allMinN~(\allMinNMotif) to \allMaxN~(\allMaxNMotif~groups). Wider sections of the violin plot represent a higher probability that members of the population will take on the given value; the skinnier sections represent a lower probability.
    \label{fig:violin}
\end{figure*}

Furthermore, \eqref{eq:yw} provides another way to similarly (and more explicitly) encode the nature of the problem into our choice of representation. Given a descriptor that appropriately distinguishes between periodic crystals and molecules, a regression model can predict their energies using the same regression weights, $\bw = \bwc = \bwm$. Substituting this into \eqref{eq:yw},
\begin{align}  
\byc &= \bxc \bw - \bxcg\bw + \berror \\
     &= \bxcr \bw + \berror
 \label{eq:bxcrw}
 \end{align}
where we define $\bxcr \equiv \left(\bxcc -\bxcg\right)$ as the so-called ``remnant'' descriptor and $\berror$ again denotes the residual errors. Explicitly adapting our representation $\bxcr$ to the nature of the lattice energy results in a yet better result to learning on $\lbrace\bxcc, \bxcg\rbrace$: \rhrrmse \unitz{}, despite being in a smaller feature space.
Conceptually, this descriptor still encodes the 3-body correlation between an environment and its neighbors but explicitly incorporates the change in molecular geometry upon crystallization and reduces the weights of atomic triplets whose interactions are primarily intramolecular and/or the same in gas and solid phase. 

\paragraph*{Extension to non-linear models} This result is mirrored in non-linear regression models, where again, a superior result is obtained by either constructing a kernel on $\bxcr$ or taking the difference of non-linear feature vectors (see Appendix C2). An optimized RBF kernel on the remnant descriptors yields a similar RMSE to the linear model, likely due to the restricted dataset size and diversity. We get some improvement (by $\sim$0.06 \unitz{} compared to the best linear model) by taking the difference of the non-linear features defined by the kernels of the crystalline and molecular descriptors. This result further emphasizes the rationale behind the remnant approach, and suggests that one can improve accuracy by combining non-linear feature constructions to mimic the mathematical formulation of target properties. To show that kernel optimization has little impact on this conclusion, we have also included corroborating results using a parameter-free kernel, also in Appendix C2.

When the molecular geometry is known \textit{a priori}, these results suggest that linear and non-linear regressions for the lattice energy should be built on descriptors conceptually akin to $\bxcr$, rather than $\bxc$, as has been common practice in the literature\cite{seko_representation_2017,bereau_transferable_2015,musil_machine_2018,wengert_data-efficient_2021}. Thus, in the remainder of the text, we will employ ML fingerprints and models based on the remnant descriptor.

\subsection{Estimating the contributions of molecular motifs}
\label{sec:filter}
\paragraph*{Regularizing the Atomic Contributions} With our target-adapted regression model, we can assign effective contributions to each atomic environment, where we take the remnant descriptor of each atomic environment and compute
\begin{equation}
\bya = \bxar \bw. 
\label{eq:bya}
\end{equation}
Despite the mathematical logic behind this step, the lack of physical underpinnings for this decomposition may result in energy being arbitrarily partitioned between neighboring atoms. This leads to disproportionately large contributions of opposite size being assigned, not dissimilar to how a regression may overfit by assigning large regression weights.
To ease this effect, we can apply a Gaussian filter to each $\bya$. For the $i^{th}$ atom, this results in
\begin{equation}
\tilde{\by}_{i} = \sum_{j} \by_{j} \frac{f(i, j)}{\sum_k f(j, k)} 
\label{eq:tbya}
\end{equation}
where $\sum_{j}$ runs over all neighbors of $i$ and $\sum_k$ runs over all neighbors of $j$ 
(defined by a cutoff of 2\AA). For neighbors $a$ and $b$ and interatomic distance $d_{ab}$, 
$f(a, b) = \exp\left[d_{ab}^2 / 2\varsigma^2\right]$. 
This procedure, introduced for the electronic density of states in \citet{ben_mahmoud_learning_2020}, has the effect of regularizing the decomposition while maintaining the regression results, \ie $\byc = \frac{1}{n_c}\sum_{a \in c} \bya = \frac{1}{n_c}\sum_{a \in c} \tbya.$
We show this effect of the filter on the distribution of atomic contributions in Appendix B2.
It is worth to compare our data-driven decomposition with one based on an empirical model of interactions, or with one of the many atoms-in-molecules decompositions of the energy computed by quantum-chemical calculations. 
On one hand, our approach makes it harder to explicitly interpret the stabilizing power of a motif in terms of physical terms (electrostatics, dispersion....). On the other, in many cases forcefields and energy decompositions have a high degree of arbitrariness, and the accurate prediction of the total binding energy comes from a cancellation of errors in the individual components. 
The atomic contributions~\eqref{eq:tbya} are obtained with the only requirement of being smooth, and (since they are built using a remnant descriptor) to correlate with the structural features associated with the crystal-forming process. As we shall see, their nature allows one to recognize the role played by collective effects - such as steric hindrance, or molecular distortions - contributing to our goal of a holistic view of lattice stability. 

\paragraph*{Visualizing the Contributions of Different Motifs}
Taking the \ntrainmol molecules from our training set, we use SMARTS descriptors\cite{noauthor_notitle_1997} and RDKit Substructure Matching \cite{noauthor_rdkit_nodate} to identify the atoms belonging to common molecular motifs, finding \allN~motifs. Details of this procedure and our table of SMARTS strings are given in Appendix B3 and Table S3, respectively.
For each motif, we determine the effective cohesive interaction $\tbysig$ as\footnote{The lattice energy of the crystal is not the sum of these motif contributions, as 1) both are averaged quantities, and 2) a single crystal may have overlapping motifs.}
\begin{equation}
\tilde{\by}_{\sigma} = \frac{1}{n_\sigma}\sum_{a \in \sigma} \tilde\by_{a}
\label{eq:stbya}
\end{equation}
We plot the span of lattice energy contributions for motifs with greater than 200 instances in the dataset in \figref{fig:violin}. The functional groups are arranged in order of increasing average cohesive interactions. Nearly all functional groups, on average, are stabilizing, although we see a clear trend in the nature of the functional groups from left to right. On the left (the motifs leading to the strongest intermolecular interactions), there are groups typically associated with hydrogen bonding (\eg carboxyls and waters). As we move to the right, the molecular motifs are, on average, weakly binding, with the largest range of interactions coming from the most broadly-defined groups, including the alkanes, alkenes, and benzene-like rings.

\begin{figure}[!ht]
    \centering
    \includegraphics[width=\linewidth]{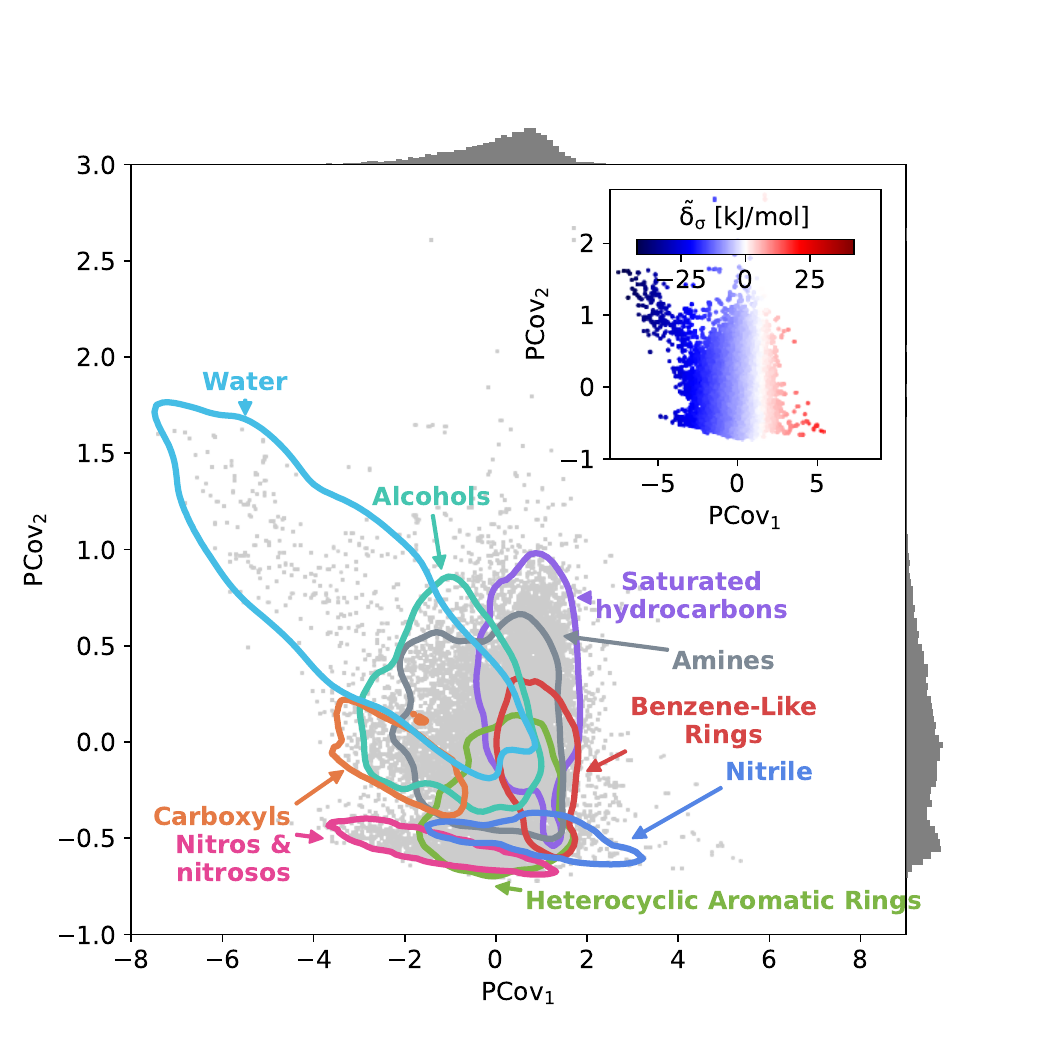}
    \caption{\textbf{Principal Covariates Regression (PCovR) Map of the Interactions of Molecular Motifs.} A structure-property map of molecular motifs, denoting major classes of motifs and outlining the regions where the 90\textsuperscript{th} percentile of these motifs occur. (inset) the same map colored by the cohesive interactions, ranging from blue (strongly attractive) to white (neutral) to red (strongly resistant). Histograms on the upper and right borders show the distribution of motifs along the covariates.}
    \label{fig:pcovr_all}
\end{figure}

This trend is further demonstrated by plotting the structure-property map of all motifs using \textit{Principal Covariates Regression} (PCovR), a hybrid supervised-unsupervised dimensionality reduction technique first introduced in \citet{de_jong_principal_1992} and adapted to chemical systems in \citet{helfrecht_structure-property_2020} This technique produces a latent-space mapping that arranges different motif classes based on their structural similarity and correlation to a set of target properties. In \figref{fig:pcovr_all}, we show a map using the average remnant descriptor for each motif and their average energy contribution, using contour lines to show where 90\% of such motifs fall on the PCovR map. 
One sees that, in this case, the first axis of this plot (PCov$_1$) correlates strongly with the (learnable) cohesive interactions. The second axis (PCov$_2$) allows us to resolve structural differences between motifs with similar energetic contributions. In this mapping, we can learn from the spread of each group. For example, the \waterN~water molecules (light blue in \figref{fig:pcovr_all}) span the greater portion of the left-hand side of the figure, highlighting the chemical diversity of intermolecular water interactions. Juxtapose this with the 2'627 nitro and nitroso groups (pink in \figref{fig:pcovr_all}) that span a smaller region in PCovR space, implying a narrower range of intermolecular interactions. Here we have combined several groups for visual simplicity; however, we have included plots highlighting each functional group in Figs. S6-S9, including the sample sizes and range of contributions.

\begin{figure*}[t]
    \centering
    \includegraphics[width=0.9\linewidth]{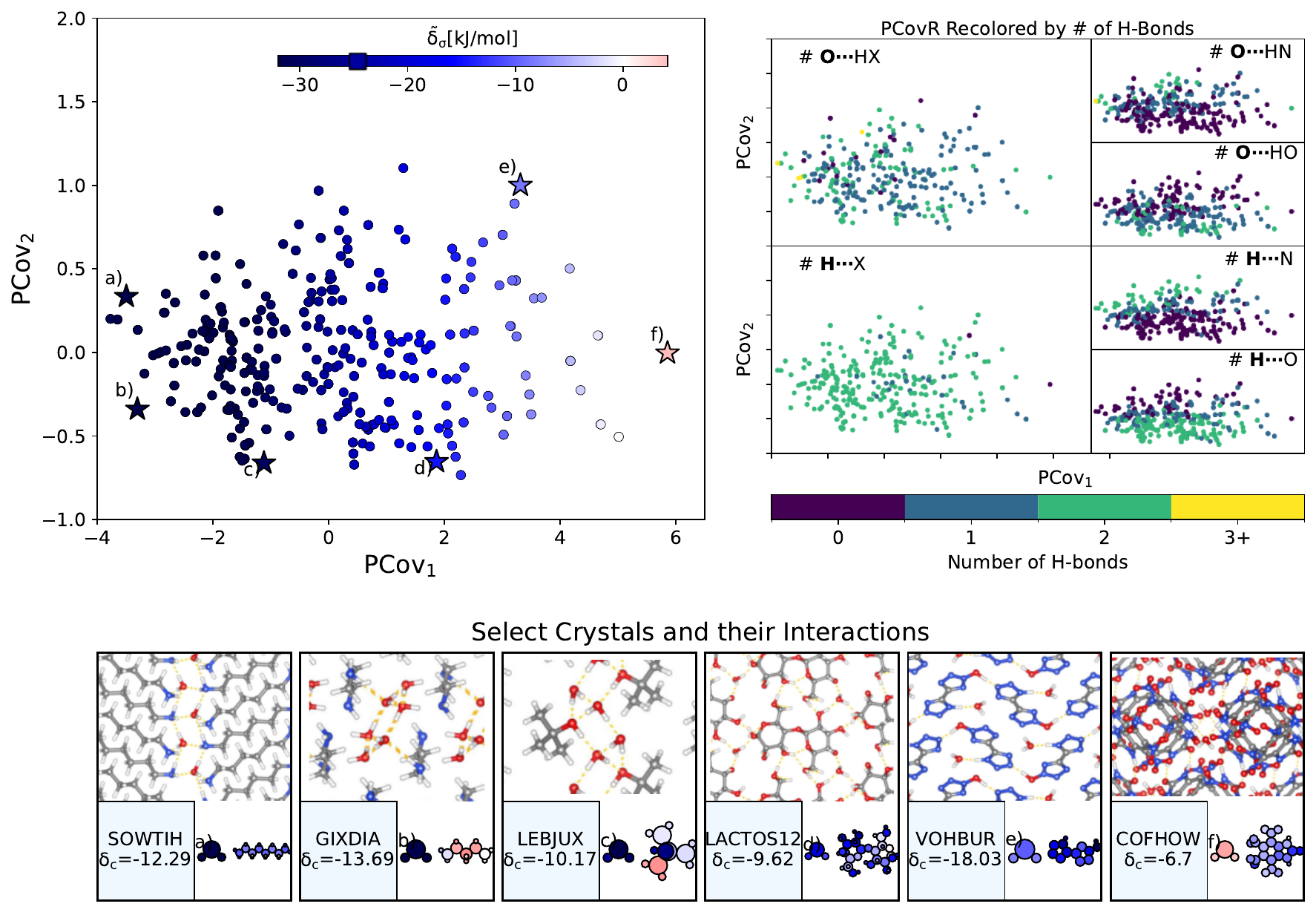}
    \caption{\textbf{The Interactions of Water Molecules.} (left) Principal Covariates Regression (PCovR) map, where the color of each point denotes the estimated cohesive interaction of that motif and a marker on the color bar denotes the average value for all waters. (right) The PCovR map recolored by the number of hydrogen bonds (H-bonds), separating those donated to the oxygen atom (top) from those donated by the hydrogen atoms (bottom). The insets on the bottom visualize several extremal or interesting environments. Select crystalline configurations and energy assignments: CSD Refs.~(a)\ccdc{SOWTIH}, (b)\ccdc{GIXDIA}, (c) \ccdc{LEBJUX}, (d) \ccdc{LACTOS12}, (e) \ccdc{VOHBUR}, and (f) \ccdc{COFHOW}. In each panel, the bottom row shows the total lattice energy of the crystal (in \unitz{}) and the corresponding molecules where the atoms have been recolored by their estimated lattice energy contribution (on the same scale as on the PCovR map).
    }
    \label{fig:water}
\end{figure*}

The PCovR framework also provides a blueprint for analyzing the interactions of different structural motifs -- given a single motif type, what characteristics of a molecular environment lead to a more stabilizing interaction? In the following sections, we will take a look at the stabilizing environments for a few classes of functional groups, starting with the well-known stabilizing interactions 
of water and
carboxylic groups, then moving onto two groups with a wide range of intermolecular interactions, 6-membered aromatic carbon rings and nitro groups. With each functional group, we generate a new PCovR only using the averaged remnant descriptors and effective interactions for the instances of that group, such that the structural diversity embedded in the map reflects the diversity of \emph{interactions}, rather than the diversity of the molecules.
We have included similar maps for all other molecular motifs in an online data repository\cite{matcloud_entry,talirz_materials_2020}.

\subsubsection{Waters}
\label{sec:water}
We begin with a ubiquitous molecular crystal stabilizer: water. The estimated contributions of the \waterN~water molecules in this dataset span a range of \mbox{\waterMinEnergy} to \mbox{\waterMaxEnergy}\unitz{}(e of atoms), with the majority of interaction strengths occurring at around \mbox{\waterMeanEnergy}$\pm$\waterStdEnergy\unitz{}.
We generate a new PCovR shown in the left panel of \figref{fig:water}. On the bottom of \figref{fig:water}, we show the crystalline conformation and the molecules recolored by $\tbya$.

\begin{figure*}[t]
    \centering
    \includegraphics[width=0.9\linewidth]{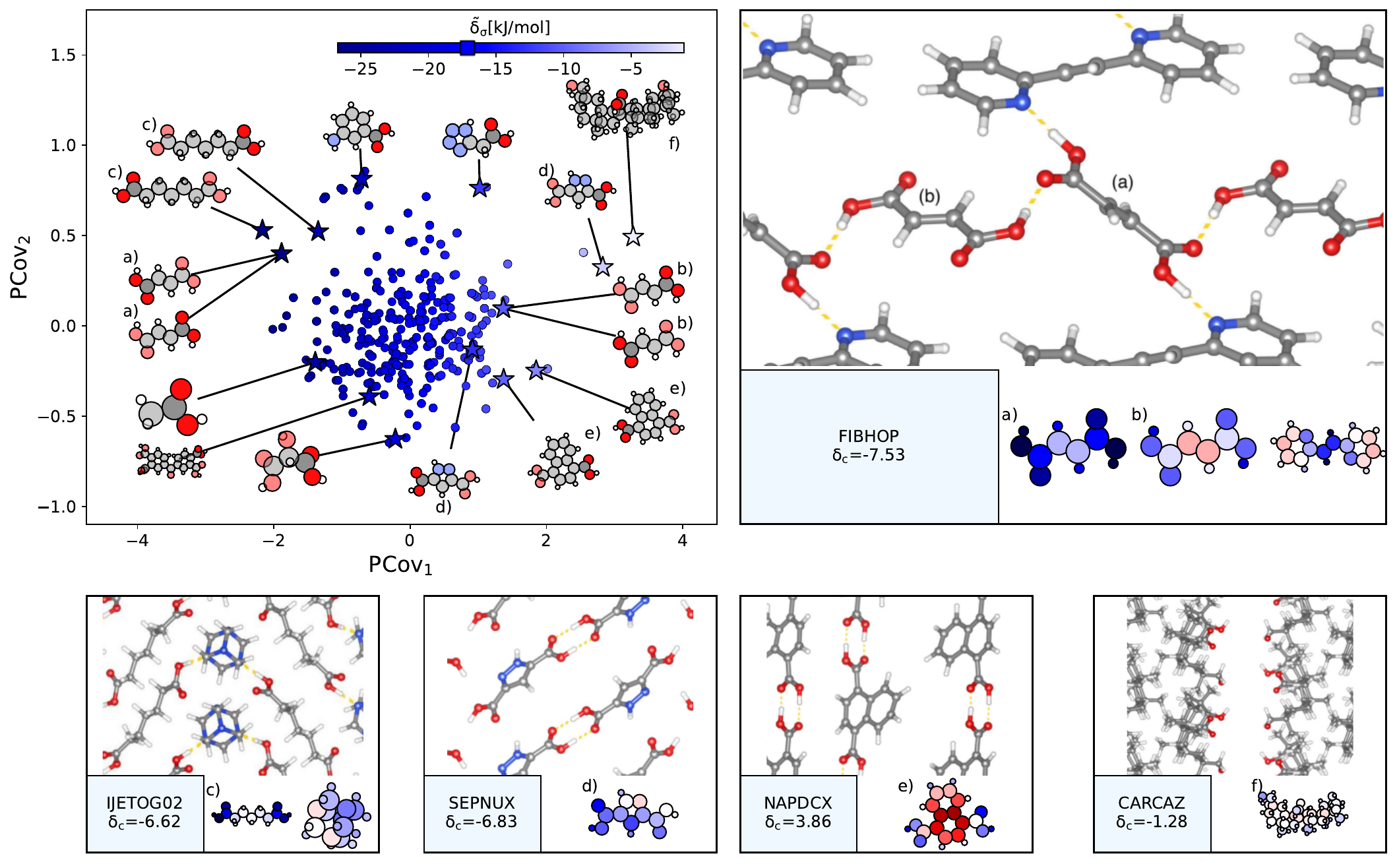}
    \caption{\textbf{The Interactions of Carboxylic Acid Groups.} (left) Principal Covariates Regression (PCovR) Map, where the color of each point denotes the estimated cohesive interaction of that motif, and a marker on the color bar denotes the average value for all carboxylic acid groups. The insets visualize several extremal or interesting motifs. (right) Select crystalline configurations and energy assignments: CSD Refs.~(a,b) \ccdc{FIBHOP}, (c) \ccdc{IJETOG02}, (d) \ccdc{SEPNUX}, (e) \ccdc{NAPDCX}, and (f) \ccdc{CARCAZ}. In each panel on the right, the bottom row shows the total lattice energy of the crystal (in \unitz{}) and the corresponding molecules where the atoms have been recolored by their estimated lattice energy contribution.}
    \label{fig:carb}
\end{figure*}

First, we look at a common parameter for measuring the stabilizing effect of water: hydrogen bonding (H-bonding). Here, we have calculated H-bonds based on when the O$\cdots$H or H$\cdots$X distance is less than 2.5\AA~and the dihedral angle of O$\cdots$H-X or OH$\cdots$X is greater than $150^\circ$. From the right side of \figref{fig:water}, we see that the number of H-bonds donated to the water molecule (O$\cdots$H) does not correlate with the cohesive interaction of the water molecules. There is some qualitative correlation/anti-correlation between the nature of these donated H-bonds and the second principal covariate (Pearson Correlation Coefficient, or PCC, =0.49,-0.59 for the number of O$\cdots$H-N and O$\cdots$H-O, respectively). There is a mild anti-correlation between the number of H-bonds the water itself donates (OH$\cdots$X) and the first covariate, with a PCC of -0.33. The second principal covariate is strongly correlated and anti-correlated with the number of OH$\cdots$N and OH$\cdots$O interactions, achieving a PCC of 0.69 and -0.73, respectively. Waters with primarily OH$\cdots$N-type hydrogen bonds are at the top of the map (\eg \figref{fig:water}(e), \CCDC{VOHBUR}), with OH$\cdots$O-type at the bottom of the map (\eg \figref{fig:water}(c-d), CSD Refs. \ccdc{LEBJUX} and \ccdc{LACTOS12}).

This analysis emphasizes that the number of hydrogen bonds does not fully capture all of the nuances of water stabilization -- the majority of water molecules participate in 2-3 such interactions, and the energy of these bonds can span a wide range. In O$\cdots$H-X interactions, there is little energetic difference based on whether the acceptor is a nitrogen or oxygen atom -- both types of hydrogen bonds span the full range of energies. The nature of the acceptor is encoded in the covariate orthogonal to the chemical features most correlated with interaction strength (\ie the nature of the acceptor is primarily correlated with the second covariate).

We see that the strongest water interactions in 1,6-Diaminohexane monohydrate (\CCDC{SOWTIH}, \figref{fig:water}(a)) and 1,3-Diaminopropane trihydrate (\CCDC{GIXDIA}, \figref{fig:water}(b)), where the water molecules associate with other water molecules and the amine group of their co-crystalline molecule. Our weakest contribution, by far, occurs in 4,5,6,7-Tetranitro-1,3-dihydro-2H-benzimidazol-2-one hemihydrate (\CCDC{COFHOW}, \figref{fig:water}(f)), where the water molecules sit interstitial to the imidazole molecules, prohibited from forming hydrogen bonds and potentially interfering with the stabilization of the imidazole clusters.

\subsubsection{Carboxylic Acid Groups}
\label{sec:carb}
As a strong electron donor, carboxylic acids are considered a key motif in molecular crystallization\cite{etter_encoding_1990, etter_hydrogen_1991,spackman_fingerprinting_2002}, which is supported by their strong negative lattice energy contribution, here ranging from \mbox{\COOHMinEnergy} \unitz{} to \mbox{\COOHMaxEnergy} \unitz{}, with the majority of interaction strengths occurring in the \COOHMeanEnergy$\pm$\COOHStdEnergy \unitz{} range. 
Taking the \COOHN~carboxylic acid groups, we generate a new PCovR shown in the left panel of \figref{fig:carb}. On the right and bottom of \figref{fig:carb}, we have included panels showing, for select motifs, the crystalline conformation and molecules recolored by $\tbya$. 

The strongest contributions are found in 1,2-Di(2-pyridyl)ethylene (\CCDC{FIBHOP}) in a succinic acid molecule (\figref{fig:carb}(a)) that forms two sets of supramolecular synthons: one homosynthon with the other succinic acid (\figref{fig:carb}(b)), and one heterosynthon with the pyridine group (consistent with the literature on the strength of carboxylic-pyridine interactions\cite{vishweshwar_recurrence_2002,shattock_hierarchy_2008,chen_preferred_2016}). Interestingly, this crystal also contains one of the most weakly interacting groups (\figref{fig:carb}(b)), in the second succinic acid molecule that only participates in the single homosynthon. 

\begin{figure*}
    \centering
    \includegraphics[width=0.9\linewidth]{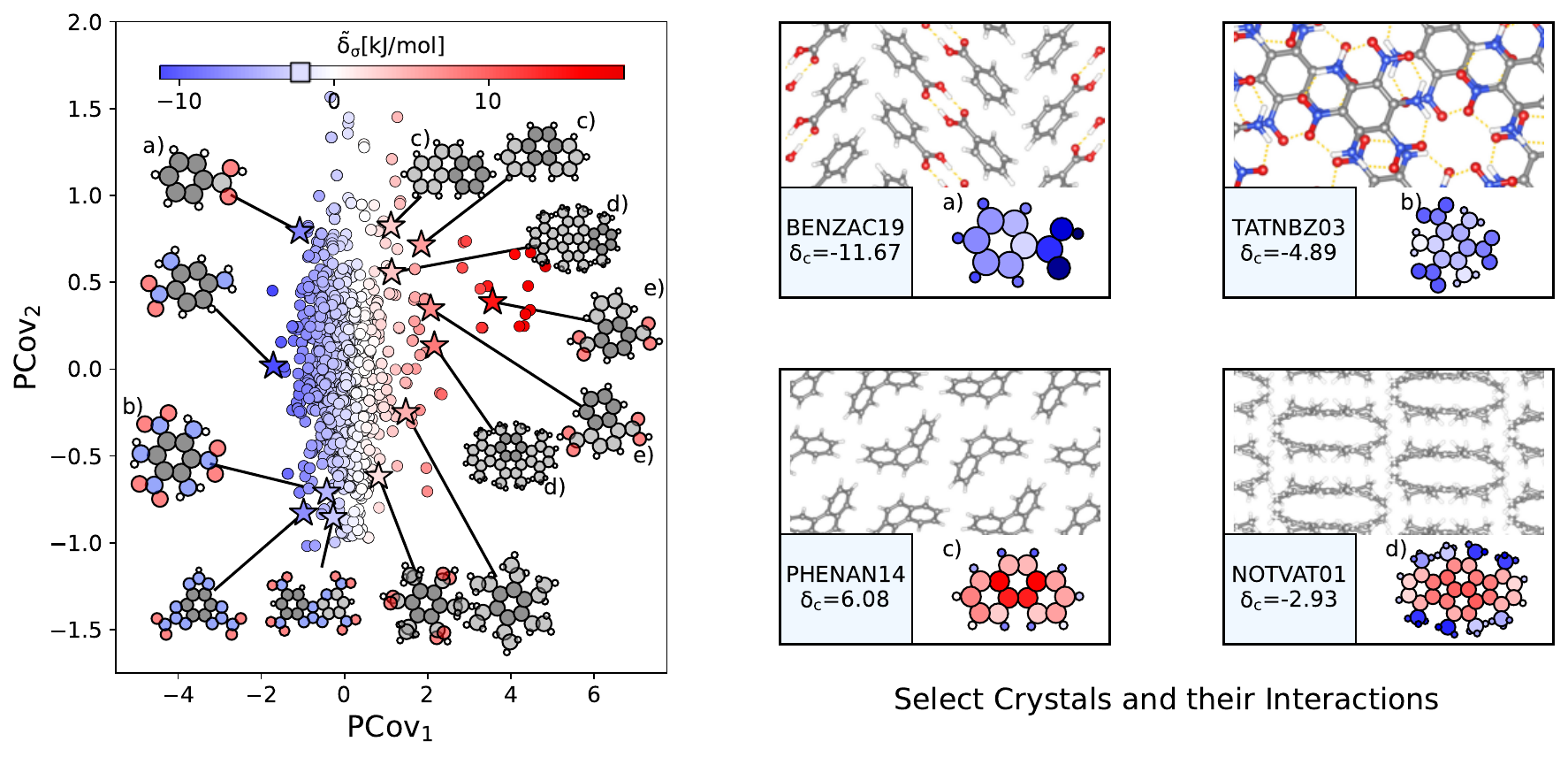}
    \caption{\textbf{The Interactions of Benzene-like Rings.} (left) Principal Covariates Regression (PCovR) Map, where the color of each point denotes the estimated cohesive interaction of that motif and a marker on the color bar denotes the average value of benzene-like rings. The insets visualize several extremal or interesting motifs. (right) Select crystalline configurations and energy assignments: CSD Refs.~(a)\ccdc{TATNBZ03}, (b)\ccdc{BENZAC19}, (c) \ccdc{PHENAN14}, and (d) \ccdc{NOTVAT01}. We also highlight the benzene-like motif from \figref{fig:carb}(f) in (e). In each panel on the right, the bottom row shows the total lattice energy of the crystal (in \unitz{}) and the corresponding molecules where the atoms have been recolored by their estimated lattice energy contribution (on the same scale as on the left panel).
    }
    \label{fig:benz}
\end{figure*}

\begin{figure}
    \centering
    \includegraphics[width=0.875\linewidth]{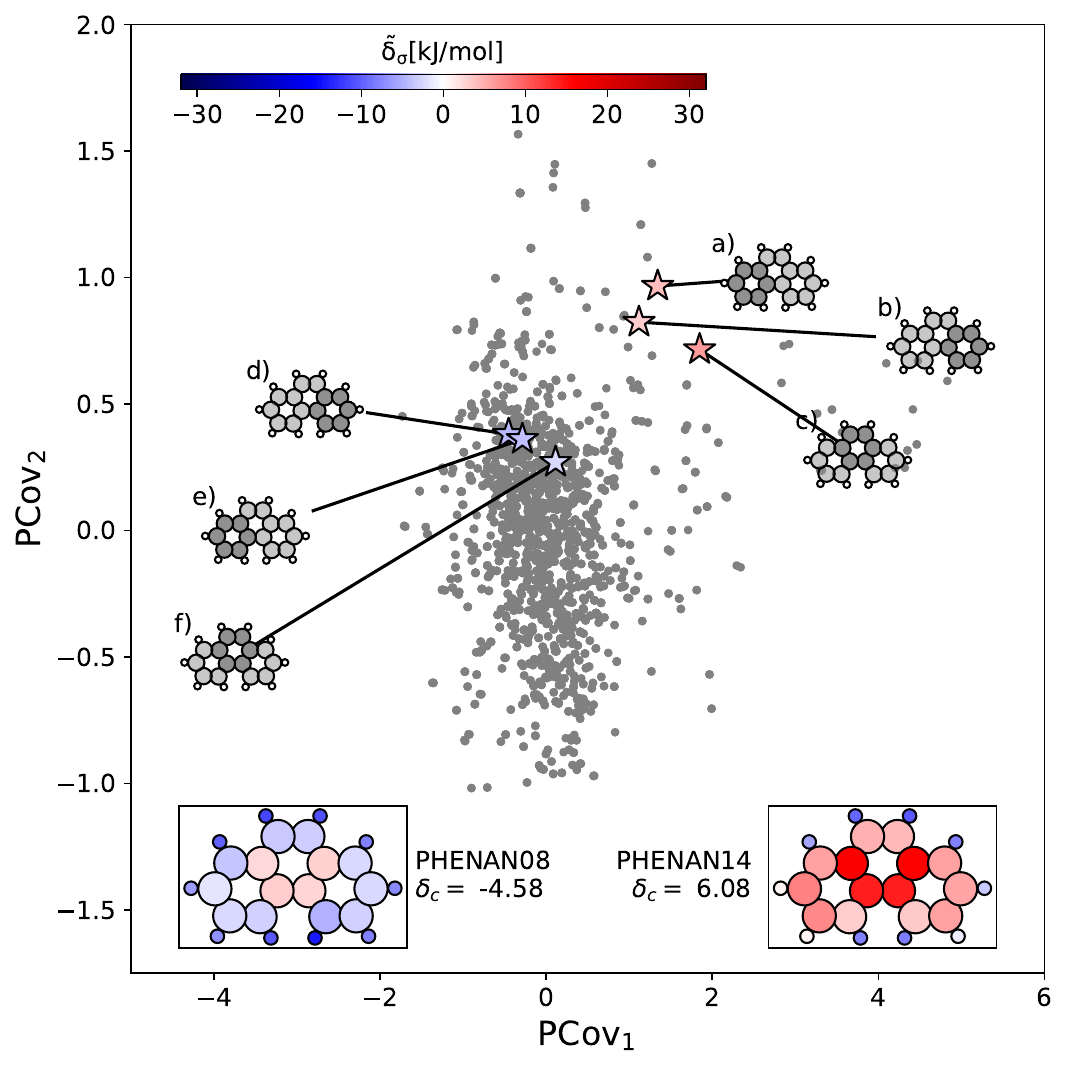}
    \caption{
    \textbf{Comparing the Motifs in Polymorphs of Phenanthrene.} Here we project two distinct polymorphs of phenanthrene onto the PCovR map shown in \figref{fig:benz}. At ambient conditions, one polymorph is stable (\ccdc{PHENAN08}), while the other is unstable (\ccdc{PHENAN14}, also shown in \figref{fig:benz}(c)). Looking at the same motifs in the unstable and stable phases, we see a shift leftwards as the motifs go from resisting crystallization to weakly binding. In the lower insets, we have recolored the atoms of the phenanthrene molecule based upon their contribution to the lattice energy in the different polymorphs. Note that the bicyclic carbon atoms, while no longer distorted, weakly resist crystallization, as they prevent the auxiliary hydrogens from more closely interacting with neighboring $\pi$ bonds by distorting the molecule.
    }
    \label{fig:phenan}
\end{figure}
Carboxylic acids form the strongest cohesive interactions when participating in multiple synthons, particularly heterosynthons (and typified by \figref{fig:carb}(a) and (c), and noted in earlier literature \cite{leiserowitz_molecular_1976}). Moving to the right, we see the contribution decrease commensurate to the number of interactions. For example, in 3,5-Pyrazoledicarboxylic acid (\CCDC{SEPNUX}, \figref{fig:carb}(d)), there are two carboxylic acid groups that have drastically different energy contributions -- one that forms a doublet homosynthon and the other is without close contacts. 
In an extreme case (\CCDC{CARCAZ}, \figref{fig:carb}(f)), the carboxylic acid group is prevented from interacting due to the bulkiness of the overall molecule, leading to a neutral contribution. 

An interesting success of this energy assignment is the ability to identify stabilizing motifs in otherwise unstable or metastable crystals. This is the case for \CCDC{NAPDCX} (\figref{fig:carb}(e)), an unstable 1,4-Naphthalene-dicarboxylic acid that has an overall positive lattice energy at ambient pressure and temperature\footnote{Of the approximately 3'200 crystals studied in this article, only 23 have a positive lattice energy. This is covered in more detail in Appendix A3}. Despite this instability, we can clearly identify a binding interaction between carboxylic acid groups.

\begin{figure*}
    \centering
    \includegraphics[width=0.9\linewidth]{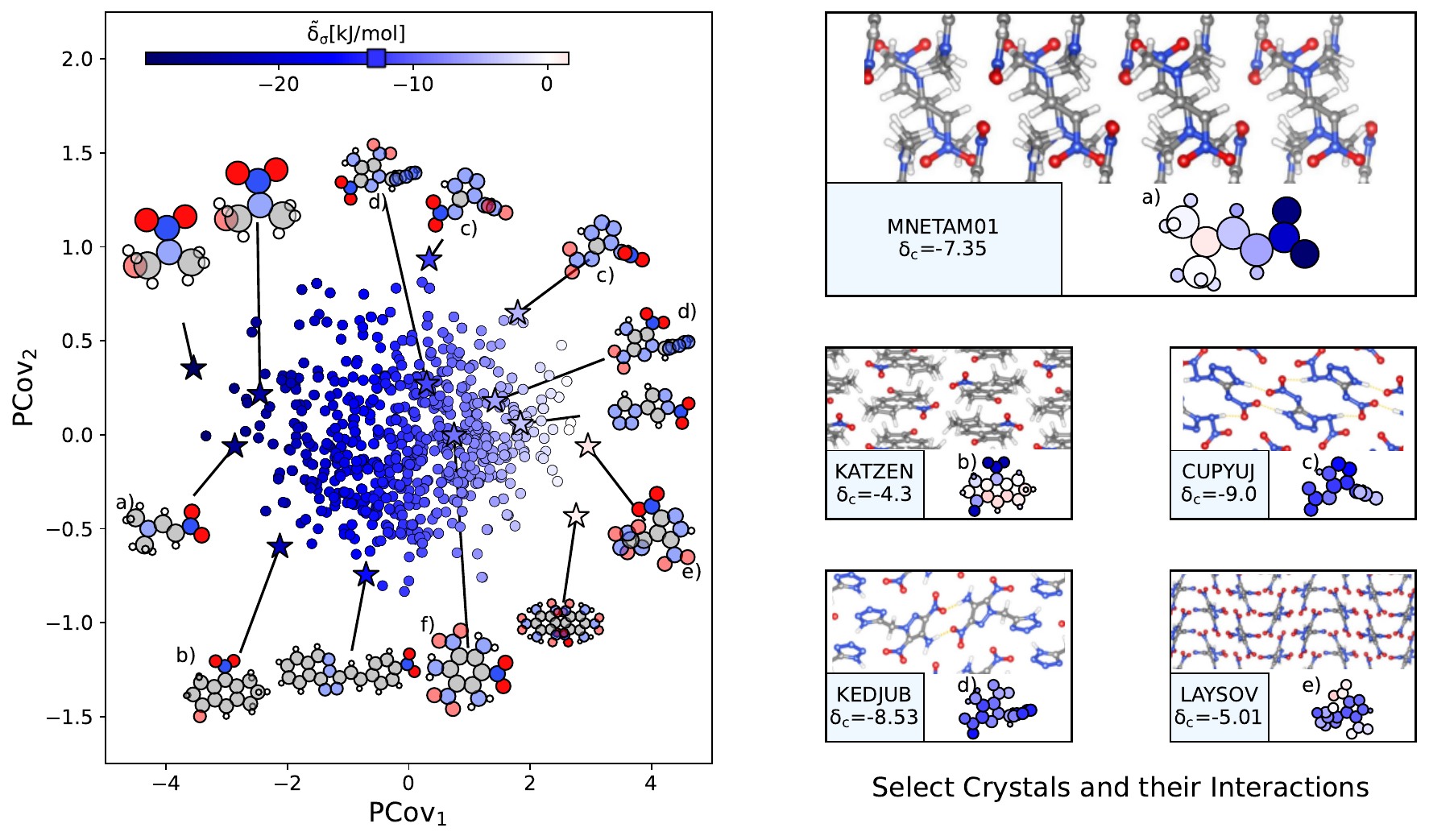}
    \caption{\textbf{The Interactions of Nitro Groups.} (left) Principal Covariates Regression (PCovR) Map, where the color of each point denotes the estimated cohesive interaction of that motif. (right) Select crystalline configurations and energy assignments: CSD Refs.~(a)\ccdc{TIJKEC}, (b) \ccdc{KATZEN}, (c) \ccdc{CUPYUJ}, (d) \ccdc{KEDJUB}, and (e) \ccdc{LAYSOV}. 
    We also highlight the nitro group of \ccdc{TATNBZ03} from \figref{fig:benz}(b) in (f). In each panel on the right, the bottom row shows the total lattice energy of the crystal (in \unitz{}) and the corresponding molecules where the atoms have been recolored by their estimated lattice energy contribution (on the same scale as on the left panel).
}
    \label{fig:nitro}
\end{figure*}
\subsubsection{6-Membered Unsaturated Carbon Rings}
6-member unsaturated carbon rings (consistent with benzene molecules but more broadly-defined to include branched rings) show weak intermolecular interactions ranging from \mbox{\benzeneMinEnergy} \unitz{} to \benzeneMaxEnergy \unitz{}, with the majority of interactions occurring in the \benzeneMeanEnergy$\pm$\benzeneStdEnergy \unitz{} range. Similar to \secref{sec:carb}, we generate a new PCovR using the averaged remnant descriptors and effective interactions using the \benzeneN~benzene-like motifs, as shown in the left panel of \figref{fig:benz}. Again, we have included a panel on the right showing the crystalline conformation and molecules colored by $\tbya$ for select configurations. 

The most strongly-binding benzene-like motifs occur in molecules where 1) the ring is functionalized by strongly interacting groups, 2) the interactions of these groups facilitate planar molecular geometry, and 3) stacking occurs between the benzene-like rings with these auxiliary groups. We see this in 2,4,6-trinitrobenzene-1,3,5-triamine (\CCDC{TATNBZ03}, \figref{fig:benz}(b)), where the aromatic carbon ring stacks above the primarily \emph{intra}molecular nitro-amine interaction and in \figref{fig:benz}(a) (\CCDC{BENZAC19}), where they stack above the carboxylic acid homosynthon.

There are various reasons for weakly-binding benzene-like motifs, including weak stacking and steric hindrance. As is evident from \figref{fig:benz}(c-d), rings will \emph{resist} crystallization when the interactions of the end groups lead to deformation of the ring geometry. 
Take for example phenanthrene (\CCDC{PHENAN14}, \figref{fig:benz}(c)), a high-pressure polymorph that is unstable at ambient conditions (therefore has an overall positive lattice energy for the DFT reference used). Interestingly, we can pinpoint the localization of this deformation by looking at the atoms with the strongest positive contribution.
While the keen reader may infer that this is solely due to the remnant descriptor reflecting the difference in strained and relaxed molecular geometry, we will note that a large difference in these representations can also coincide with a wealth of stabilizing intermolecular interactions, demonstrating that this simple linear model can differentiate molecular deformation from the introduction of new interactions.

This is further supported by comparing the motifs of this polymorph with its ambient-pressure, stable counterpart (\CCDC{PHENAN08}) to see how the nature of the same molecule changes based upon the interactions in the crystal. Both polymorphs adopt a similar herringbone crystal structure; however, the decreased molecular distortion and increased interactions between the auxiliary hydrogens and neighboring aromatic rings in \ccdc{PHENAN08} results in a significantly lower lattice energy of $\byc = -4.58$. In \figref{fig:phenan}, we project the motifs of \ccdc{PHENAN08} and \ccdc{PHENAN14} onto our PCovR map from \figref{fig:benz}, we see this reflected by a left-shift of the motifs on the map, where the center ring moves from strongly resisting crystallization (\figref{fig:phenan}(c)) to weakly interacting (\figref{fig:phenan}(f)) and the periphery rings move from weakly resisting crystallization (\figref{fig:phenan}(a,b)) to weakly binding (\figref{fig:phenan}(e,f)). It is worth noting that \ccdc{PHENAN08} is an out-of-sample data point ($\epsilon = 0.2$\unitz{}), demonstrating that the analysis in \figref{fig:benz} is applicable beyond the initial reference set. We have included images of the \ccdc{PHENAN08} crystal configuration in Figure S5.

\subsubsection{Nitro Groups}
Nitro groups, defined as a nitrogen atom bonded to two terminal oxygen atoms, range in cohesive contributions from \mbox{\nitroMinEnergy} \unitz{} to \mbox{\nitroMaxEnergy} \unitz{}, with most interaction strengths being \mbox{\nitroMeanEnergy}$\pm$\nitroStdEnergy\unitz{}. 
Similar to our previous examples, we generate a new PCovR using the averaged remnant descriptors and effective interactions of the \nitroN~nitro groups, as shown in the left panel of \figref{fig:nitro}. Again, we have included a panel on the right showing the crystalline conformation and constituent molecules colored by $\tbya$.
Unlike carboxyl and benzene-like groups, the chemical diversity of nitro interactions is limited -- this is either due to the chemical nature of nitro interactions or the availability of nitro-containing crystals in CSD.

The resonant or partial charge of the oxygen atoms leads to strong binding in hydrogen-rich environments, supported by the results in \figref{fig:nitro}. This is best typified by trans-N,N-Dimethyl-2-nitrovinylamine (\CCDC{MNETAM01}), a molecule where the nitro group is strongly interacting with the CH$_3$ end groups with some potential $\pi$-hole stacking\cite{bauza_directionality_2015} between the nitrogen moieties, as shown in \figref{fig:nitro}(a). The strength of these binding interactions lessens with the strength of the electron donors, with smaller contributions in crystals where the primary $O\cdots H$ interaction is with amine donors (\eg \figref{fig:nitro}(c, d), CSD Refs.~\ccdc{CUPYUJ}, \ccdc{KEDJUB}). In some of these cases, the binding is likely weakened by intramolecular interactions, similar to the contributions of the nitro groups in 2,4,6-trinitrobenzene-1,3,5-triamine (\CCDC{TATNBZ03}, \figref{fig:nitro}(f), seen earlier in \figref{fig:benz}(b)). Finally, to the right of the map, we see the strongest repulsive interactions from nitro groups in proximity to other nitro or aromatic nitrogen groups, such as the nitro-oxidiazole interaction in 3-(3,5-Dinitro-1H-pyrazol-4-yl)-4-nitro-1,2,5-oxadiazole (\CCDC{LAYSOV}, \figref{fig:nitro}(f)).

\begin{figure}
    \centering
    \includegraphics[width=\linewidth]{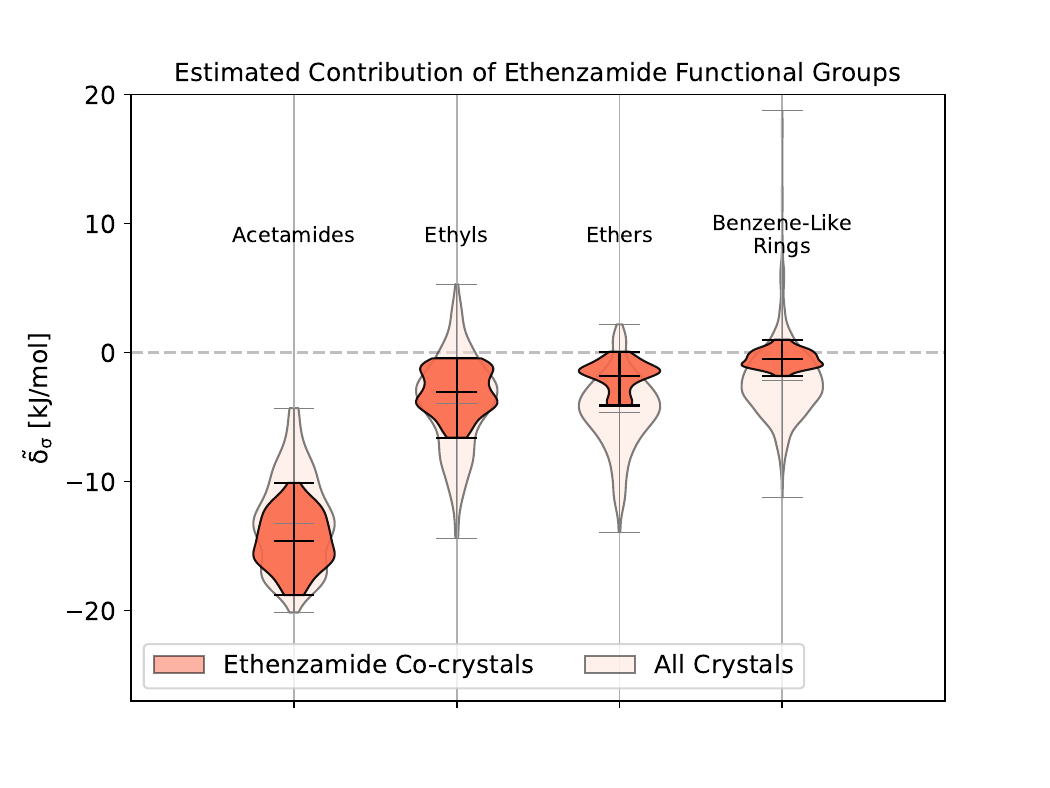}
    \caption{\textbf{Distribution of Energetic Contributions for the Functional Groups of Ethenzamide.} Following the procedure outlined in  \ref{sec:filter}, we have computed the estimated contribution to the binding energy $\tbysig$. Similar to \figref{fig:violin}, we have arranged the functional groups in order of average contribution, and the lines on each plot denote each group's extreme and mean contributions. Wider sections of the violin plot represent a higher probability that members of the population will take on the given value; the skinnier sections represent a lower probability. Here, darker sections refer to the distribution in the functional groups of the ethenzamide molecules, with lighter sections showing the distribution for the same functional group from \figref{fig:violin}.}
    \label{fig:ethenz_violin}
\end{figure}

\subsection{A Case Study: Ethenzamide Co-crystals}

We conclude by demonstrating how these models and methods can be used in the more practical context of crystal design.
Ethenzamide is a common analgesic that has been the subject of numerous co-crystallization studies \cite{JIFHAK,ODICUC-ODIDEN,FENQEX-ODIDEN01,QULLUF-QULLUF01-QULLUF02,REHSAA-REHSII-REHSUU-REHTAB,ORIKIL-ORIKOR-ORIKUX-ORILAE-ORILEI,WUZHOP-WUZHOP01-WUZJEH-WUZJIL-WUZJOR-WUZJUX-WUZKAE-WUZKEI, TIWPIB-TIWPOH, VAKTOS-VAKTOS01,VUHFIO-VUHFIO01} due to the poor solubility of its homocrystalline form\cite{DUKXAJ}. On the Cambridge Structure Database, there are currently 47 reported co-crystals of ethenzamide, of which there are 29 crystals that fit within the scope of this study and contain complete crystallographic information. The co-forming molecules in these co-crystals are primarily hydrobenzoic acids, nitrobenzoic acids, and dicarboxylic acids, as well as a 3-toluic acid co-crystal\cite{ORIKIL-ORIKOR-ORIKUX-ORILAE-ORILEI} and two saccharin co-crystals\cite{VUHFIO-VUHFIO01}. A list of these crystals with their experimental and computed properties is given in Appendix A4.

We first compute the relaxed energies of the co-crystals and their molecular components, following the procedures outlined in Appendix A to obtain the reference geometries and binding energies of each crystal. For reference, our previous model built using \eqref{eq:bxcrw} achieves an RMSE of 0.45\unitz{} and an MAE of 0.35\unitz{} more than sufficient to distinguish between the different categories of co-forming molecules, yet unable to provide any guidance in isomeric contexts (we have included a labeled parity plot in Figure S10). Following the procedure outlined in Appendix B, we identify the functional groups within the ethenzamide and estimate the contribution of their interactions to the molecular binding.

\begin{figure}[ht!]
    \centering
    \includegraphics[width=\linewidth]{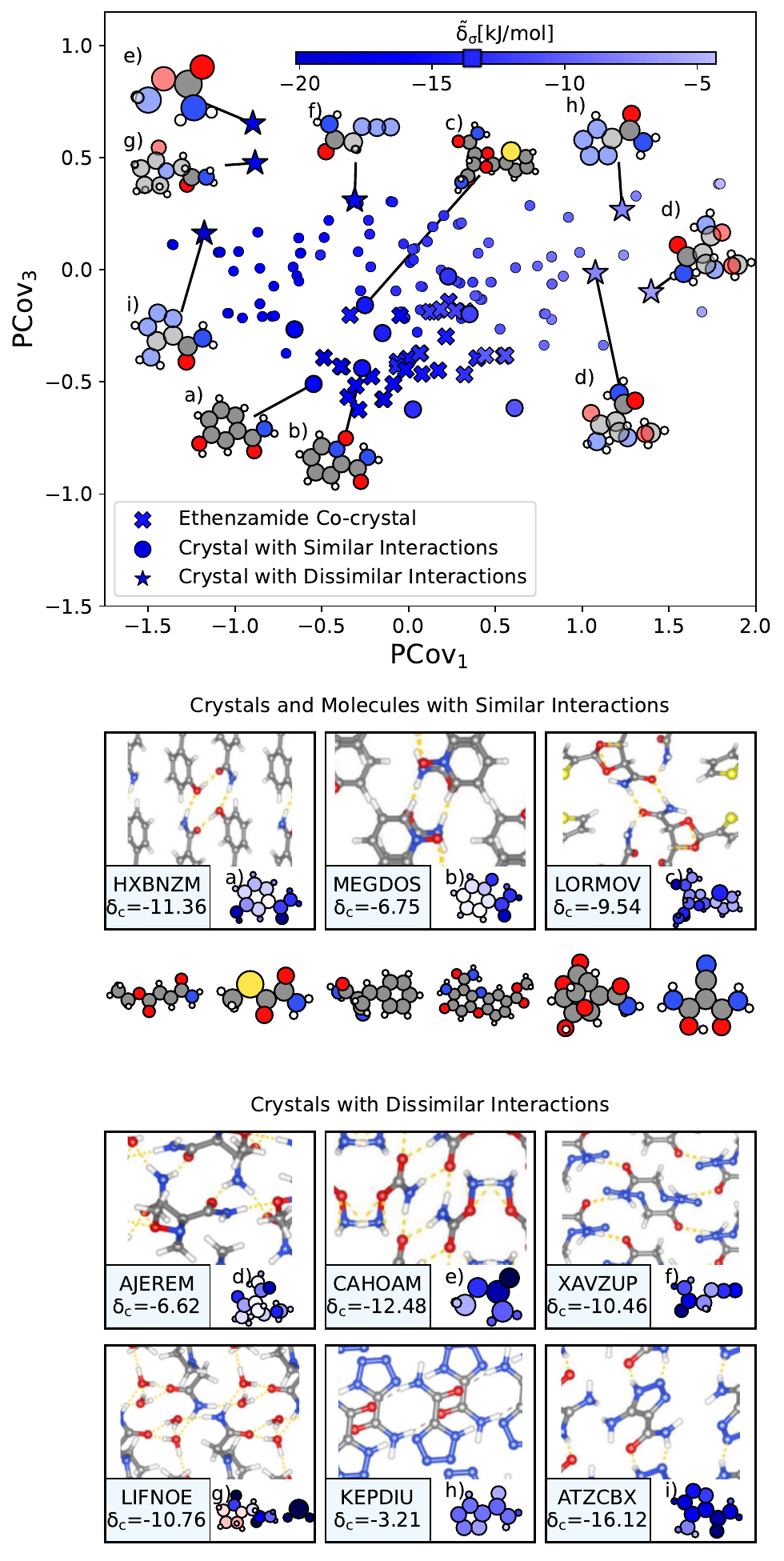}
    \caption{\textbf{The Interactions of Acetamide Groups.} (top) Principal Covariates Regression (PCovR) Map, where the color of each point denotes the estimated cohesive interaction of that motif. Here we have plotted along the first and third covariates to best distinguish the acetamide groups of the known ethenzamide co-crystals from other groups. The markers correspond to: (x) known ethenzamide co-crystals, (o) crystals with similar acetamide interactions, and (*) crystals with dissimilar acetamide interactions. (middle) Crystals and molecules with similar acetamide interactions as those in the known ethenzamide co-crystals, including CSD Refs.~(a) \ccdc{HXBNZM}, (b) \ccdc{MEGDOS}, and (c) \ccdc{LORMOV}. (bottom) Crystals dissimilar acetamide interactions to the known ethenzamide co-crystals: CSD Refs.~(d) \ccdc{AJEREM}, (e) \ccdc{CAHOAM}, (f) \ccdc{XAVZUP}, (g) \ccdc{LIFNOE}, (h) \ccdc{KEPDIU} and (i) \ccdc{ATZCBX}. Insets have been ordered from highest to lowest similarity to the interactions in the known ethenzamide co-crystals.\vfill}
    \label{fig:ethenz}
\end{figure}

As shown in \figref{fig:ethenz_violin}, most of the binding interactions unsurprisingly occur due to the acetamide group in the ethenzamide, with a 3.5\unitz{} difference between the weakest contributing acetamide motif and the most strongly contributing ethyl group, which is beyond the error in the overall model. From \figref{fig:ethenz_violin}, we can also see that, while the ethyl and benzene-like rings behave similarly to other similar motifs across the entire dataset, the acetamide and ether groups are generally more and less stabilizing, respectively, than their counterparts at large. With the ether groups, this is reasonable -- the geometry of the ether prevents much intermolecular interaction. With the acetamide group, this demonstrates that there is a large range of engineering that can happen to affect crystalline stability, which might be beneficial when considering molecular solubility.

From here, we use the PCovR of acetamide groups to identify other acetamide motifs that behave similarly, or dissimilarly, to those we see in the known ethenzamide co-crystals, as highlighted in \figref{fig:ethenz}. We first train our PCovR model on the acetamide groups in the training set, and project those from the ethenzamide dataset into the corresponding latent space. Because the interactions across the training set are much more diverse than within the ethenzamide set, we plot along the first and \emph{third} covariate to show the best distinction between the two datasets\footnote{There is an interactive map of the first four covariates within our online repository at \citet{matcloud_entry}}. We define similarity based upon the Euclidean distance in PCovR space -- that is, acetamide groups that appear at a similar place on the map in \figref{fig:ethenz}. Note that because we compute the distance using all covariates, some points that seem extremal in \figref{fig:ethenz} are not, as they are closer or further from the ethenzamide motifs in other dimensions.

We highlight the molecules that form the most similar acetamide networks to those in the ethenzamide dataset in \figref{fig:ethenz} using an (o) marker and showing the molecule below. Those closest in PCovR space are molecules that form single acetamide homosynthons (\eg \figref{fig:ethenz}(b,c), CSD Refs. \ccdc{MEGDOS} and \ccdc{LORMOV}) or heterosynthons with a carboxylic acid group (\eg \figref{fig:ethenz}(a), CSD Refs. \ccdc{HXBNZM}).

Perhaps more interesting are the acetamide groups that form different network interactions, highlighted in the right panel in \figref{fig:ethenz}. These groups give insight into the other supramolecular synthons that form with ethenzamide across a range of stabilizing and destabilizing contributions. We see strong interactions in triazole-5-carboxaldehyde (\figref{fig:ethenz}(i), \CCDC{ATZCBX}), where the acetamide group forms a heterosynthon with the triazole group, and in O-carbamoylhydroxylamine (\figref{fig:ethenz}(e), \CCDC{CAHOAM}), where the small size of the molecule facilitates both multiple homosynthons between acetamide groups as well as heterosynthons with the oxygen of the hydroxylamine groups. In 2-Oxopyrrolidineacetamide dihydrate  (\figref{fig:ethenz}(g), \CCDC{LIFNOE}), a network of hydrogen bonds is formed between acetamide groups and water molecules. In azidoacetamide  (\figref{fig:ethenz}(f), \CCDC{XAVZUP}), we see an acetamide homosynthon formed at an offset so that the azide group can stack directly above the NH$\cdots$O interaction. We see weaker interactions in tetrazole-5-carboxamide (\figref{fig:ethenz}(h), \CCDC{KEPDIU}), where the acetamide group is interacting with the azole group, which, when compared with triazole-5-carboxaldehyde (\figref{fig:ethenz}(i)), demonstrates the range of acetamide-azole synthon binding. Finally, in 1-Methoxyaziridine-2,2-dicarboxamide (\figref{fig:ethenz}(d), \CCDC{AJEREM}), despite multiple acetamide interactions, we see a weakened acetamide network, likely due to the geometry of the molecule itself.

We do not suggest that these molecules could be used directly as co-formers; the train set was obtained with diversity as the primary goal, with no regard for availability, toxicity, ease of synthesis, or stability. Instead, each of these related and unrelated crystals gives insight into the types of interactions that may beget new ethenzamide co-crystals. The molecules shown in \figref{fig:ethenz} can be used as inspiration to identify co-former candidates from libraries of biocompatible compounds and to guide future crystallization studies.

\section{Conclusions}

Molecular crystallization is a complex, multi-faceted process, that poses tremendous challenges to both quantitative modeling, and to the derivation of qualitative design principles. 
In this work we propose a data-driven strategy to build a database of the interaction motifs that are found in a diverse set of molecular crystals, to determine semi-quantitatively their contribution to the lattice energy, and to generate a library of molecular motifs that can be used to interpret the stability of known crystals and to assist the design of new ones.

In doing so, we have to strike a balance between several conflicting goals. 
By using a dataset that is constructed by selecting structures from the CSD while maximizing their structural diversity we ensure that we cover a broad range of chemical and packing motifs, while remaining focused on structures that are known to be experimentally realizable. 
By using a general-purpose, atom-centered structural representation that is capable of describing arbitrary structural correlations, we ensure that our data analysis is flexible and that it does not incorporate pre-conceived notions about molecular bonding. At the same time, we ensure that the model focuses on the features that are most relevant to determine crystal stability by building a remnant descriptor that mimics the definition of the lattice energy as a difference between the total energies of the crystal and its constituents. 

The resulting models achieve a respectable mean absolute error of about 0.4\unitz{} in predicting the atomic contributions to crystal stability using these descriptors that gives us a semi-quantitative estimate of the contribution of each atomic environment to the lattice energy and to compare between different co-crystals or between polymorphs that are stable at very different conditions.
In order to translate these atomic contributions in a language that can be useful to crystal chemistry, we then assemble them to estimate the stabilizing power of traditional chemical groups (carboxylic acids, amines, ...) and build data-driven maps that facilitate the comparison of different chemical environments, by expressing the greatest amount of structural variability and simultaneously the best correlation with the lattice energy contribution. 
For each chemical moiety we provide an interactive map (on Materials Cloud\cite{matcloud_entry}) that allows to juxtapose different types of crystal environments, to identify structural patterns that are either stabilizing or destabilizing, and to contrast them with conventional motifs (e.g. hydrogen-bonding), demonstrated here for a few selected cases. With these tools, we aim to guide those designing molecular co-crystals in identifying suitable co-formers, which we demonstrate for the analgesic ethenzamide. As we demonstrate for phenanthrene, it is also possible to use these maps to compare polymorphs of the same molecule, and to analyze molecular motifs for a structure that is not part of our original reference set.

We hope that this library of molecular motifs will prove useful to applications to specific crystal-design problems. More broadly, we believe that the general ML protocol that we follow, combining regression of the ultimate target property with unsupervised analysis of molecular motifs, can inspire similar applications to the study of other classes of materials, ranging from metal and covalent organic frameworks to self-assembled monolayers and biological systems.

\section{Author Contributions}
RKC and MC designed the study and wrote the manuscript. 
RKC computed the molecular energies and geometries, built the machine learning models, and designed the figures.
MP separated the crystals into molecular components, screened the dataset before relaxation calculations, started the molecular energy calculations, and edited the manuscript.
EAE advised on the dataset provenance and curation and edited the manuscript.
\section{Acknowledgments}

\noindent This project was funded by NCCR Marvel Inspire Fellowship (MP), NCCR Marvel (RKC \& MC), Trinity College (EAE), and ERC Grant 677013-HBMAP (RKC \& MC).

The authors would like to acknowledge Federico Giberti, Andrea Anelli, and Guillaume Fraux for fruitful conversations at the study's start and culmination.

\section*{Conflicts of interest}
There are no conflicts to declare.

\bibliographystyle{rsc}
\providecommand*{\mcitethebibliography}{\thebibliography}
\csname @ifundefined\endcsname{endmcitethebibliography}
{\let\endmcitethebibliography\endthebibliography}{}

  \begin{@twocolumnfalse}
      \includepdf[fitpaper, pages={-}, pagecommand={\thispagestyle{empty}\clearpage}]{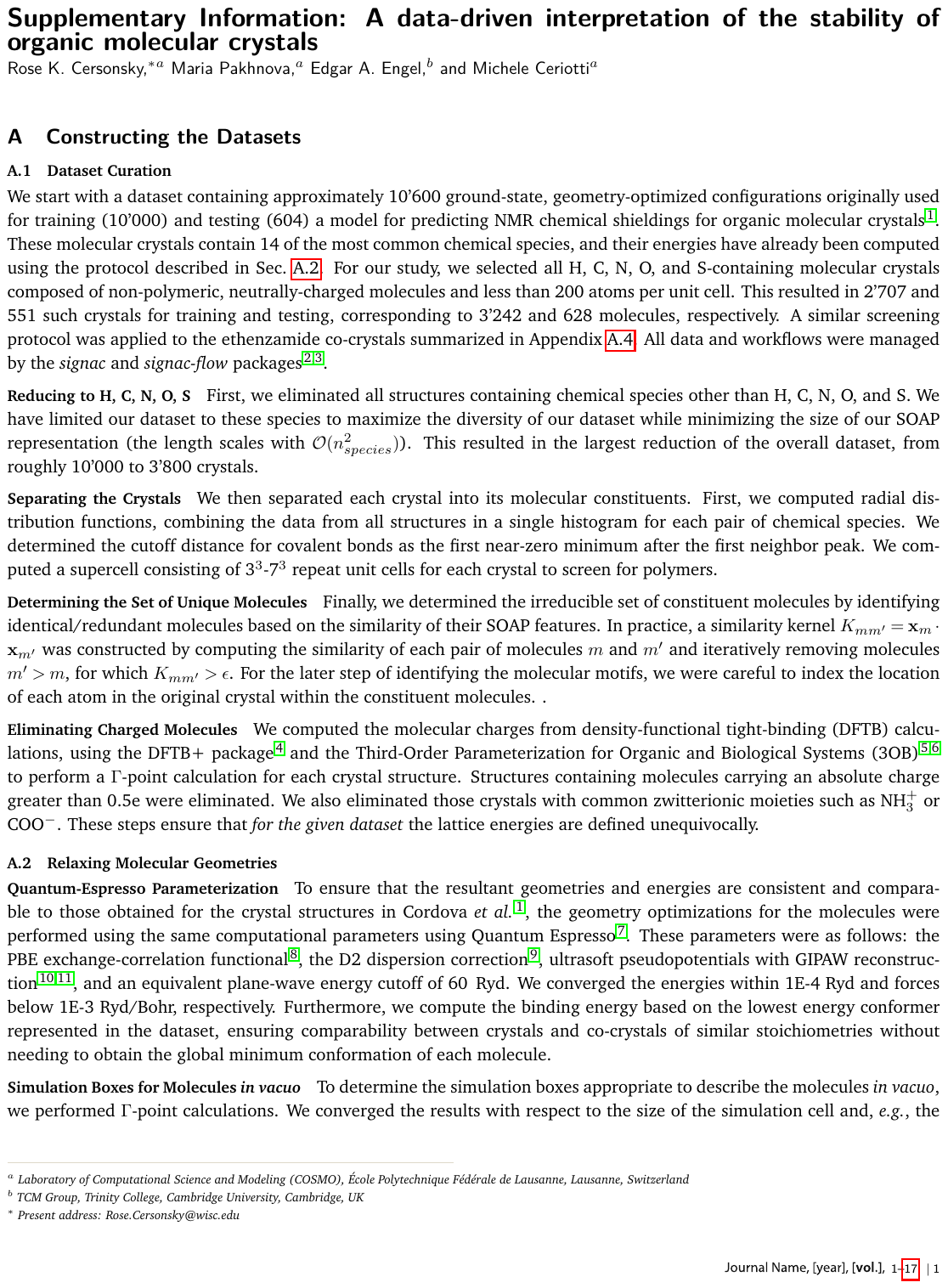}
 \end{@twocolumnfalse}
\end{document}